\documentclass[9pt,twocolumn,twoside]{pnas-new}

\setboolean{displaywatermark}{false} 
\fancypagestyle{firststyle}{}
\templatetype{pnasresearcharticle} 
\usepackage{hyperref}
\usepackage{siunitx}
\usepackage{mathtools}
\usepackage{xspace}
\usepackage{enumitem}

\usepackage{xr-hyper}
\makeatletter
\newcommand*{\addFileDependency}[1]{
  \typeout{(#1)}
  \@addtofilelist{#1}
  \IfFileExists{#1}{}{\typeout{No file #1.}}
}
\makeatother
\newcommand*{\myexternaldocument}[1]{%
    \externaldocument{#1}%
    \addFileDependency{#1.tex}%
    \addFileDependency{#1.aux}%
}
\myexternaldocument{SI}

\newcommand{\ttt}{\ensuremath{\Delta^{*}}\xspace}
\newcommand{\ttttr}{\ensuremath{\Delta^{*}_{\textrm{true}}}\xspace}
\newcommand{\tttpr}{\ensuremath{\Delta^{*}_{\textrm{pred}}}\xspace}
\newcommand{\ttrans}{\ensuremath{t^{*}}\xspace}

\begin{document}

\title{Interpretable Early Warnings using Machine Learning in an Online Game-experiment\thanks{Published in \textit{Proc.\ Natl.\ Acad.\ Sci.\ U.S.A.}\ \textbf{123}(1), e2503493122 (2026). \url{https://doi.org/10.1073/pnas.2503493122}}} 

\author[a,b,c,1,2]{Guillaume Falmagne}
\author[a,b,1]{Anna B. Stephenson}
\author[a,b]{Simon A. Levin}

\affil[a]{High Meadows Environmental Institute, Princeton University, Princeton, NJ 08544, United States}
\affil[b]{Department of Ecology and Evolutionary Biology, Princeton University, Princeton, NJ 08544, United States}
\affil[c]{Centre for Sociology Of Humans And Machines, Research Hubs, Technological University Dublin, Grangegorman Lower, Dublin 7, D07 C972, Ireland}

\leadauthor{Falmagne}

\authorcontributions{G.F., A.B.S., and S.A.L. designed the research. G.F. and A.B.S. analyzed data and contributed new analytic tools. G.F., A.B.S., and S.A.L. wrote the paper.}
\authordeclaration{The authors declare no competing interest.}
\equalauthors{\textsuperscript{1}These authors contributed equally to the work.}
\correspondingauthor{\textsuperscript{2}To whom correspondence may be addressed. E-mail: guillaume.falmagne@TUDublin.ie}

\begin{abstract}

Stemming from physics and later applied to other fields such as ecology, the theory of critical transitions suggests that some regime shifts are preceded by statistical early warning signals. 
Reddit's r/place experiment, a large-scale social game, provides a unique opportunity to test these signals consistently across thousands of subsystems undergoing critical transitions. In r/place, millions of users collaboratively created \textit{compositions}, or pixel-art drawings, in which transitions occur when one composition rapidly replaces another. We develop a machine-learning-based early warning system that combines the predictive power of multiple system-specific time series via gradient-boosted decision trees with memory-retaining features. Our method significantly outperforms standard early warning indicators. Trained on the 2022 r/place data, our algorithm detects half of the transitions occurring within 20 minutes at a false positive rate of just 3.6\%. Its performance remains robust when tested on the 2023 r/place event, demonstrating generalizability across different contexts. 
Using SHapley Additive exPlanations (SHAP) for interpreting the predictions, we investigate the underlying drivers of warnings, which could be relevant to other complex systems, especially online social systems. We reveal an interplay of patterns preceding transitions, such as critical slowing down or speeding up, a lack of innovation or coordination, turbulent histories, and a lack of image complexity. These findings show the potential of machine learning indicators in socio-ecological systems for predicting regime shifts and understanding their dynamics.
\end{abstract}

\dates{This manuscript was compiled on \today}
\doi{\url{www.pnas.org/cgi/doi/10.1073/pnas.XXXXXXXXXX}}

\maketitle

\thispagestyle{firststyle}

\ifthenelse{\boolean{shortarticle}}{\ifthenelse{\boolean{singlecolumn}}{\abscontentformatted}{\abscontent}}{}





\dropcap{A}brupt transitions from one state to another are modeled as critical transitions, bifurcations, tipping points, or regime shifts. In social and ecological systems, shifts such as fishery collapses~\cite{jacksonHistoricalOverfishingRecent2001, petrieStructureStabilityExploited2009}, disease outbreaks~\cite{lagorioQuarantinegeneratedPhaseTransition2011, orozco-fuentesEarlyWarningSignals2019}, stock market crashes~\cite{vandewalleCrashOctober19871998}, or the tipping of Earth subsystems~\cite{armstrongmckayExceeding15degCGlobal2022} directly impact people’s lives. Detecting early signs of critical transitions could enable the relevant actors to prepare for, mitigate, or even avoid such transitions. Yet, finding robust early warning signals is no small task. Empirical data on many transitions in a system is needed to validate models of transitions or learn the patterns preceding them. 

Online social systems can offer such detailed observations of many subsystems with intercomparable complex dynamics, and one platform particularly well-suited to studying critical transitions is Reddit's r/place~\cite{Place2024}. In this event, which was part game, part social experiment, millions of users created collaborative art by changing the colors of pixels on a 
shared canvas, inspiring a host of new research to understand the resulting collective behaviors~\cite{armstrongCoordinationPeerProduction2018, rappazLatentStructureCollaboration2018, vachherUnderstandingCommunityLevelConflicts2020, litherlandInstructionVsEmergence2021, israeliFlyingColorsPredicting2023,mullerCompressionCulturalEvolution2018, wuLargescaleCollectiveDynamics2024, botelhoArtExpandedField2024}. On the canvas, conflict broke out as users fought over the limited space,
making r/place analogous to a socio-ecological system. Groups created \textit{compositions}\footnote{In other works discussing r/place, compositions, pixels, and pixel changes are sometimes called artworks, tiles, and updates.}---images representing the interests of their community---sometimes by attacking other compositions to rapidly replace their pixels, which we regard as transitions. In this work, we build a warning system for transitions in the social world of r/place, providing a new framework for predicting transitions and understanding their dynamics in large-scale complex systems. 

Critical slowing down, the slowing return of the system’s state after perturbations, has been widely investigated as a broadly applicable early warning signal~\cite{wisselUniversalLawCharacteristic1984, vannesSlowRecoveryPerturbations2007, schefferEarlywarningSignalsCritical2009, boettigerEarlyWarningSignals2013, boettigerPatternsPredictions2013}. 
Critical slowing down was shown in ecological experiments more than a decade ago~\cite{carpenterEarlyWarningsRegime2011, daiGenericIndicatorsLoss2012, dakosMethodsDetectingEarly2012}, and 
in theory decades earlier~\cite{wisselUniversalLawCharacteristic1984}. However, its theoretical grounds require the state to closely follow an equilibrium that moves due to a slow parameter change, such as in bifurcations~\cite{schefferEarlywarningSignalsCritical2009}, spinodal instabilities~\cite{levinPhaseTransitionsTheory2023}, continuous or second-order
phase transitions, and other reversible, non-catastrophic transitions~\cite{kefiEarlyWarningSignals2013}. 
Critical slowing down may not apply to transitions triggered by a fast parameter change, large perturbations~\cite{boettigerNoEarlyWarning2013}, or rate-induced tipping~\cite{ritchieEarlywarningIndicatorsRateinduced2016}; the opposite signal, critical speeding up, even appears in some systems~\cite{titusCriticalSpeedingEarly2020, pomeauCriticalSpeedvsCritical2011, gietkaSqueezingCriticalSpeeding2022}. 
Although other indicators have been proposed (see recent review in Ref.~\citenum{georgeEarlyWarningSignals2023}), notably based on network~\cite{wunderlingHowMotifsCondition2020,suweisEarlyWarningSigns2014}, dynamical~\cite{xuNonequilibriumEarlywarningSignals2023}, and spatial ~\cite{chenEigenvaluesCovarianceMatrix2019} properties, different warnings have been shown to precede different types of transitions. For some transitions, however, there may be no warnings at all~\cite{boettigerEarlyWarningSignals2013,schefferAnticipatingCriticalTransitions2012}, including when long transient dynamics take place before the transition~\cite{hastingsTransientPhenomenaEcology2018}. In the absence of a data-validated model, an empirical system cannot be assumed to show any generic warning. Since building empirically sound models is challenging for many complex systems, model-free methods to find early warning indicators are greatly needed.

Early warning indicators leveraging machine learning methods~\cite{georgeEarlyWarningSignals2023, 
obrienEWSmethodsPackageForecast2023} are model-independent and have been shown to outperform generic indicators in some systems~\cite{choiEarlyWarningCritical2022}. Because real-world systems often lack the large amount of data needed for training, some researchers have trained on simulated data~\cite{qiUsingMachineLearning2020, brettDynamicalFootprintsEnable2020, grassiaMachineLearningDismantling2021, buryDeepLearningEarly2021, debMachineLearningMethods2022, dylewskyUniversalEarlyWarning2023,miryDeepLearningDisease2025}. 
However, algorithms using simulated data can only discover predictive features that are embedded in the chosen simulated models, which may not capture all the relevant warning properties of a real-world transition. 
Models trained on synthetic data and then fine-tuned to empirical data might identify more warning signals~\cite{liuEarlyPredictorOnset2024}, but could still overlook features absent in the simulations and too subtle in the sparse tuning data.
Another approach is to find systems that provide large amounts of observational data to train on, which has been successful in medicine~\cite{tapakComparativeEvaluationTime2019, gaoMachineLearningBased2020, hylandEarlyPredictionCirculatory2020, kobylarzribeiroMachineLearningEarly2020}, economics ~\cite{samitasMachineLearningEarly2020, barthelemyEarlyWarningSystem2024} and engineering~\cite{maDatadrivenPowerSystem2018, lassetterUsingCriticalSlowing2021}. 
Nevertheless, techniques that train on large real-world datasets have not yet been applied to predict transitions in socio-ecological systems. 
Moreover, a standard issue with machine learning predictions is that they rarely provide more understanding of the system to experts or decision makers~\cite{rudinStopExplainingBlack2019}. This ``black-box'' aspect is not addressed in most studies predicting transitions with machine learning, which focus on the performance of the predictions at the expense of insights into the determinants of the transition.

We use gradient-boosted decision trees to predict the time remaining until impending transitions in Reddit’s r/place; this time is set to 12 h when there is no incoming transition or it is far in the future. Critically, we analyze the resulting warnings with SHapley Additive exPlanations (SHAP)~\cite{lundbergUnifiedApproachInterpreting2017} to understand the mechanisms that drive these transitions. Our warning system predicts transitions well, outperforming each of the generic signals we tested using thresholding and Kendall's $\tau$~\cite{chenPracticalGuideUsing2022}. Training on time series of system-specific variables with a 7-hour memory for thousands of compositions, we obtain predictive power up to a few hours before transition. Our predictions are still significant when training on the 2022 event and testing on the 2023 event, which demonstrates that the algorithm exploits genuine and generalizable properties. 
By examining how different input features contribute to the predictions, we highlight a complex mixture of dynamical properties of compositions near transition, including critical slowing down or speeding up, a lack of innovation or coordination, a turbulent past, and image simplicity. To our knowledge, this analysis provides the first machine learning predictions of critical transitions using detailed observational data in a large-scale complex social system. It reveals drivers of transitions and showcases a framework for building and understanding data-driven warning signals that could be deployed in other socio-ecological scenarios. 

\begin{figure*}[hbtp]
\centering
\includegraphics{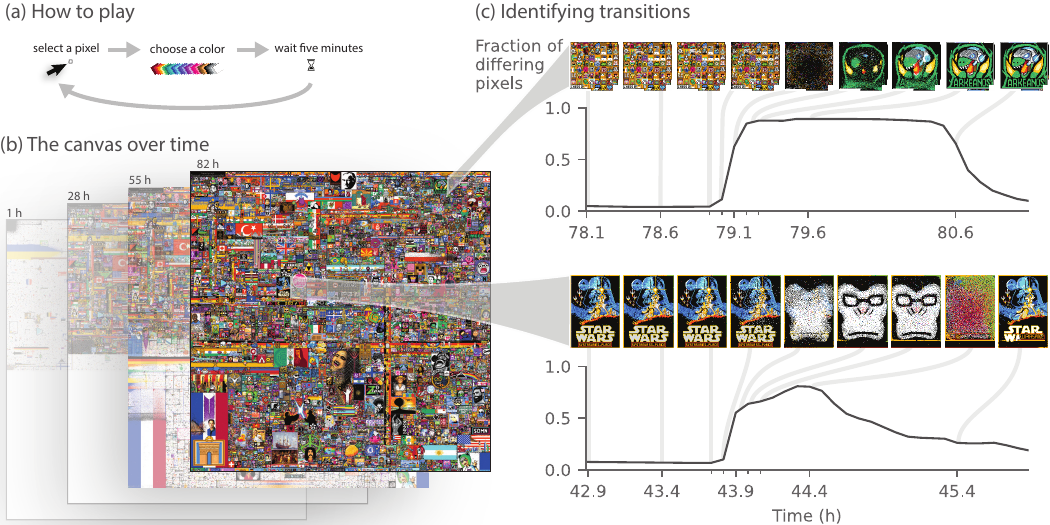}
\caption{Description of the r/place game, its compositions, and the transitions they undergo. \textbf{(a)} The rules of the game for a given user. \textbf{(b)} Snapshots of the full 2022 canvas at multiple points in time; some parts of this canvas were available only later in the game. \textbf{(c)} Fraction of pixels differing from the reference image (\texttt{diff pixels reference}) for the ``Chessboard'' and ``Star Wars: Episode IV -- A New Hope'' compositions as they undergo transition. Insets show snapshots of the compositions at different points in time. Time is measured from the beginning of the event.}
\label{fig:rplace}
\end{figure*}

\matmethods{
\label{section:trans-and-vars}

\begin{figure}[htbp]
\centering
\includegraphics{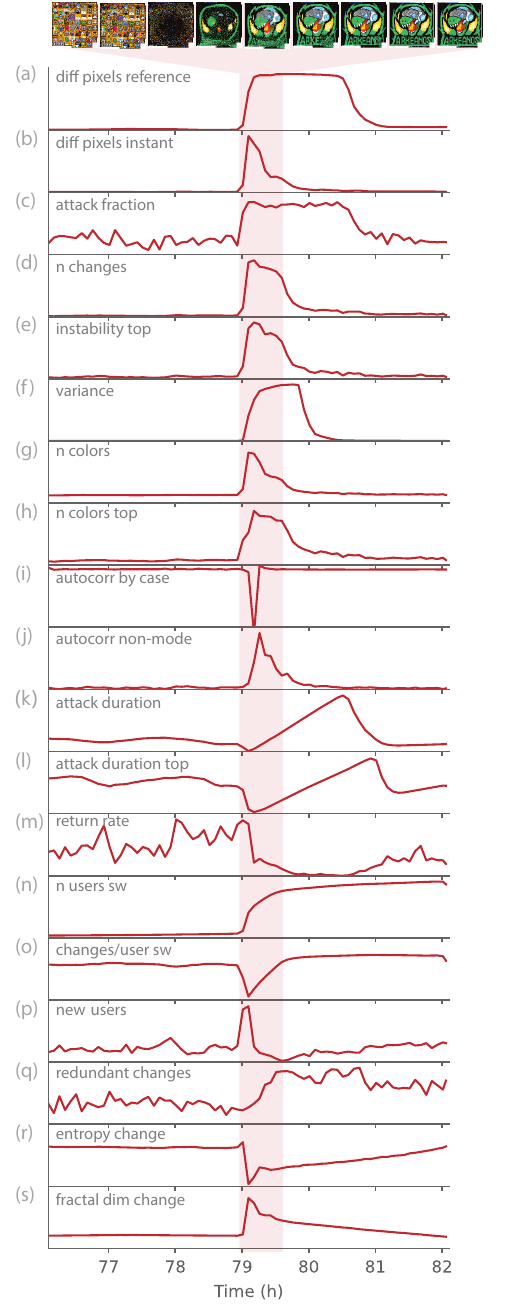}
\caption{\textbf{(a-s)} The time-dependent variables used in the training of the algorithm, for the  ``Chessboard'' composition. See text 
for explanations of all variables.}
\label{fig:ews-vars}
\end{figure}

\subsection*{The r/place game-experiment}

To predict transitions, we choose r/place~\cite{Place2024} as our study system, as it provides data on complex interactions of millions of participants. Presented as a game without a specific goal, r/place allowed registered Reddit users to select any pixel on a canvas, choose a new color from a palette, and then wait for a 5-minute \textit{cooldown} before changing another pixel (see Fig.~\ref{fig:rplace}a). We use the term \textit{event} to refer to each r/place game-experiment that Reddit hosted, which happened in 2017, 2022 and 2023. Each event lasted several days and followed the same basic rules. The striking complexity of the resulting collective art is evident in the snapshots of the full 2022 canvas shown in Fig.~\ref{fig:rplace}b. Various aspects of the 2017 experiment have been studied~\cite{armstrongCoordinationPeerProduction2018, mullerCompressionCulturalEvolution2018, rappazLatentStructureCollaboration2018, vachherUnderstandingCommunityLevelConflicts2020, litherlandInstructionVsEmergence2021, israeliFlyingColorsPredicting2023, wuLargescaleCollectiveDynamics2024, botelhoArtExpandedField2024}, while the 2022 and 2023 events have been discussed only once~\cite{wuLargescaleCollectiveDynamics2024}. 

We use the full data of pixel changes publicly released by Reddit, which includes anonymized user identifiers for each pixel change, for the 2022 and 2023 events. The 2022 event welcomed 10.4 million participants who changed pixel colors 160 million times over 3.5 days, while the 2023 event saw 8.6 million participants making 134 million changes over 5.4 days. We use the 2022 event as our primary system because it was the most popular, with ten times more participants than in 2017, and we use the 2023 event to test the generality of our predictions. The canvas started with 1 million pixels and 16 color choices (8 in 2023) and was later expanded in discrete steps to 4 million pixels (6 million in 2023) with 32 color choices---see SI Table~\ref{tabSI:extensions}. Detailed rules, statistics, and data provenance on the r/place events are given in Section~\ref{secSI:rplacedata} of the Supporting Information (SI), along with global canvas distributions in Fig.~\ref{figSI:rplace}.

In 2022, users organized to create about 10,900 compositions on the canvas, compared to 6,700 in 2023. These culturally meaningful drawings were often made by communities that previously existed on Reddit or other social media; some also spontaneously formed for the occasion.
We identify these compositions using the Atlas dataset~\cite{haagmansPlaceAtlasInitiative2024}. This crowdsourced map of the canvas specifies, for each composition, the borders of the image, the birth and death times, the communities involved, and descriptions of the depicted content. We minimally clean the Atlas and separate some compositions so that each is continuous in time and in canvas space (see SI Section~\ref{secSI:compodata}).

\subsection*{State variable and reference image}
We define transitions in compositions using thresholds on a state variable, which must be carefully chosen. To this end, we first define concepts used throughout this text. The \textit{sliding window} is the 3-hour window preceding a certain time point. The \textit{mode color} is the color that held the longest over a time period, either a time step or the sliding window. At any given time, the \textit{reference image} contains, for each pixel within the composition borders, the \textit{reference color}, which is the mode color over the sliding window. 
The state variable is then the \textit{fraction of pixels differing from the reference image}; it indicates recent changes, or \textit{attacks}, to the ``usual'' image at any time point. \textit{Defense} changes restore attacked pixels to the reference color.

\subsection*{Definition of transitions} 
Our definition of a transition in a composition represents what a user perceives as a sudden change in its image, which translates into a high fraction of differing pixels. Therefore, we define the transition time \ttrans as the moment this state variable both exceeds 0.35 and is 6 times higher than its average value over the sliding window. Changing these parameters, including the sliding window length, does not substantially affect our performance, which we show in a sensitivity analysis (SI Section~\ref{secSI:sensitivity} and Fig.~\ref{figSI:sensitivity}a). These thresholds ensure that a large portion of the image changes, and that this portion is much larger than usual. The relative threshold of 6 selects for sudden transitions, which are harder to predict and of more interest to the warning signals literature, while requiring a stable system to transition from. We ran an alternative training with only a relative threshold of 2: this effectively includes smoother transitions, and performance is increased (see Fig.~\ref{figSI:sensitivity}c), but these transitions are harder to interpret and less relevant to real-world warnings. We also reject transitions that start from a patchwork of compositions (see examples in Fig.~\ref{figSI:notrans}). Because these transitions do not start from a single, coherent composition, this region of the canvas would not be monitored and defended before the new composition exists. Valid transitions include those where the composition returns to its pre-transition state as well as those where a new image takes over (see Fig.~\ref{fig:rplace}c for examples of each). More context and justification for how we define transitions is given in SI Section~\ref{secSI:transitions}. Examples of compositions whose evolving images may resemble transitions but do not pass our criteria are shown in Fig.~\ref{figSI:notrans}.

\subsection*{Time series for each composition} 
\phantomsection 
\label{subsec:vars}
To construct the training data for our algorithm to predict transitions, we first calculate variables that reveal how compositions evolve. Each variable's time series captures different properties of the composition that we expect to have predictive power, including the dynamics of the image, pixel changes, and user activity. 
We compute each variable for each composition in 5-minute increments---a natural time scale considering the game's 5-minute cooldown rule---resulting in about 1000 time steps in 2022, and 1500 in 2023. Most variables are averaged over the time-step duration. The variables related to a per-pixel quantity are averaged over all active pixels, apart from those which we average over the decile of pixels with the highest values. We denote $t$ and $t-1$ as the current and previous time steps. The variables, plotted for an example transition in Fig.~\ref{fig:ews-vars}, are:

\setlist[description]{font=\normalfont\ttfamily}
\begin{description}[noitemsep]
    \item[diff pixels reference] (Fig.~\ref{fig:ews-vars}a). Fraction of pixels differing between the instantaneous image at $t$ and the reference image, as described above.
    \item[diff pixels instant] (Fig.~\ref{fig:ews-vars}b). Fraction of pixels differing between the instantaneous image at $t$ and the image composed of the mode colors in step $t-1$.
    \item[attack fraction] (Fig.~\ref{fig:ews-vars}c). Fraction of  pixel changes to a color different from the reference. 
    \item[n changes] (Fig.~\ref{fig:ews-vars}d). Number of pixel changes.
    \item[instability top] (Fig.~\ref{fig:ews-vars}e). Fraction of this time step spent in a non-mode color, averaged over the top decile of pixels.
    \item[variance] (Fig.~\ref{fig:ews-vars}f). $\frac{1}{2(n-1)}\sum_{i=0}^{n-1} D(t-i,\,\, t-i-1)^2$ where $n=10$ 
    time steps 
    and $D(t_1,t_2)$ is the fraction of pixels differing between the instantaneous images at times $t_1$ and $t_2$. This equals the standard variance
    in the gaussian and large-statistics limit. 
    \item[n colors] and \texttt{n colors top} (Fig.~\ref{fig:ews-vars}g,h). Number of colors used in a pixel during this time step, averaged over all pixels or over the top decile.
    \item[autocorr by case] (Fig.~\ref{fig:ews-vars}i). The relative change in the autocorrelation between $t$ and $t-1$, defined case-by-case with positive
    values when pixels change from the reference to the same color, and negative values when pixels change to different colors. 
    \item[autocorr non-mode] (Fig.~\ref{fig:ews-vars}j). The autocorrelation $\sum_{\textrm{colors}} c_t \, c_{t-1}$ where $c_{t}$ is the time spent in color $c$ in step $t$, except we set $c_{t}=0$ for the mode color of step $t$.
    \item[attack duration] and \texttt{attack duration top} (Fig.~\ref{fig:ews-vars}k,l). Time since the attack of a pixel currently in a non-reference color or duration of the most recent attack of a pixel, averaged over pixels attacked within the sliding window and over the top decile.
    \item[return rate] (Fig.~\ref{fig:ews-vars}m). Fraction of pixels attacked at $t-1$ that returned to the reference at $t$.
    \item[n users sw] (Fig.~\ref{fig:ews-vars}n). Number of users who made changes in the sliding window. 
    \item[changes/user sw] (Fig.~\ref{fig:ews-vars}o). Number of changes per user in the sliding window.
    \item[new users] (Fig.~\ref{fig:ews-vars}p). Fraction of users in step $t$ that were not active in the preceding sliding window.
    \item[redundant changes] (Fig.~\ref{fig:ews-vars}q). Fraction of pixel changes to the color that the pixel was already in. Users likely performed these changes to have their username attached to this pixel.
    \item[entropy change] (Fig.~\ref{fig:ews-vars}r). Change relative to the sliding window of an entropy proxy, the ratio of the compressed file size of the image to the image area, as introduced in Ref.~\citenum{martinianiQuantifyingHiddenOrder2019}. 
    \item[fractal dim change] (Fig.~\ref{fig:ews-vars}s). Change relative to the sliding window of the fractal dimension, calculated using the reticular box counting method separated by color as described in Ref.~\citenum{bisoiCalculationFractalDimension2001}.
\end{description}

We also define five variables that are trivially time-dependent for a given composition:

\begin{description}[noitemsep]
    \item[area.] The logarithm of the number of active pixels.
    \item[age.] The time from the birth of the composition in the Atlas to $t$.
    \item[canvas quadrant.] Which of the 3 (7 in 2023) canvas extensions the composition is placed in.
    \item[entropy.] The entropy averaged over the sliding window.
    \item[border corner center.] Desirability of locations on the canvas: $1$ for a composition close to a border or a center of the available canvas, $2$ for one close to a corner, $0$ otherwise.
\end{description}

Details, justifications and normalizations for the above variables are in SI Section~\ref{secSI:variables}. Note that some variables explicitly relate to critical slowing down, such as the variance, autocorrelation, return rate, and duration of last attack; they are used to assess the presence of this generic signal. 

\label{sec:extendedvars} 
Large correlations make some variables redundant in the training, which is a criterion to discard a number of variables (see SI Section~\ref{secSI:featureselec}). The correlation matrix of all the variables used in the training is shown in Fig.~\ref{figSI:correlations}. The variables that were implemented but rejected from the final training dataset are also described in SI Section~\ref{secSI:variables}. A subset of these variables is related to additional spatial properties of the image dynamics: \texttt{levenshtein complexity}, \texttt{multiscale complexity}, and those described in SI Section~\ref{secSI:spatialvars} (SI Table~\ref{tabSI:additionalvariables} after pruning); these variables are included in an extended set of variables on which we train the algorithm. While these variables individually have predictive power (see Fig.~\ref{figSI:shapdistr}b and c), including them does not improve on the total performance of the nominal training, even after pruning the added variables as detailed in SI Section~\ref{secSI:featureselec} (see Fig.~\ref{figSI:sensitivity}b). Since they also add complexity---and thus decrease interpretability---these variables are excluded from the training.

\subsection*{Machine learning warning system}

\begin{figure*}[htbp]
\vspace*{-3mm} 
\centering
\includegraphics{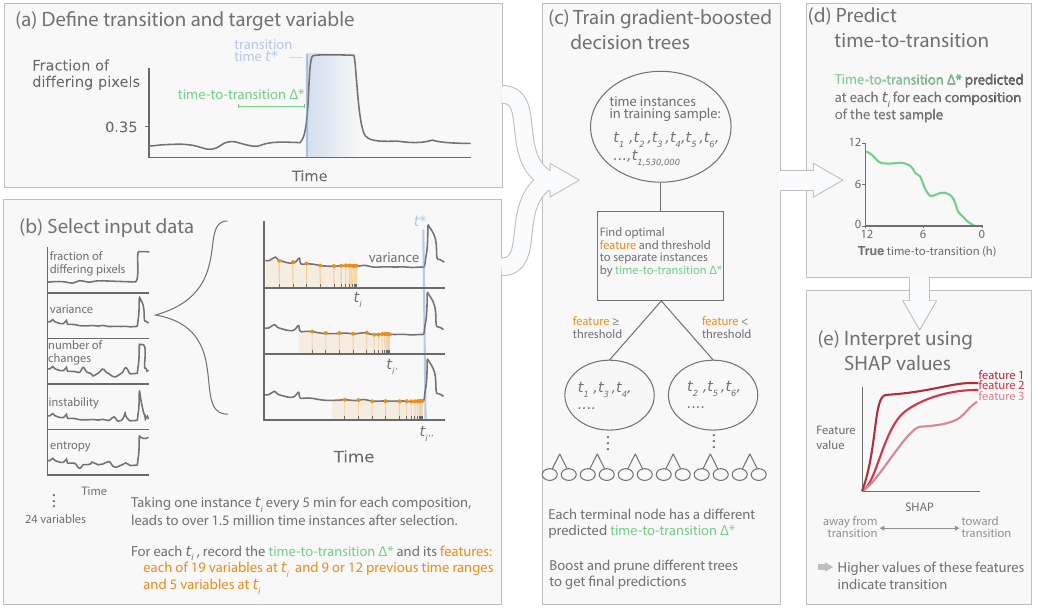}
\caption{Workflow of our machine learning warning system. \textbf{(a)} Transitions at time $t^*$ and the associated target variable time-to-transition (\ttt) are identified based on changes in \texttt{diff pixels reference}. \textbf{(b)} The features for each time instance consist of each of 19 input variables recorded over 9 or 12 time ranges of a \SI{7}{\hour} memory, as well as 5 variables without memory. \textbf{(c)} Gradient-boosted decision trees are trained to predict the time-to-transition. \textbf{(d)} Predicted and true values of the time-to-transition are compared in the test sample. \textbf{(e)} The drivers of predictions are analyzed based on SHAP values at a given feature value.
}
\label{fig:ml-algo}
\end{figure*}

We build our warning system by training gradient-boosted decision trees with XGBoost~\cite{chenXGBoostScalableTree2016} to  predict the time-to-transition, $\ttt = t^* - t$. Our training data consists of 1.53 million  \textit{time instances}, each containing \textit{feature} values that provide predictive information for a given time step and composition. These features consist of the variables listed in the previous section, computed at an instance's time step or averaged over preceding time periods to emulate a 7-hour memory. 

\paragraph{Selection of time instances.} 

We select time instances to avoid spurious signals. In early versions of this work, we discovered that the transition requirement of a relative threshold on \texttt{diff pixels reference} caused the algorithm to exploit stable pre-transition periods for prediction, rather than using genuine warning signals. Transitions can only be defined on a system in a stable or quasi-stable state, meaning this relative threshold is necessary. To prevent the optimization for stable periods, we require the past sliding-window average of \texttt{diff pixels reference} to be less than $0.35/6$ for all time instances. The effect of removing this filter on the performance is shown in the ``no stable times'' curve of Fig.~\ref{figSI:sensitivity}c. This reduces the size of our input dataset by a third. Other requirements on time instances are explained in SI Section~\ref{secSI:filter}. Of the 14,160 compositions in 2022, only 6,289 have at least one time instance meeting our requirements, compared to 3,546 compositions out of 6,930 in 2023.

\paragraph{Target value: time-to-transition.}

The time-to-transition, or \ttt, which we aim to predict, is the time remaining until a transition, and is recorded for each instance (Fig.~\ref{fig:ml-algo}a). Accurate predictions are not expected beyond a few hours before transition; we therefore modify the time-to-transition to favor times close to transition. Specifically, we: (a) set $\ttt=$~\SI{12}{\hour} for all instances with $\ttt>$~\SI{12}{\hour} or in compositions with no transition; (b) assign progressively lower weights when $\ttt>$~\SI{3}{\hour}; (c) apply a logarithmic transform to \ttt (see SI Section~\ref{secSI:target} on these three items); (d) use a training loss term in the objective function based on the \textit{relative} rather than the \textit{absolute} deviation between the predicted and true targets (see SI Section~\ref{secSI:algo}). 

\paragraph{Feature building.}

To account for past dynamics, each of 19 time series is recorded with a 7-hour memory for each time instance. This memory consists of 9 or 12 features (see SI Section~\ref{secSI:memory} for explanation of the number of time features) containing the variable averaged over adjacent time ranges preceding a time instance (see Fig.~\ref{fig:ml-algo}b and SI Section~\ref{secSI:memory}). The five trivially time-dependent variables are recorded as features without memory. Variables used for training are summarized in SI Table~\ref{tabSI:variables}, and the exact time ranges of the memory are listed in SI Table~\ref{tabSI:timeranges}.

\paragraph{Training the model.}

The resulting input data contains 175 features and one target value for 1.53 million time instances in 2022 and 1.34 million in 2023. When using 2022 data only, we attribute 80\% of the compositions, with all their instances, to the training dataset and the rest to the test dataset. We also separately train on the full 2022 data to test on the 2023 data. The algorithm outputs a predicted time-to-transition, $\tttpr$, for each instance of each test composition (Fig.~\ref{fig:ml-algo}d). The $\tttpr$ is then calibrated through a monotonic transformation, so that its average is closer to the true time-to-transition, $\ttttr$ (see SI Section~\ref{secSI:calib}). Details on the model hyperparameters are given in SI Section~\ref{secSI:algo}. 

\subsection*{Testing predictions in binary warning systems}

To compare our warning system to standard warning signals, we convert our predicted time-to-transition into a binary indicator of warning or no warning. A warning is issued if $\tttpr$ is within a set \textit{warning range} of, for example, $0$ to \SI{20}{\minute}. This warning is a true positive if \ttttr is below a certain threshold. We sweep over this threshold to create the receiver operator characteristic (ROC) and precision-recall (PR) curves, as explained in SI Section~\ref{secSI:ROCPR}. We calculate ROC and PR curves for four warning ranges: \SI{20}{\minute}, \SI{1}{\hour}, \SI{3}{\hour} and \SI{6}{\hour} (see Figs.~\ref{figSI:ROC}a and~\ref{figSI:PR}a). We also compute binary warning signals for individual transitions (see SI Section~\ref{secSI:testing} and Fig.~\ref{figSI:ROC}b).

We next build binary warnings based on the standard indicators of autocorrelation, return rate, and variance, using two methods. In the first, we use the Kendall's $\tau$~\cite{chenPracticalGuideUsing2022}, characterizing the increasing or decreasing trend of a time series in a $50$-minute ($10$ time step) time window. A warning is issued when $\tau$ reaches a threshold, which we sweep over to calculate the ROC and PR curves. In the second method, warnings are issued when the variable itself reaches the threshold. 

\subsection*{SHAP values to understand algorithm predictions}

SHAP values measure, for each instance, the contribution of a certain feature to the target value predicted by the algorithm~\cite{lundbergUnifiedApproachInterpreting2017}. The distributions of SHAP values over instances for each variable are shown in Fig.~\ref{figSI:shapdistr} (panel (a) for the nominal variable selection). When averaged over instances, SHAP absolute values measure feature importance. We use this along with correlations between the features values and correlations between their SHAP values to prune variables and features (SI Section~\ref{secSI:featureselec}).

Importantly, SHAP values describe how the algorithm relates a given instance and its feature values to the proximity of transition. Therefore, they can point to characteristics of data that the algorithm associates with incoming transitions. We examine average SHAP values as a function of feature values; these curves for all 175 features are shown in Fig.~\ref{figSI:SHAPall} and ~\ref{figSI:SHAPnotime}. Because some feature distributions are fat-tailed, we use the percentiles of feature values among instances. 
We interpret the trends of these curves as various warning indicators identified by the algorithm. To ensure that these trends genuinely relate to transition-like dynamics, we only report the trends that are robust against changes in the pre-transition stable period requirement (see SI Section~\ref{secSI:shap}). 

\begin{figure*}[h!]
\centering
\includegraphics{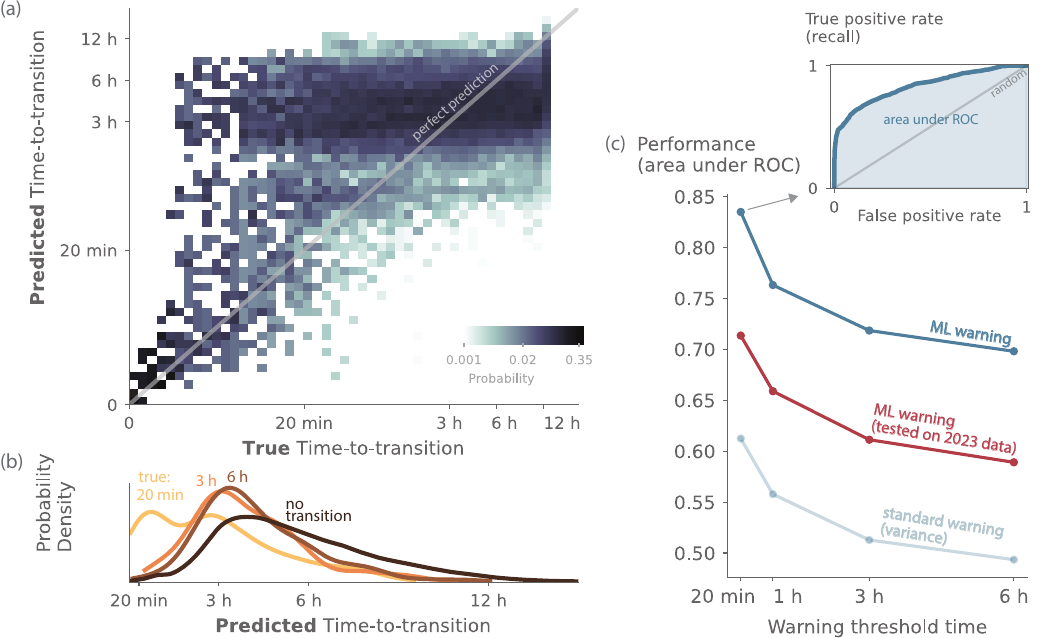}
\caption{Time-to-transition predictions. \textbf{(a)} Predicted time-to-transition \tttpr versus the true values \ttttr. The color shows the probability at a given \ttttr to predict a certain \tttpr value. To accommodate display on the log axis, \SI{100}{s} is added to all time-to-transition values. Perfect predictions would align with the grey line. 
\textbf{(b)} Probability distribution functions, constructed by kernel density estimation, of \tttpr at four different $\ttttr\pm $ \SI{5}{\minute} values. The darkest curve includes all instances for which there is no incoming transition. \textbf{(c)} ROC area under the curve (AUC) as a function of the warning range for the machine learning warning signal (dark blue), a single-variable standard warning signal using \texttt{variance} (light blue), and the machine learning warning signal tested on 2023 r/place data (red). The inset shows, as an example, the ROC curve used to compute the ROC AUC for the \SI{20}{\minute} warning range.}
\label{fig:true-v-pred}
\end{figure*}

} 
\showmatmethods{}

\section*{Results}


\subsection*{Predicted time-to-transition} 

Our trained XGBoost model shows predictive power for the time-to-transition \ttt up to hours ahead of the transition. For lower \ttttr values, predictions are closer to the perfect-prediction line (Fig.~\ref{fig:true-v-pred}a): at $\ttttr=\SI{20}{\minute}$, the mode prediction is \SI{44}{\minute}; at $\ttttr=\SI{3}{\hour}$, the mode is \SI{3}{\hour} \SI{6}{\minute}, and at $\ttttr=\SI{6}{\hour}$, it is \SI{3}{\hour} \SI{22}{\minute}. As \ttttr grows, a peak in the distribution of \tttpr emerges around \SIrange{3}{4}{\hour}, corresponding to a bulk of instances the algorithm cannot classify (Fig.~\ref{fig:true-v-pred}b). This peak is already evident for $\ttttr=\SI{20}{\minute}$. While perfect predictions would sharply peak at \SI{12}{\hour} for the ``no transition'' curve, the mode prediction is actually \SI{3}{\hour} \SI{56}{\minute}. Nonetheless, performance is determined by the ordering and separation of the predicted values, as predictions are scaled by calibration. The monotonic increase of prediction modes with increasing \ttttr up to the maximum value is remarkable, as it shows that the algorithm succeeds in discriminating times \SI{3}{\hour} from transition from times \SI{6}{\hour} away. 

\subsection*{Performance against standard warning signals} 

Our machine learning warning system trained on many variables far outperforms single-variable, standard warning indicators, as shown by the higher area under the curve (AUC) for the ROC and PR curves (see Figs.~\ref{fig:true-v-pred}c,~\ref{figSI:ROC}a, and~\ref{figSI:PR}). Compared to the Kendall's $\tau$ warning calculated using our \texttt{variance} variable, which was the best-performing of the standard warning variables we computed, our algorithm has drastically more predictive power for all warning ranges.

\subsection*{Testing on data from the following year}

To further evaluate the performance of our early warning system, we trained on the entire 2022 r/place dataset and tested on the 2023 one. The 2023 predictions still hold significant predictive power, decreasing only slightly in performance compared to the 2022 predictions, as shown in the AUC values of the ROC and PR curves (Figs.~\ref{fig:true-v-pred}c and~\ref{figSI:PR}b). 

\subsection*{Evaluation by composition and transition}
\begin{figure}[tbh]
\centering
\includegraphics{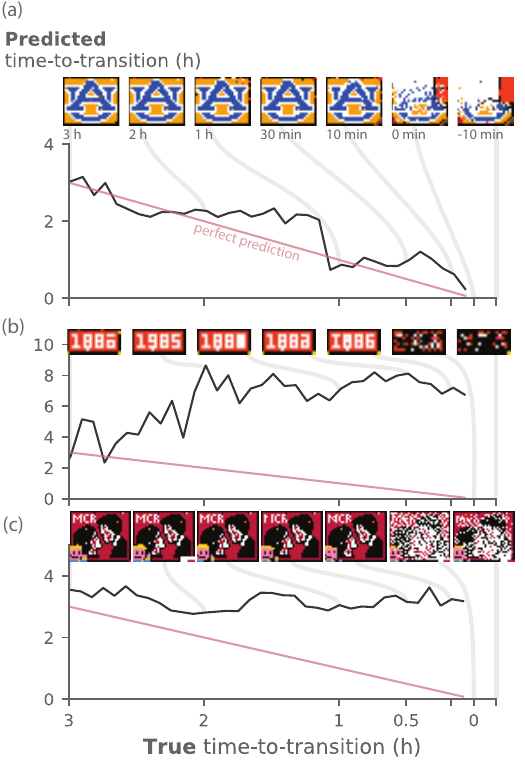}
\caption{Trajectories of predicted versus true time-to-transition for three compositions. The red line indicates a perfect prediction. The insets show the composition at different points in time. \textbf{(a)} The ``Auburn University'' composition gives predictions that trend downward and are occasionally close to the perfect-prediction line. \textbf{(b)} ``1886'' gives low-accuracy predictions with oscillations. \textbf{(c)} ``Three Cheers For Sweet Revenge'' gives low-accuracy predictions with relatively constant values.}
\label{fig:traj}
\end{figure}

The above evaluation results are computed over all time instances of all compositions. However, we expect our algorithm to perform better for some transitions than others. We examine the per-composition predictions by calculating a ROC AUC for each transition in the 2022 test set, as discussed in SI Section~\ref{secSI:percompo}. For a warning range of \SI{20}{\minute}, $44\%$ of the transitions have a ROC AUC greater than 0.8, compared to  $27\%$  at a range of \SI{1}{\hour} (Fig.~\ref{figSI:ROC}b). At the 20-minute range, $27$\% perform at or worse than random, compared to $41$\%  at a range of \SI{1}{\hour}. While our algorithm performs well on a significant portion of transitions, it is unable to predict some of them. 

For some compositions, our predictions provide useful warnings where predictions are reasonably accurate near transition (Fig.~\ref{fig:traj}a). For compositions where the algorithm performs poorly and \tttpr oscillates in time (Fig.~\ref{fig:traj}b), there may be continually high activity. When there are no detectable warning signals, \tttpr should remain relatively constant (Fig.~\ref{fig:traj}c).  

To avoid many subsequent warning alarms, which could be inconvenient for a real-world transition, we implement a system with a \textit{warning cooldown} (see SI Section~\ref{secSI:cooldownWarn}). Here, the system must ``cool down'' before issuing another warning, as in Ref.~\citenum{hylandEarlyPredictionCirculatory2020}. Requirements for such warnings are stricter, which inevitably leads to a lower performance for our cooldown warning system (see the ROC curve in Fig.~\ref{figSI:continuous-warn}).  

\subsection*{Interpretations of the warning system}

\begin{figure*}[tbh]
\centering
\includegraphics{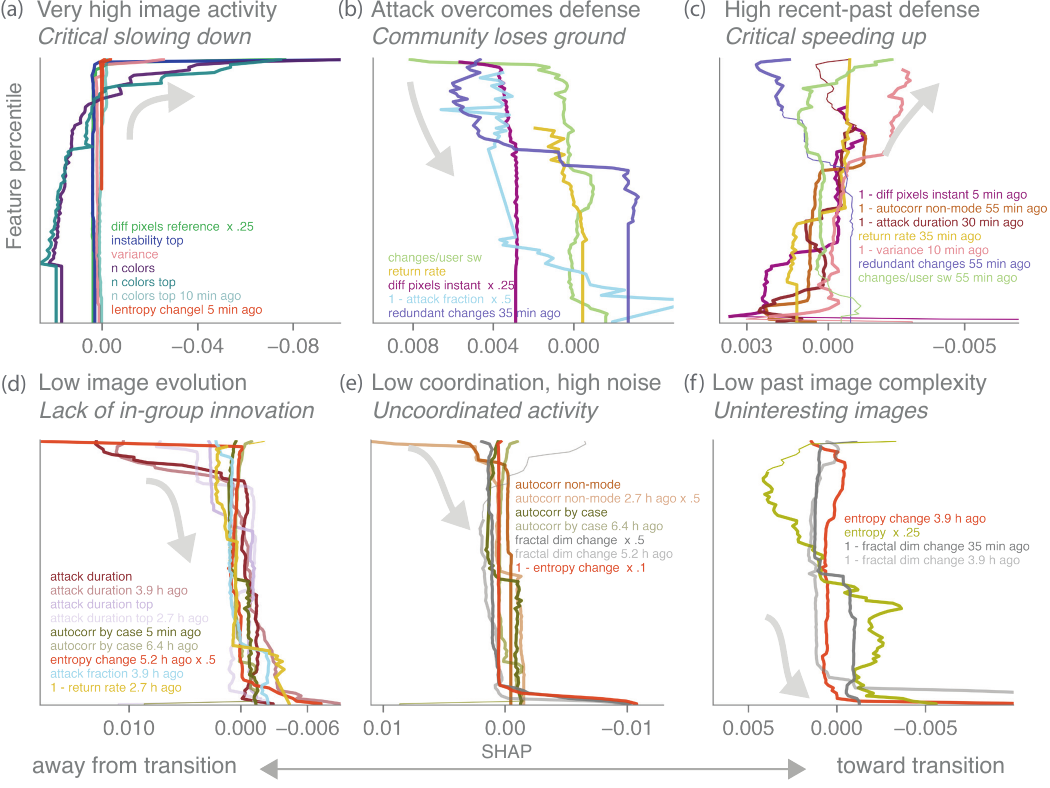}
\caption{Interpretation of model predictions with SHAP values. \textbf{(a-f)} Feature percentiles versus mean SHAP values, classified based on curve trends into six pre-transition behaviors, which are described in the panel titles. Top titles describe the signal of a coming transition, and italicized subtitles provide an interpretation of the associated dynamics of the users and of the image of the composition. Grey arrows indicate the qualitative trend of the curves of a panel. Each legend label describes the curve of the same color as the label's text. Curves are drawn thinner when they show a trend that is not the focus of their respective panel.
}
\label{fig:shap}
\end{figure*}

We analyze the algorithm predictions to uncover mechanisms driving transitions in r/place, which may be transferrable to other complex systems. This approach helps mitigate the often overlooked ``black-box'' nature of machine learning predictions. 

Uncovering mechanisms starts with examining the 175 SHAP trends as a function of feature value (see Figs.~\ref{figSI:SHAPall} and~\ref{figSI:SHAPnotime}). Our SHAP plots represent time progressing toward the right, so that the right side of the plot (low SHAP) corresponds to feature behaviors that are typical close to transitions. A \textit{low} SHAP therefore designates instances and features values that the algorithm associates with \textit{closeness} to transition; a \textit{high} SHAP (left side of the plot) shifts predictions \textit{away} from transition. By grouping features that describe the same  trend and aspect of dynamics, we build evidence for distinct behaviors which can be considered potential warning signals. To find these behaviors, we first classify the variables for their role in the composition dynamics: strength of the attack or defense changes, image activity, user activity and engagement, image complexity, innovation, and coordination versus noise (see SI Section~\ref{secSI:shap} for details on the classification). We uncover a set of 12 behaviors of compositions close to transition, shown in Figs.~\ref{fig:shap} and~\ref{figSI:SHAPinterp}. 

These behaviors span different aspects of activity and image dynamics and relate to recent or past features. First, a strong critical slowing down shows as a very high image activity (Fig.~\ref{fig:shap}a), as well as a dominating attack force against the defending community (Fig.~\ref{fig:shap}b) and a high user recruitment (Fig.~\ref{figSI:SHAPinterp}a). However, this slowing down coexists with a mild critical speeding up, reflected by high defense in the recent past (Fig.~\ref{fig:shap}c). 

Trends for past time features provide additional signals. A high past image activity suggests a large presence of past attacks (Fig.~\ref{figSI:SHAPinterp}b), which we refer to as a turbulent past. Past high user engagement (Fig.~\ref{figSI:SHAPinterp}c) could also indicate substantial past attacks that the defense reacted against. Though a turbulent or highly engaged past signals an approaching transition, low past activity can indicate a neglected composition, which could attract opportunistic attacks (Fig.~\ref{figSI:SHAPinterp}d). In contrast, a very low past image activity may indicate a general lack of interest (Fig.~\ref{figSI:SHAPinterp}e), which makes the composition safer.

Attacked pixels left in their new color can indicate evolution of the image supported by the community, which we refer to as innovation. A lack of innovation is indicative of transitions (Fig.~\ref{fig:shap}d), which may suggest that a more innovative community is safer from transition. Coordinated activity rather than noisy activity, whether from the defending community or not, also indicates stronger compositions (Fig.~\ref{fig:shap}e). Unsurprisingly, low defense activity in the past is
a transition signal (Fig.~\ref{figSI:SHAPinterp}f), which could mean that the defending community is unable or unmotivated to mobilize. Finally, images of higher complexity are also typically safer (Fig.~\ref{fig:shap}f).  

\section*{Discussion}
This work is, to our knowledge, the first machine-learning-based early warning indicator for transitions in a large-scale human social system. Some machine learning prediction systems using large empirical datasets have outperformed ours in fields such as medicine~\cite{gaoMachineLearningBased2020, hylandEarlyPredictionCirculatory2020}. However, social systems present unique challenges because human behavior depends on many interacting factors and lacks comprehensive datasets. Decision trees using the r/place dataset tackle both these challenges while providing data-driven insights on the complexity of social dynamics that analytical models cannot offer.

We extract human-readable warning signals from the SHAP analysis of predictions. Many dynamical behaviors, rather than single, generic early warning signals, are found to indicate incoming transitions. These behaviors might apply to transitions in other socio-ecological systems and inform their managers. Some of our findings confirm intuitions from other approaches; for example, innovation has been discussed as contributing to adaptation and sustainability transitions~\cite{makitieDigitalInnovationsContribution2023}, but also as being insufficient preceding the collapse of ancient societies~\cite{tainterCollapseComplexSocieties1988, mascarenoCriticalTransitionsEcosystems2022}. Closer to the theoretical origins of critical slowing down~\cite{levinPhaseTransitionsTheory2023, wisselUniversalLawCharacteristic1984}, the higher risk of transition for simpler images also evokes the reduction of dimensionality due to power law dynamics close to second-order critical transitions. 

These insights capture the diversity of warning signs in a real-world system because we use a large and high-resolution empirical dataset. Forecasting abilities using deep learning have been shown in some ecology~\cite{debMachineLearningMethods2022, buryDeepLearningEarly2021} and epidemiology~\cite{miryDeepLearningDisease2025} studies with training data simulated from simple models inspired by disciplinary tradition. Unlike our algorithm using system-specific time series with features encoding their memory, these methods cannot reveal pre-transition phenomena that are not present in the simulated models. 
Using empirical data also improves the applicability to related systems, which is illustrated by the striking validity of our 2022 warning system when applied in 2023. Though the r/place experiments of 2022 and 2023 had similar rules, contextual differences altered the 2023 dynamics: for example, differences in expansions of the canvas size and available colors, total duration, cooldown times, moderator and bot activity, level of preparedness of communities, player motivations, and popularity of the game. 

While different compositions may show varied warning behaviors, some compositions may show no warnings at all. In fact, unpredictable transitions make up a significant part of our data, as shown by the band of poor predictions shown in Fig.~\ref{fig:true-v-pred}a and the broadly distributed performance per transition in Fig.~\ref{figSI:ROC}b. This cross-subsystem variation is illustrated by the diverse pre-transition behaviors of Figs.~\ref{fig:shap} and~\ref{figSI:SHAPinterp}. It is also supported by recent literature~\cite{boettigerEarlyWarningSignals2013}, which has proposed a number of indicators of varying generalizability~\cite{georgeEarlyWarningSignals2023}. A key advantage of a machine learning warning system is its potential to uncover any warning signals present in real-world data, even when they were not hypothesized in \textit{a priori} models. Some signals may seem contradictory when seen in the same or similar subsystems, as in the co-occurrence of large critical slowing down and mild critical speeding up in our results. Slowing down typically occurs very close to transition, while speeding up is an earlier, subtle signal, and compositions could show none, one, or both. While slowing down is well-established, speeding up is less conventional but has been discussed~\cite{titusCriticalSpeedingEarly2020, pomeauCriticalSpeedvsCritical2011, gietkaSqueezingCriticalSpeeding2022}. Interestingly, a simple logistic growth model~\cite{maySimpleMathematicalModels1976} displays a direct link between the increase of the reaction rate and the departure from a stable equilibrium. However, these and other analytic indicators all rely on separate, often incompatible models. 

The lack of a common framework to describe early warning indicators complicates comparisons between our r/place results and simple indicators from the literature. Different indicators apply to different types of transitions, and identifying the types is not trivial. For example, critical slowing down assumes small perturbations around an equilibrium that slowly moves towards a bifurcation (sometimes called \textit{B-tipping}), while some transitions are instead triggered by large perturbations that push the state outside the basin of attraction of the initial equilibrium (\textit{S-} for \textit{state-tipping})~\cite{boettigerBifurcationStateTipping2020}. For sufficiently large and fast perturbations, warning signals may~\cite{drakeEarlyWarningSignals2013} or may not~\cite{boettigerEarlyWarningSignals2012, boettigerNoEarlyWarning2013} be present. 
Interestingly, the theory of critical slowing down applies only to second-order, continuous phase transitions 
and to spinodal instabilities~\cite{levinPhaseTransitionsTheory2023}, but is not expected in abrupt, first-order transitions that are usually more alarming in real-world systems. Depending on how the transition is modeled, tipping can also be conceived as occurring when the state fails to track a fast-changing equilibrium (\textit{rate-induced}, or \textit{R-tipping})~\cite{ashwinTippingPointsOpen2012}; critical slowing down might exist but be delayed in these cases~\cite{ritchieEarlywarningIndicatorsRateinduced2016}. 
In addition, systems like ours with stochasticity, multiple timescales, and more dimensions than typical dynamical models may be more prone to transients, meaning long quasi-stable periods that eventually transition, potentially with no warning signals~\cite{hastingsTransientPhenomenaEcology2018}.
The transitions in our compositions might be of any type, or even mixed types; for example, coordinated attacks can be seen as a sudden perturbation (S-tipping) or as a fast change of an attack parameter (R-tipping). Our machine learning system does not rely on explicit modeling and can therefore account for this variety of transition dynamics.

As a consequence, a warning signal in our system might not fit into the framework of standard warning signals. This issue materializes as challenges and ambiguities in defining a state variable and its equilibrium. Transitioning systems in ecology, medicine, or physics are usually described by a variable summarizing their state. However, our subsystems are images; the information in all their pixels is hardly reducible to a scalar state variable. Critical slowing down relies on trends in scalar variables, such as variance, autocorrelation, or return rate. In r/place compositions, these quantities could be defined in many valid ways (see SI Section~\ref{secSI:variables} for an exhaustive list of variables we computed). As an example, we computed variance variables that fall into three categories: deviations from the long-term reference image (like our \texttt{diff pixels reference}), instantaneous changes in the image (like our \texttt{diff pixels instant}), or the variance of instantaneous deviations from a fast-adapting equilibrium (like our \texttt{variance}, a proxy of the variance of \texttt{diff pixels instant}). A covariance matrix describing the colors of all pixels would be the closest to a true variance but would be too high-dimensional to compute, interpret, and compare to the literature. As another example, attributing a sign to autocorrelation is difficult because color differences from the reference image lack a direction, as colors are unordered. 

More generally, we navigate multiple issues that do not have clear solutions. The first challenge is identifying the equilibrium from which the system departs, including the timescale over which the equilibrium should be evaluated, or even knowing whether an equilibrium exists\footnote{When compositions are out of equilibrium, the reference image is only an average of out-of-equilibrium states.}. We varied the timescale of the equilibrium by training on datasets with different sliding window lengths (see Fig.~\ref{figSI:sensitivity}a); the sample of instances and their feature values differ, but the resulting changes in performance are marginal. Second, noise, which consists of small perturbations around equilibrium, is difficult to distinguish from coordinated attacks, which actively shift the equilibrium. While our results might in principle depend on these choices, the multiple options we compute often result in highly correlated variables, showing that these options reflect similar dynamics. 

To shed light on the impact of model design on observed warnings, and to illustrate the coexistence of slowing down and speeding up in a system, we design a toy model describing hypothetical dynamics in r/place compositions, presented in SI Section~\ref{secSI:toymodel}. To extract meaningful conclusions about pre-transition dynamics, we must carefully determine the state variable, the type of transition, and the size of perturbations. In our model, the state variable is the deviation from the defended image (similar to our \texttt{diff pixels reference}), and the order parameters are the attack rate\footnote{The attacks could also be seen as perturbations external to the model, and not as a parameter.} and defense strength. We incorporate both an increase of defense user recruitment at medium-high deviations from the reference image---representing the mobilization of defense when the attack is not too abrupt---and a desertion of defense at very high deviations. Two types of transitions emerge: a B-tipping transition when decreasing the defense strength, and a second, R-tipping transition when increasing the attack rate. To probe pre-transition dynamics, we apply a small perturbation to the equilibrium, which has a non-zero value of the image deviations. When approaching the bifurcation-like transition, we observe a decrease of return rate, or slowing down; we instead see an increase, or speeding up, preceding the rate-induced transition. These observations show that seemingly contradictory warning signals, associated with different types of transitions, can exist within a single system. 

Approaching the rate-induced transition, the trend of the variance depends on which \textit{pseudo-mean} the deviations are computed from, as the mean and equilibrium are ill-defined in our r/place images. The pseudo-mean can be taken as the equilibrium that the state fails to track; but in r/place compositions, this theoretical equilibrium is unknown. Alternatively, the variance can be computed using a short-timescale sample mean---which is only approximate in our compositions, but we use it as the main \texttt{variance} variable. Lastly, the pseudo-mean of the fraction of differing pixels can simply be zero, which corresponds to using the reference image of our compositions. These three variance definitions can lead to different interpretations in the toy model, which underscores the consequences of model design. 

Though the r/place data offers unprecedented insights into large-scale human behavior, there are certain limitations. First, we rely on the user-annotated Atlas~\cite{haagmansPlaceAtlasInitiative2024} to classify the pixel change data into compositions. Omissions or inaccuracies from contributors can impact the image borders of compositions, as well as their birth and death times. We rectify some of these mistakes with minimal corrections to composition borders (see SI Section~\ref{secSI:compodata}); these corrections only impact a small fraction of pixels and do not significantly modify the time series. The collective work of the Atlas contains crucial cultural information that automatic methods such as clustering~\cite{wuLargescaleCollectiveDynamics2024} or edge detection~\cite{cannyComputationalApproachEdge1986} cannot compensate for. Another issue is pixel changes that violated the 5-minute cooldown time. Moderators were not subject to the cooldown rule, and some users had multiple accounts or programmed bots to change pixels automatically (see SI Section~\ref{secSI:rplacedata}). These pixel changes are hard to eliminate systematically. However, they contribute to the canvas dynamics observed by users and are therefore part of the experiment; they also constitute less than 1\% of changes in 2022.

Learning from empirical data also requires careful study design, since prediction performance can be artificially inflated if we ignore the constraints of a real-world warning system. We tested how much removing these precautions increases ROC and PR performance (see SI Section~\ref{secSI:sensitivity} and Fig.~\ref{figSI:sensitivity}c). First, including compositions that emerge from patchworks---areas where multiple compositions meet and would not be tracked as a single unit---raises performance because the algorithm exploits artifacts that are specific to these patchworks and would therefore not exist in a live setting. 
Second, if we include the time instances that are too unstable to pass the relative threshold for transitions, the ROC AUC increases because these instances are easily categorized as being far from transition, although the PR AUC decreases due to a substantially smaller signal fraction. In addition, in unstable systems, defining a composition to preserve is ambiguous. Third, using a lower relative threshold of $2$ also increases performance, as slower transitions are treated as signal and provide earlier warnings. However, our focus is on abrupt transitions, which are both more difficult and more critical to detect in real-world systems. It is ultimately up to the system manager to decide which systems they wish to preserve against what transitions. In summary, to approximate a live warning system, we reject compositions that are not yet monitored, use only information preceding the assessed time instance, use only instances preceded by a stable period, and separate whole compositions between train and test samples (see Materials and Methods and SI Sections~\ref{secSI:transitions} and~\ref{secSI:filter}). 
Without such safeguards, prediction results may show high performance at the expense of real-world applicability. 

Lastly, sorting through the large amount of information encoded in the SHAP curves of Fig.~\ref{figSI:SHAPall} is challenging, and our interpretations involve a degree of conjecture. We group the SHAP trends into distinct readable signals using our understanding of the r/place social system, but applying such interpretations to other systems would require comparable domain expertise. Moreover, SHAP values comes with inherent drawbacks, as their computation assumes feature independence and can be affected by overfitting (see SI Section~\ref{secSI:shap}). Despite these difficulties, SHAP remains one of the few tools available for interpreting machine learning models and reveals insights into our system and possibly related ones.

Future work could test these insights in other socio-ecological systems where communities emerge and compete to achieve their goals. In addition, our predictive and interpretable approach is applicable not only to community-driven online platforms, but also to other systems that provide large and dense datasets. Another direction is to train a deep learning algorithm on individual pixel features rather than pixel-averaged features. This approach would bypass the difficulty of defining variables analogous to those in the early warning signals literature and may perform better, though these predictions would be significantly more difficult to interpret. However, adding complexity in the training dataset with additional spatial variables does not improve our performance (see Materials and Methods and SI Section~\ref{secSI:additionalVars}). This might be due to inherent unpredictability of some human behaviors, which is illustrated in Fig.~\ref{figSI:ROC}b: transitions in some compositions are easily predicted, while the performance of the algorithm is null for other compositions where transitions could be due to unstrategic, and thus unpredictable, attacks. By refining our warning systems and broadening their applicability, our interdisciplinary community can improve detection and understanding of transitions in complex systems. 

\section*{Conclusion}
We both predict transitions using a machine-learning-based warning system and translate its predictions into meaningful warning behaviors. Leveraging the large empirical dataset of r/place and emulating real-world constraints, we build a warning system suitable for practical monitoring, with a method transferrable to systems that provide similar data. Furthermore, the human-readable interpretations provide insights into pre-transition dynamics in an online social system. This is the first time that interpretable warnings have been developed with machine learning in an empirical system driven by human behavior. 

Unlike other warning systems, ours predicts a time-to-transition. This allows us to determine not just whether a transition is coming, but also when, which is key to an efficient response to a warning. We predict half of the incoming transitions within 20 minutes with only $3.6\%$ false positives and a ROC AUC of $0.835$. We also show predictive power up to 6 hours before the transition with a ROC AUC of $0.698$. Our algorithm far outperforms standard early warning signals. It also performs well when trained on 2022 data and tested on the 2023 experiment even though conditions in the two years differed, which demonstrates the generalizability of our results. 

We then interpret the algorithm's predictions using SHAP values to find human-readable warning signals. Our analysis reveals 12 behaviors that precede transitions. These include turbulent past image activity, simpler images, and lack of innovation or coordination. Strong defense reactions against a ramping-up attack force can also lead to both critical slowing down and critical speeding up. When used in a live setting, these human-readable warnings could suggest ways to protect a vulnerable composition. These insights may readily apply as qualitative warning indicators in socio-ecological systems with similar dynamics. 

We combine the high predictive performance of machine learning trained on empirical data with the explainability provided by SHAP values to extract readable warning signals. This work effectively improves the accuracy, applicability, and readability of warnings of transitions in complex adaptive systems. Our approach could be a model for the design of warning systems in multiple domains, from online social systems to earth systems and ecology. 

\subsection*{Data availability} 
All figure data is available on Zenodo~\cite{stephenson_anna_2025}. The code reproducing the results of this paper is available on Github at \url{https://github.com/AnnieStephenson/r-place-emergence/tree/EWSpaper}.


\acknow{G.F. and A.B.S. are supported by a gift from William H. Miller III. A.B.S. is supported by funding from the Princeton University Dean for Research, and the High Meadows Environmental Institute. We thank Giuseppe Ferro, Chris Kempes, Mikhail Kuleshov, Nusrat Molla, Andrew Romans, and Emma Zajdela for helpful discussions.}

\showacknow{} 

\bibsplit[3]

\makeatletter\@input{xx.tex}\makeatother

\bibliography{references}

@mastersthesis{armstrongCoordinationPeerProduction2018,
  title = {Coordination in a {{Peer Production Platform}}: {{A}} Study of {{Reddit}}'s /r/{{Place}} Experiment},
  shorttitle = {Coordination in a {{Peer Production Platform}}},
  author = {Armstrong, Ben},
  year = {2018},
  month = oct,
  urldate = {2022-11-30},
  langid = {english},
  school = {University of Waterloo}
}

@article{armstrongmckayExceeding15degCGlobal2022,
  title = {Exceeding 1.5{$^\circ$}{{C}} Global Warming Could Trigger Multiple Climate Tipping Points},
  author = {Armstrong McKay, David I. and Staal, Arie and Abrams, Jesse F. and Winkelmann, Ricarda and Sakschewski, Boris and Loriani, Sina and Fetzer, Ingo and Cornell, Sarah E. and Rockstr{\"o}m, Johan and Lenton, Timothy M.},
  year = {2022},
  month = sep,
  journal = {Science},
  volume = {377},
  number = {6611},
  pages = {eabn7950},
  publisher = {American Association for the Advancement of Science},
  doi = {10.1126/science.abn7950},
  urldate = {2023-10-20}
}

@article{ashwinTippingPointsOpen2012,
  title = {Tipping Points in Open Systems: Bifurcation, Noise-Induced and Rate-Dependent Examples in the Climate System},
  shorttitle = {Tipping Points in Open Systems},
  author = {Ashwin, Peter and Wieczorek, Sebastian and Vitolo, Renato and Cox, Peter},
  year = {2012},
  month = mar,
  journal = {Philosophical Transactions of the Royal Society A: Mathematical, Physical and Engineering Sciences},
  volume = {370},
  number = {1962},
  pages = {1166--1184},
  publisher = {Royal Society},
  doi = {10.1098/rsta.2011.0306},
  urldate = {2024-05-23}
}

@article{barthelemyEarlyWarningSystem2024,
  title = {Early Warning System for Currency Crises Using Long Short-Term Memory and Gated Recurrent Unit Neural Networks},
  author = {Barth{\'e}l{\'e}my, Sylvain and Gautier, Virginie and Rondeau, Fabien},
  year = {2024},
  journal = {Journal of Forecasting},
  volume = {43},
  number = {5},
  pages = {1235--1262},
  issn = {1099-131X},
  doi = {10.1002/for.3069},
  urldate = {2024-12-11},
  copyright = {{\copyright} 2024 The Authors. Journal of Forecasting published by John Wiley \& Sons Ltd.},
  langid = {english}
}

@article{bisoiCalculationFractalDimension2001,
  title = {On Calculation of Fractal Dimension of Images},
  author = {Bisoi, Ajay Kumar and Mishra, Jibitesh},
  year = {2001},
  month = may,
  journal = {Pattern Recognition Letters},
  volume = {22},
  number = {6-7},
  pages = {631--637},
  issn = {01678655},
  doi = {10.1016/S0167-8655(00)00132-X},
  urldate = {2023-07-18},
  langid = {english}
}

@article{boettigerBifurcationStateTipping2020,
  title = {Bifurcation or State Tipping: Assessing Transition Type in a Model Trophic Cascade},
  shorttitle = {Bifurcation or State Tipping},
  author = {Boettiger, Carl and Batt, Ryan},
  year = {2020},
  month = jan,
  journal = {Journal of Mathematical Biology},
  volume = {80},
  number = {1-2},
  pages = {143--155},
  issn = {0303-6812, 1432-1416},
  doi = {10.1007/s00285-019-01358-z},
  urldate = {2024-04-19},
  langid = {english}
}

@article{boettigerEarlyWarningSignals2012,
  title = {Early Warning Signals and the Prosecutor's Fallacy},
  author = {Boettiger, Carl and Hastings, Alan},
  year = {2012},
  month = dec,
  journal = {Proceedings of the Royal Society B: Biological Sciences},
  volume = {279},
  number = {1748},
  pages = {4734--4739},
  publisher = {Royal Society},
  doi = {10.1098/rspb.2012.2085},
  urldate = {2022-09-22}
}

@article{boettigerEarlyWarningSignals2013,
  title = {Early Warning Signals: {{The}} Charted and Uncharted Territories},
  shorttitle = {Early Warning Signals},
  author = {Boettiger, Carl and Ross, Noam and Hastings, Alan},
  year = {2013},
  month = may,
  journal = {Theoretical Ecology},
  volume = {6},
  doi = {10.1007/s12080-013-0192-6}
}

@article{boettigerNoEarlyWarning2013,
  title = {No Early Warning Signals for Stochastic Transitions: Insights from Large Deviation Theory},
  shorttitle = {No Early Warning Signals for Stochastic Transitions},
  author = {Boettiger, Carl and Hastings, Alan},
  year = {2013},
  month = sep,
  journal = {Proceedings of the Royal Society B: Biological Sciences},
  volume = {280},
  number = {1766},
  pages = {20131372},
  issn = {0962-8452, 1471-2954},
  doi = {10.1098/rspb.2013.1372},
  urldate = {2024-04-19},
  langid = {english}
}

@article{boettigerPatternsPredictions2013,
  title = {From Patterns to Predictions},
  author = {Boettiger, Carl and Hastings, Alan},
  year = {2013},
  month = jan,
  journal = {Nature},
  volume = {493},
  number = {7431},
  pages = {157--158},
  issn = {0028-0836, 1476-4687},
  doi = {10.1038/493157a},
  urldate = {2023-02-21},
  langid = {english}
}

@inproceedings{botelhoArtExpandedField2024,
  title = {Art as an {{Expanded Field}}: {{The Case}} of the {{R}}/{{Place Social Experiment}}},
  shorttitle = {Art as an {{Expanded Field}}},
  booktitle = {{{ArtsIT}}, {{Interactivity}} and {{Game Creation}}},
  author = {Botelho, Marcela Jatene Cavalcante and Oliveira, Hosana Celeste},
  editor = {Brooks, Anthony L.},
  year = {2024},
  pages = {123--134},
  publisher = {Springer Nature Switzerland},
  address = {Cham},
  doi = {10.1007/978-3-031-55319-6_9},
  isbn = {978-3-031-55319-6},
  langid = {english}
}

@article{brettDynamicalFootprintsEnable2020,
  title = {Dynamical Footprints Enable Detection of Disease Emergence},
  author = {Brett, Tobias S. and Rohani, Pejman},
  year = {2020},
  month = may,
  journal = {PLoS Biology},
  volume = {18},
  number = {5},
  pages = {e3000697},
  issn = {1544-9173},
  doi = {10.1371/journal.pbio.3000697},
  urldate = {2023-11-02},
  pmcid = {PMC7239390},
  pmid = {32433658}
}

@article{buryDeepLearningEarly2021,
  title = {Deep Learning for Early Warning Signals of Tipping Points},
  author = {Bury, Thomas M. and Sujith, R. I. and Pavithran, Induja and Scheffer, Marten and Lenton, Timothy M. and Anand, Madhur and Bauch, Chris T.},
  year = {2021},
  month = sep,
  journal = {Proceedings of the National Academy of Sciences},
  volume = {118},
  number = {39},
  pages = {e2106140118},
  publisher = {Proceedings of the National Academy of Sciences},
  doi = {10.1073/pnas.2106140118},
  urldate = {2023-03-24}
}

@article{cannyComputationalApproachEdge1986,
  title = {A {{Computational Approach}} to {{Edge Detection}}},
  author = {Canny, John},
  year = {1986},
  month = nov,
  journal = {IEEE Transactions on Pattern Analysis and Machine Intelligence},
  volume = {PAMI-8},
  number = {6},
  pages = {679--698},
  issn = {1939-3539},
  doi = {10.1109/TPAMI.1986.4767851},
  urldate = {2023-12-12}
}

@article{carpenterEarlyWarningsRegime2011,
  title = {Early {{Warnings}} of {{Regime Shifts}}: {{A Whole-Ecosystem Experiment}}},
  shorttitle = {Early {{Warnings}} of {{Regime Shifts}}},
  author = {Carpenter, S. R. and Cole, J. J. and Pace, M. L. and Batt, R. and Brock, W. A. and Cline, T. and Coloso, J. and Hodgson, J. R. and Kitchell, J. F. and Seekell, D. A. and Smith, L. and Weidel, B.},
  year = {2011},
  month = may,
  journal = {Science},
  volume = {332},
  number = {6033},
  pages = {1079--1082},
  issn = {0036-8075, 1095-9203},
  doi = {10.1126/science.1203672},
  urldate = {2023-02-21},
  langid = {english}
}

@article{chenEigenvaluesCovarianceMatrix2019,
  title = {Eigenvalues of the Covariance Matrix as Early Warning Signals for Critical Transitions in Ecological Systems},
  author = {Chen, Shiyang and O'Dea, Eamon B. and Drake, John M. and Epureanu, Bogdan I.},
  year = {2019},
  month = feb,
  journal = {Scientific Reports},
  volume = {9},
  number = {1},
  pages = {2572},
  issn = {2045-2322},
  doi = {10.1038/s41598-019-38961-5},
  urldate = {2024-04-19},
  langid = {english}
}

@article{chenPracticalGuideUsing2022,
  title = {Practical Guide to Using {{Kendall}}'s {$\tau$} in the Context of Forecasting Critical Transitions},
  author = {Chen, Shiyang and Ghadami, Amin and Epureanu, Bogdan I.},
  year = {2022},
  month = jul,
  journal = {Royal Society Open Science},
  volume = {9},
  number = {7},
  pages = {211346},
  publisher = {Royal Society},
  doi = {10.1098/rsos.211346},
  urldate = {2023-11-02}
}

@inproceedings{chenXGBoostScalableTree2016,
  title = {{{XGBoost}}: {{A Scalable Tree Boosting System}}},
  shorttitle = {{{XGBoost}}},
  booktitle = {Proceedings of the 22nd {{ACM SIGKDD International Conference}} on {{Knowledge Discovery}} and {{Data Mining}}},
  author = {Chen, Tianqi and Guestrin, Carlos},
  year = {2016},
  month = aug,
  eprint = {1603.02754},
  primaryclass = {cs},
  pages = {785--794},
  doi = {10.1145/2939672.2939785},
  urldate = {2023-12-10},
  archiveprefix = {arXiv}
}

@article{choiEarlyWarningCritical2022,
  title = {Early Warning for Critical Transitions Using Machine-Based Predictability},
  author = {Choi, Jaesung and Kim, Pilwon and Choi, Jaesung and Kim, Pilwon},
  year = {2022},
  journal = {AIMS Mathematics},
  volume = {7},
  number = {math-07-11-1112},
  pages = {20313--20327},
  issn = {2473-6988},
  doi = {10.3934/math.20221112},
  urldate = {2023-12-07},
  copyright = {2022 The Author(s)},
  langid = {english}
}

@article{daiGenericIndicatorsLoss2012,
  title = {Generic {{Indicators}} for {{Loss}} of {{Resilience Before}} a {{Tipping Point Leading}} to {{Population Collapse}}},
  author = {Dai, Lei and Vorselen, Daan and Korolev, Kirill S and Gore, Jeff},
  year = {2012},
  journal = {Science},
  volume = {336},
  number = {6085},
  pages = {1175--1177},
  doi = {10.1126/science.1219805},
  langid = {english}
}

@article{dakosMethodsDetectingEarly2012,
  title = {Methods for {{Detecting Early Warnings}} of {{Critical Transitions}} in {{Time Series Illustrated Using Simulated Ecological Data}}},
  author = {Dakos, Vasilis and Carpenter, Stephen R. and Brock, William A. and Ellison, Aaron M. and Guttal, Vishwesha and Ives, Anthony R. and K{\'e}fi, Sonia and Livina, Valerie and Seekell, David A. and van Nes, Egbert H. and Scheffer, Marten},
  year = {2012},
  month = jul,
  journal = {PLOS ONE},
  volume = {7},
  number = {7},
  pages = {e41010},
  publisher = {Public Library of Science},
  issn = {1932-6203},
  doi = {10.1371/journal.pone.0041010},
  urldate = {2023-06-12},
  langid = {english}
}

@article{debMachineLearningMethods2022,
  title = {Machine Learning Methods Trained on Simple Models Can Predict Critical Transitions in Complex Natural Systems},
  author = {Deb, Smita and Sidheekh, Sahil and Clements, Christopher F. and Krishnan, Narayanan C. and Dutta, Partha S.},
  year = {2022},
  month = feb,
  journal = {Royal Society Open Science},
  volume = {9},
  number = {2},
  pages = {211475},
  publisher = {Royal Society},
  doi = {10.1098/rsos.211475},
  urldate = {2023-11-02}
}

@incollection{dixonRipleysFunction2014,
  title = {Ripley's {{K Function}}},
  booktitle = {Wiley {{StatsRef}}: {{Statistics Reference Online}}},
  author = {Dixon, Philip M.},
  year = {2014},
  publisher = {John Wiley \& Sons, Ltd},
  doi = {10.1002/9781118445112.stat07751},
  urldate = {2025-09-28},
  copyright = {Copyright {\copyright} 2013 John Wiley \& Sons, Ltd. All rights reserved.},
  isbn = {978-1-118-44511-2},
  langid = {english}
}

@article{drakeEarlyWarningSignals2013,
  title = {Early Warning Signals of Stochastic Switching},
  author = {Drake, John M.},
  year = {2013},
  month = sep,
  journal = {Proceedings of the Royal Society B: Biological Sciences},
  volume = {280},
  number = {1766},
  pages = {20130686},
  issn = {0962-8452, 1471-2954},
  doi = {10.1098/rspb.2013.0686},
  urldate = {2024-04-19},
  langid = {english}
}

@article{dylewskyUniversalEarlyWarning2023,
  title = {Universal Early Warning Signals of Phase Transitions in Climate Systems},
  author = {Dylewsky, Daniel and Lenton, Timothy M. and Scheffer, Marten and Bury, Thomas M. and Fletcher, Christopher G. and Anand, Madhur and Bauch, Chris T.},
  year = {2023},
  month = apr,
  journal = {Journal of The Royal Society Interface},
  volume = {20},
  number = {201},
  pages = {20220562},
  publisher = {Royal Society},
  doi = {10.1098/rsif.2022.0562},
  urldate = {2023-12-07}
}

@article{gaoMachineLearningBased2020,
  title = {Machine Learning Based Early Warning System Enables Accurate Mortality Risk Prediction for {{COVID-19}}},
  author = {Gao, Yue and Cai, Guang-Yao and Fang, Wei and Li, Hua-Yi and Wang, Si-Yuan and Chen, Lingxi and Yu, Yang and Liu, Dan and Xu, Sen and Cui, Peng-Fei and Zeng, Shao-Qing and Feng, Xin-Xia and Yu, Rui-Di and Wang, Ya and Yuan, Yuan and Jiao, Xiao-Fei and Chi, Jian-Hua and Liu, Jia-Hao and Li, Ru-Yuan and Zheng, Xu and Song, Chun-Yan and Jin, Ning and Gong, Wen-Jian and Liu, Xing-Yu and Huang, Lei and Tian, Xun and Li, Lin and Xing, Hui and Ma, Ding and Li, Chun-Rui and Ye, Fei and Gao, Qing-Lei},
  year = {2020},
  month = oct,
  journal = {Nature Communications},
  volume = {11},
  number = {1},
  pages = {5033},
  publisher = {Nature Publishing Group},
  issn = {2041-1723},
  doi = {10.1038/s41467-020-18684-2},
  urldate = {2023-12-07},
  copyright = {2020 The Author(s)},
  langid = {english}
}

@article{georgeEarlyWarningSignals2023,
  title = {Early Warning Signals for Critical Transitions in Complex Systems},
  author = {George, Sandip V. and Kachhara, Sneha and Ambika, G.},
  year = {2023},
  month = jun,
  journal = {Physica Scripta},
  volume = {98},
  number = {7},
  pages = {072002},
  publisher = {IOP Publishing},
  issn = {1402-4896},
  doi = {10.1088/1402-4896/acde20},
  urldate = {2023-11-02},
  langid = {english}
}

@article{gietkaSqueezingCriticalSpeeding2022,
  title = {Squeezing by Critical Speeding up: {{Applications}} in Quantum Metrology},
  shorttitle = {Squeezing by Critical Speeding Up},
  author = {Gietka, Karol},
  year = {2022},
  month = apr,
  journal = {Physical Review A},
  volume = {105},
  number = {4},
  pages = {042620},
  publisher = {American Physical Society},
  doi = {10.1103/PhysRevA.105.042620},
  urldate = {2024-02-05}
}

@article{grapsIntroductionWavelets1995,
  title = {An Introduction to Wavelets},
  author = {Graps, A.},
  year = {1995},
  journal = {IEEE Computational Science and Engineering},
  volume = {2},
  number = {2},
  pages = {50--61},
  issn = {1558-190X},
  doi = {10.1109/99.388960},
  urldate = {2025-09-28}
}

@article{grassiaMachineLearningDismantling2021,
  title = {Machine Learning Dismantling and Early-Warning Signals of Disintegration in Complex Systems},
  author = {Grassia, Marco and De Domenico, Manlio and Mangioni, Giuseppe},
  year = {2021},
  month = aug,
  journal = {Nature Communications},
  volume = {12},
  number = {1},
  pages = {5190},
  publisher = {Nature Publishing Group},
  issn = {2041-1723},
  doi = {10.1038/s41467-021-25485-8},
  urldate = {2023-06-27},
  copyright = {2021 The Author(s)},
  langid = {english}
}

@misc{haagmansPlaceAtlasInitiative2024,
  title = {Place {{Atlas Initiative}}},
  author = {Haagmans, Stephano},
  year = {Accessed 2024-12-13},
  urldate = {2023-12-12},
  howpublished = {https://place-atlas.stefanocoding.me/},
  langid = {english}
}

@article{hastingsTransientPhenomenaEcology2018,
  title = {Transient Phenomena in Ecology},
  author = {Hastings, Alan and Abbott, Karen C. and Cuddington, Kim and Francis, Tessa and Gellner, Gabriel and Lai, Ying-Cheng and Morozov, Andrew and Petrovskii, Sergei and Scranton, Katherine and Zeeman, Mary Lou},
  year = {2018},
  month = sep,
  journal = {Science},
  volume = {361},
  number = {6406},
  pages = {eaat6412},
  publisher = {American Association for the Advancement of Science},
  doi = {10.1126/science.aat6412},
  urldate = {2022-11-07}
}

@article{hylandEarlyPredictionCirculatory2020,
  title = {Early Prediction of Circulatory Failure in the Intensive Care Unit Using Machine Learning},
  author = {Hyland, Stephanie L. and Faltys, Martin and H{\"u}ser, Matthias and Lyu, Xinrui and Gumbsch, Thomas and Esteban, Crist{\'o}bal and Bock, Christian and Horn, Max and Moor, Michael and Rieck, Bastian and Zimmermann, Marc and Bodenham, Dean and Borgwardt, Karsten and R{\"a}tsch, Gunnar and Merz, Tobias M.},
  year = {2020},
  month = mar,
  journal = {Nature Medicine},
  volume = {26},
  number = {3},
  pages = {364--373},
  issn = {1078-8956, 1546-170X},
  doi = {10.1038/s41591-020-0789-4},
  urldate = {2023-11-16},
  langid = {english}
}

@misc{israeliFlyingColorsPredicting2023,
  title = {With {{Flying Colors}}: {{Predicting Community Success}} in {{Large-scale Collaborative Campaigns}}},
  shorttitle = {With {{Flying Colors}}},
  author = {Israeli, Abraham and Tsur, Oren},
  year = {2023},
  month = jul,
  number = {arXiv:2307.09650},
  eprint = {2307.09650},
  primaryclass = {cs},
  publisher = {arXiv},
  urldate = {2024-07-19},
  archiveprefix = {arXiv},
  langid = {english}
}

@article{jacksonHistoricalOverfishingRecent2001,
  title = {Historical {{Overfishing}} and the {{Recent Collapse}} of {{Coastal Ecosystems}}},
  author = {Jackson, Jeremy B. C. and Kirby, Michael X. and Berger, Wolfgang H. and Bjorndal, Karen A. and Botsford, Louis W. and Bourque, Bruce J. and Bradbury, Roger H. and Cooke, Richard and Erlandson, Jon and Estes, James A. and Hughes, Terence P. and Kidwell, Susan and Lange, Carina B. and Lenihan, Hunter S. and Pandolfi, John M. and Peterson, Charles H. and Steneck, Robert S. and Tegner, Mia J. and Warner, Robert R.},
  year = {2001},
  month = jul,
  journal = {Science},
  volume = {293},
  number = {5530},
  pages = {629--637},
  issn = {0036-8075, 1095-9203},
  doi = {10.1126/science.1059199},
  urldate = {2024-07-29},
  langid = {english}
}

@article{kefiEarlyWarningSignals2013,
  title = {Early Warning Signals Also Precede Non-Catastrophic Transitions},
  author = {K{\'e}fi, Sonia and Dakos, Vasilis and Scheffer, Marten and Van Nes, Egbert H. and Rietkerk, Max},
  year = {2013},
  journal = {Oikos},
  volume = {122},
  number = {5},
  pages = {641--648},
  issn = {1600-0706},
  doi = {10.1111/j.1600-0706.2012.20838.x},
  urldate = {2024-07-29},
  langid = {english}
}

@inproceedings{kobylarzribeiroMachineLearningEarly2020,
  title = {A {{Machine Learning Early Warning System}}: {{Multicenter Validation}} in {{Brazilian Hospitals}}},
  shorttitle = {A {{Machine Learning Early Warning System}}},
  booktitle = {2020 {{IEEE}} 33rd {{International Symposium}} on {{Computer-Based Medical Systems}} ({{CBMS}})},
  author = {Kobylarz Ribeiro, Jhonatan and {dos Santos}, Henrique D.P. and Barletta, Felipe and {Cichelero da Silva}, Mateus and Vieira, Renata and M. P. Morales, Hugo and {da Costa Rocha}, Cristian},
  year = {2020},
  month = jul,
  pages = {321--326},
  issn = {2372-9198},
  doi = {10.1109/CBMS49503.2020.00067},
  urldate = {2023-12-07}
}

@article{lagorioQuarantinegeneratedPhaseTransition2011,
  title = {Quarantine-Generated Phase Transition in Epidemic Spreading},
  author = {Lagorio, C. and Dickison, M. and Vazquez, F. and Braunstein, L. A. and Macri, P. A. and Migueles, M. V. and Havlin, S. and Stanley, H. E.},
  year = {2011},
  month = feb,
  journal = {Physical Review E},
  volume = {83},
  number = {2},
  pages = {026102},
  publisher = {American Physical Society},
  doi = {10.1103/PhysRevE.83.026102},
  urldate = {2024-07-29}
}

@inproceedings{lassetterUsingCriticalSlowing2021,
  title = {Using {{Critical Slowing Down Features}} to {{Enhance Performance}} of {{Artificial Neural Networks}} for {{Time-Domain Power System Data}}},
  booktitle = {2021 {{IEEE}} 9th {{International Conference}} on {{Smart Energy Grid Engineering}} ({{SEGE}})},
  author = {Lassetter, Austin and {Cotilla-Sanchez}, Eduardo and Kim, Jinsub},
  year = {2021},
  month = aug,
  pages = {117--123},
  publisher = {IEEE},
  address = {Oshawa, ON, Canada},
  doi = {10.1109/SEGE52446.2021.9535027},
  urldate = {2023-11-16},
  isbn = {978-1-6654-4094-3},
  langid = {english}
}

@incollection{levinPhaseTransitionsTheory2023,
  title = {Phase {{Transitions}} and the {{Theory}} of {{Early Warning Indicators}} for {{Critical Transitions}}},
  booktitle = {How {{Worlds Collapse}}},
  author = {Levin, Simon A. and Hagstrom, George I.},
  year = {2023},
  publisher = {Routledge},
  isbn = {978-1-003-33138-4}
}

@article{litherlandInstructionVsEmergence2021,
  title = {Instruction vs. Emergence on r/Place: {{Understanding}} the Growth and Control of Evolving Artifacts in Mass Collaboration},
  shorttitle = {Instruction vs. Emergence on r/Place},
  author = {Litherland, Kristina T. and M{\o}rch, Anders I.},
  year = {2021},
  month = sep,
  journal = {Computers in Human Behavior},
  volume = {122},
  pages = {106845},
  issn = {0747-5632},
  doi = {10.1016/j.chb.2021.106845},
  urldate = {2022-11-30},
  langid = {english}
}

@article{liuEarlyPredictorOnset2024,
  title = {Early {{Predictor}} for the {{Onset}} of {{Critical Transitions}} in {{Networked Dynamical Systems}}},
  author = {Liu, Zijia and Zhang, Xiaozhu and Ru, Xiaolei and Gao, Ting-Ting and Moore, Jack Murdoch and Yan, Gang},
  year = {2024},
  month = jul,
  journal = {Physical Review X},
  volume = {14},
  number = {3},
  pages = {031009},
  issn = {2160-3308},
  doi = {10.1103/PhysRevX.14.031009},
  urldate = {2024-07-29},
  langid = {english}
}

@inproceedings{lundbergUnifiedApproachInterpreting2017,
  title = {A {{Unified Approach}} to {{Interpreting Model Predictions}}},
  booktitle = {Advances in {{Neural Information Processing Systems}}},
  author = {Lundberg, Scott M and Lee, Su-In},
  year = {2017},
  volume = {30},
  publisher = {Curran Associates, Inc.},
  urldate = {2023-12-10}
}

@inproceedings{maDatadrivenPowerSystem2018,
  title = {Data-Driven {{Power System Collapse Predicting Using Critical Slowing Down Indicators}}},
  booktitle = {2018 {{International Conference}} on {{Power System Technology}} ({{POWERCON}})},
  author = {Ma, Junyu and Tang, Junci and Yan, Zheng and Jiang, Feng and Zeng, Hui and Fang, Chen},
  year = {2018},
  month = nov,
  pages = {1879--1884},
  publisher = {IEEE},
  address = {Guangzhou},
  doi = {10.1109/POWERCON.2018.8602265},
  urldate = {2023-11-16},
  isbn = {978-1-5386-6461-2},
  langid = {english}
}

@article{makitieDigitalInnovationsContribution2023,
  title = {Digital Innovation's Contribution to Sustainability Transitions},
  author = {M{\"a}kitie, Tuukka and Hanson, Jens and Damman, Sigrid and Wardeberg, Mari},
  year = {2023},
  month = may,
  journal = {Technology in Society},
  volume = {73},
  pages = {102255},
  issn = {0160-791X},
  doi = {10.1016/j.techsoc.2023.102255},
  urldate = {2025-01-31}
}

@article{martinianiQuantifyingHiddenOrder2019,
  title = {Quantifying {{Hidden Order}} out of {{Equilibrium}}},
  author = {Martiniani, Stefano and Chaikin, Paul M. and Levine, Dov},
  year = {2019},
  month = feb,
  journal = {Physical Review X},
  volume = {9},
  number = {1},
  pages = {011031},
  publisher = {American Physical Society},
  doi = {10.1103/PhysRevX.9.011031},
  urldate = {2023-05-22}
}

@misc{martinianiSweetsourcod2019,
  title = {Sweetsourcod},
  author = {Martiniani, Stefano and Guo, Buming},
  year = {2019, https://github.com/martiniani-lab/sweetsourcod},
  urldate = {2025-02-12}
}

@article{mascarenoCriticalTransitionsEcosystems2022,
  title = {Critical {{Transitions}} in {{Ecosystems}} and {{Society}}. {{The Contribution}} of {{Sociological Systems Theory}} to the {{Analysis}} of {{Socio-Environmental Transformations}}},
  author = {Mascare{\~n}o, Aldo},
  year = {2022},
  month = jan,
  journal = {Frontiers in Sociology},
  volume = {6},
  publisher = {Frontiers},
  issn = {2297-7775},
  doi = {10.3389/fsoc.2021.763453},
  urldate = {2025-01-31},
  langid = {english}
}

@article{maySimpleMathematicalModels1976,
  title = {Simple Mathematical Models with Very Complicated Dynamics},
  author = {May, Robert M.},
  year = {1976},
  month = jun,
  journal = {Nature},
  volume = {261},
  number = {5560},
  pages = {459--467},
  publisher = {Nature Publishing Group},
  issn = {1476-4687},
  doi = {10.1038/261459a0},
  urldate = {2025-01-31},
  copyright = {1976 Springer Nature Limited},
  langid = {english}
}

@misc{miryDeepLearningDisease2025,
  title = {Deep {{Learning}} for {{Disease Outbreak Prediction}}: {{A Robust Early Warning Signal}} for {{Transcritical Bifurcations}}},
  shorttitle = {Deep {{Learning}} for {{Disease Outbreak Prediction}}},
  author = {Miry, Reza and Chakraborty, Amit K. and Greiner, Russell and Lewis, Mark A. and Wang, Hao and Guan, Tianyu and Ramazi, Pouria},
  year = {2025},
  month = jan,
  number = {arXiv:2501.07764},
  eprint = {2501.07764},
  primaryclass = {cs},
  publisher = {arXiv},
  doi = {10.48550/arXiv.2501.07764},
  urldate = {2025-02-09},
  archiveprefix = {arXiv}
}

@article{mullerCompressionCulturalEvolution2018,
  title = {Compression in Cultural Evolution: {{Homogeneity}} and Structure in the Emergence and Evolution of a Large-Scale Online Collaborative Art Project},
  shorttitle = {Compression in Cultural Evolution},
  author = {M{\"u}ller, Thomas F. and Winters, James},
  year = {2018},
  month = sep,
  journal = {PLOS ONE},
  volume = {13},
  number = {9},
  pages = {e0202019},
  publisher = {Public Library of Science},
  issn = {1932-6203},
  doi = {10.1371/journal.pone.0202019},
  urldate = {2022-11-30},
  langid = {english}
}

@article{obrienEWSmethodsPackageForecast2023,
  title = {{{EWSmethods}}: An {{R}} Package to Forecast Tipping Points at the Community Level Using Early Warning Signals, Resilience Measures, and Machine Learning Models},
  shorttitle = {{{EWSmethods}}},
  author = {O'Brien, Duncan A. and Deb, Smita and Sidheekh, Sahil and Krishnan, Narayanan C. and Sharathi Dutta, Partha and Clements, Christopher F.},
  year = {2023},
  journal = {Ecography},
  volume = {2023},
  number = {10},
  pages = {e06674},
  issn = {1600-0587},
  doi = {10.1111/ecog.06674},
  urldate = {2023-12-07},
  copyright = {{\copyright} 2023 The Authors. Ecography published by John Wiley \& Sons Ltd on behalf of Nordic Society Oikos},
  langid = {english}
}

@article{orozco-fuentesEarlyWarningSignals2019,
  title = {Early Warning Signals in Plant Disease Outbreaks},
  author = {{Orozco-Fuentes}, S. and Griffiths, G. and Holmes, M. J. and Ettelaie, R. and Smith, J. and Baggaley, A. W. and Parker, N. G.},
  year = {2019},
  month = feb,
  journal = {Ecological Modelling},
  volume = {393},
  pages = {12--19},
  issn = {0304-3800},
  doi = {10.1016/j.ecolmodel.2018.11.003},
  urldate = {2024-07-29}
}

@article{papadimitriouModellingSpatialLandscape2009,
  title = {Modelling Spatial Landscape Complexity Using the {{Levenshtein}} Algorithm},
  author = {Papadimitriou, Fivos},
  year = {2009},
  month = jan,
  journal = {Ecological Informatics},
  volume = {4},
  number = {1},
  pages = {48--55},
  issn = {1574-9541},
  doi = {10.1016/j.ecoinf.2009.01.001},
  urldate = {2025-09-28}
}

@article{petrieStructureStabilityExploited2009,
  title = {Structure and Stability in Exploited Marine Fish Communities: Quantifying Critical Transitions},
  shorttitle = {Structure and Stability in Exploited Marine Fish Communities},
  author = {Petrie, Brian and Frank, Kenneth T. and Shackell, Nancy L. and Leggett, William C.},
  year = {2009},
  journal = {Fisheries Oceanography},
  volume = {18},
  number = {2},
  pages = {83--101},
  issn = {1365-2419},
  doi = {10.1111/j.1365-2419.2009.00500.x},
  urldate = {2024-07-29},
  langid = {english}
}

@misc{Place2024,
  title = {Place},
  year = {Accessed 2024-12-13},
  urldate = {2024-12-09},
  howpublished = {https://www.reddit.com/r/place/}
}

@misc{pomeauCriticalSpeedvsCritical2011,
  title = {Critical Speed-up vs Critical Slow-down: A New Kind of Relaxation Oscillation with Application to Stick-Slip Phenomena},
  shorttitle = {Critical Speed-up vs Critical Slow-Down},
  author = {Pomeau, Yves and Berre, Martine Le},
  year = {2011},
  month = jul,
  number = {arXiv:1107.3331},
  eprint = {1107.3331},
  primaryclass = {physics},
  publisher = {arXiv},
  doi = {10.48550/arXiv.1107.3331},
  urldate = {2024-02-05},
  archiveprefix = {arXiv}
}

@article{qiUsingMachineLearning2020,
  title = {Using Machine Learning to Predict Extreme Events in Complex Systems},
  author = {Qi, Di and Majda, Andrew J.},
  year = {2020},
  month = jan,
  journal = {Proceedings of the National Academy of Sciences},
  volume = {117},
  number = {1},
  pages = {52--59},
  issn = {0027-8424, 1091-6490},
  doi = {10.1073/pnas.1917285117},
  urldate = {2024-04-19},
  langid = {english}
}

@article{rappazLatentStructureCollaboration2018,
  title = {Latent {{Structure}} in {{Collaboration}}: {{The Case}} of {{Reddit}} r/Place},
  shorttitle = {Latent {{Structure}} in {{Collaboration}}},
  author = {Rappaz, J{\'e}r{\'e}mie and Catasta, Michele and West, Robert and Aberer, Karl},
  year = {2018},
  month = jun,
  journal = {Proceedings of the International AAAI Conference on Web and Social Media},
  volume = {12},
  number = {1},
  issn = {2334-0770},
  doi = {10.1609/icwsm.v12i1.15013},
  urldate = {2022-11-30},
  copyright = {Copyright (c) 2022 Proceedings of the International AAAI Conference on Web and Social Media},
  langid = {english}
}

@article{ritchieEarlywarningIndicatorsRateinduced2016,
  title = {Early-Warning Indicators for Rate-Induced Tipping},
  author = {Ritchie, Paul and Sieber, Jan},
  year = {2016},
  month = sep,
  journal = {Chaos: An Interdisciplinary Journal of Nonlinear Science},
  volume = {26},
  number = {9},
  pages = {093116},
  issn = {1054-1500},
  doi = {10.1063/1.4963012},
  urldate = {2024-04-22}
}

@article{rudinStopExplainingBlack2019,
  title = {Stop Explaining Black Box Machine Learning Models for High Stakes Decisions and Use Interpretable Models Instead},
  author = {Rudin, Cynthia},
  year = {2019},
  month = may,
  journal = {Nature Machine Intelligence},
  volume = {1},
  number = {5},
  pages = {206--215},
  publisher = {Nature Publishing Group},
  issn = {2522-5839},
  doi = {10.1038/s42256-019-0048-x},
  urldate = {2024-12-11},
  copyright = {2019 Springer Nature Limited},
  langid = {english}
}

@article{samitasMachineLearningEarly2020,
  title = {Machine Learning as an Early Warning System to Predict Financial Crisis},
  author = {Samitas, Aristeidis and Kampouris, Elias and Kenourgios, Dimitris},
  year = {2020},
  month = oct,
  journal = {International Review of Financial Analysis},
  volume = {71},
  pages = {101507},
  issn = {1057-5219},
  doi = {10.1016/j.irfa.2020.101507},
  urldate = {2023-12-07}
}

@article{schefferAnticipatingCriticalTransitions2012,
  title = {Anticipating {{Critical Transitions}}},
  author = {Scheffer, Marten and Carpenter, Stephen R. and Lenton, Timothy M. and Bascompte, Jordi and Brock, William and Dakos, Vasilis and {van de Koppel}, Johan and {van de Leemput}, Ingrid A. and Levin, Simon A. and {van Nes}, Egbert H. and Pascual, Mercedes and Vandermeer, John},
  year = {2012},
  month = oct,
  journal = {Science},
  volume = {338},
  number = {6105},
  pages = {344--348},
  publisher = {American Association for the Advancement of Science},
  doi = {10.1126/science.1225244},
  urldate = {2024-07-29}
}

@article{schefferEarlywarningSignalsCritical2009,
  title = {Early-Warning Signals for Critical Transitions},
  author = {Scheffer, Marten and Bascompte, Jordi and Brock, William A. and Brovkin, Victor and Carpenter, Stephen R. and Dakos, Vasilis and Held, Hermann and {van Nes}, Egbert H. and Rietkerk, Max and Sugihara, George},
  year = {2009},
  month = sep,
  journal = {Nature},
  volume = {461},
  number = {7260},
  pages = {53--59},
  publisher = {Nature Publishing Group},
  issn = {1476-4687},
  doi = {10.1038/nature08227},
  urldate = {2022-09-22},
  copyright = {2009 Macmillan Publishers Limited. All rights reserved},
  langid = {english}
}

@article{suweisEarlyWarningSigns2014,
  title = {Early {{Warning Signs}} in {{Social-Ecological Networks}}},
  author = {Suweis, Samir and D'Odorico, Paolo},
  editor = {Lambiotte, Renaud},
  year = {2014},
  month = jul,
  journal = {PLoS ONE},
  volume = {9},
  number = {7},
  pages = {e101851},
  issn = {1932-6203},
  doi = {10.1371/journal.pone.0101851},
  urldate = {2023-11-17},
  langid = {english}
}

@book{tainterCollapseComplexSocieties1988,
  title = {The {{Collapse}} of {{Complex Societies}}},
  author = {Tainter, Joseph},
  year = {1988},
  publisher = {Cambridge Univ. Press}
}

@article{tapakComparativeEvaluationTime2019,
  title = {Comparative Evaluation of Time Series Models for Predicting Influenza Outbreaks: Application of Influenza-like Illness Data from Sentinel Sites of Healthcare Centers in {{Iran}}},
  shorttitle = {Comparative Evaluation of Time Series Models for Predicting Influenza Outbreaks},
  author = {Tapak, Leili and Hamidi, Omid and Fathian, Mohsen and Karami, Manoochehr},
  year = {2019},
  month = dec,
  journal = {BMC Research Notes},
  volume = {12},
  number = {1},
  pages = {353},
  issn = {1756-0500},
  doi = {10.1186/s13104-019-4393-y},
  urldate = {2023-11-16},
  langid = {english}
}

@article{titusCriticalSpeedingEarly2020,
  title = {Critical Speeding up as an Early Warning Signal of Stochastic Regime Shifts},
  author = {Titus, Mathew and Watson, James},
  year = {2020},
  month = dec,
  journal = {Theoretical Ecology},
  volume = {13},
  number = {4},
  pages = {449--457},
  issn = {1874-1738, 1874-1746},
  doi = {10.1007/s12080-020-00451-0},
  urldate = {2024-04-19},
  langid = {english}
}

@inproceedings{vachherUnderstandingCommunityLevelConflicts2020,
  title = {Understanding {{Community-Level Conflicts Through Reddit}} r/Place},
  booktitle = {Conference {{Companion Publication}} of the 2020 on {{Computer Supported Cooperative Work}} and {{Social Computing}}},
  author = {Vachher, Prateek and Levonian, Zachary and Cheng, Hao-Fei and Yarosh, Svetlana},
  year = {2020},
  month = oct,
  series = {{{CSCW}} '20 {{Companion}}},
  pages = {401--405},
  publisher = {Association for Computing Machinery},
  address = {New York, NY, USA},
  doi = {10.1145/3406865.3418311},
  urldate = {2022-11-30},
  isbn = {978-1-4503-8059-1}
}

@article{vandewalleCrashOctober19871998,
  title = {The Crash of {{October}} 1987 Seen as a Phase Transition: Amplitude and Universality},
  shorttitle = {The Crash of {{October}} 1987 Seen as a Phase Transition},
  author = {Vandewalle, N. and Boveroux, {\relax Ph}. and Minguet, A. and Ausloos, M.},
  year = {1998},
  month = jun,
  journal = {Physica A: Statistical Mechanics and its Applications},
  volume = {255},
  number = {1},
  pages = {201--210},
  issn = {0378-4371},
  doi = {10.1016/S0378-4371(98)00115-0},
  urldate = {2024-07-29}
}

@article{vannesSlowRecoveryPerturbations2007,
  title = {Slow Recovery from Perturbations as a Generic Indicator of a Nearby Catastrophic Shift},
  author = {{van Nes}, Egbert H. and Scheffer, Marten},
  year = {2007},
  month = jun,
  journal = {The American Naturalist},
  volume = {169},
  number = {6},
  pages = {738--747},
  issn = {1537-5323},
  doi = {10.1086/516845},
  langid = {english},
  pmid = {17479460}
}

@article{wisselUniversalLawCharacteristic1984,
  title = {A Universal Law of the Characteristic Return Time near Thresholds},
  author = {Wissel, C.},
  year = {1984},
  month = dec,
  journal = {Oecologia},
  volume = {65},
  number = {1},
  pages = {101--107},
  issn = {1432-1939},
  doi = {10.1007/BF00384470},
  urldate = {2024-07-29},
  langid = {english}
}

@misc{wuLargescaleCollectiveDynamics2024,
  title = {Large-Scale {{Collective Dynamics}} in the {{Three Iterations}} of the {{Reddit}} r/Place {{Experiment}}},
  author = {Wu, Yutong and Silva, Arlei},
  year = {2024},
  month = aug,
  number = {arXiv:2408.13236},
  eprint = {2408.13236},
  primaryclass = {cs},
  publisher = {arXiv},
  doi = {10.48550/arXiv.2408.13236},
  urldate = {2024-12-13},
  archiveprefix = {arXiv}
}

@article{wunderlingHowMotifsCondition2020,
  title = {How Motifs Condition Critical Thresholds for Tipping Cascades in Complex Networks: {{Linking}} Micro- to Macro-Scales},
  shorttitle = {How Motifs Condition Critical Thresholds for Tipping Cascades in Complex Networks},
  author = {Wunderling, Nico and Stumpf, Benedikt and Kr{\"o}nke, Jonathan and Staal, Arie and Tuinenburg, Obbe A. and Winkelmann, Ricarda and Donges, Jonathan F.},
  year = {2020},
  month = apr,
  journal = {Chaos: An Interdisciplinary Journal of Nonlinear Science},
  volume = {30},
  number = {4},
  pages = {043129},
  issn = {1054-1500, 1089-7682},
  doi = {10.1063/1.5142827},
  urldate = {2022-11-17},
  langid = {english}
}

@article{xuNonequilibriumEarlywarningSignals2023,
  title = {Non-Equilibrium Early-Warning Signals for Critical Transitions in Ecological Systems},
  author = {Xu, Li and Patterson, Denis and Levin, Simon Asher and Wang, Jin},
  year = {2023},
  month = jan,
  journal = {Proceedings of the National Academy of Sciences},
  volume = {120},
  number = {5},
  pages = {e2218663120},
  publisher = {Proceedings of the National Academy of Sciences},
  doi = {10.1073/pnas.2218663120},
  urldate = {2024-12-10}
}

@article{zhangComplexity1Noise1991,
  title = {Complexity and 1/f Noise. {{A}} Phase Space Approach},
  author = {Zhang, Yi-Cheng},
  year = {1991},
  month = jul,
  journal = {Journal de Physique I},
  volume = {1},
  number = {7},
  pages = {971--977},
  issn = {1155-4304, 1286-4862},
  doi = {10.1051/jp1:1991180},
  urldate = {2025-09-28},
  langid = {english}
}

@electronic{stephenson_anna_2025,
  author      = {Stephenson, Anna B. and 
                Falmagne, Guillaume and 
                Levin, Simon},
  title       = {{Figure data for ``Interpretable Early Warnings using Machine Learning
                  in an Online Game-experiment''}},
  publisher   = {{Zenodo}},
  year        = 2025,
  note         = {DOI:10.5281/zenodo.14968895},
  url          = {https://doi.org/10.5281/zenodo.14968895}
}

\end{document}


\title{Interpretable Early Warnings using Machine Learning in an Online Game-experiment}
\author{Guillaume Falmagne, Anna B. Stephenson, and Simon A. Levin}
\correspondingauthor{Guillaume Falmagne\\E-mail: guillaume.falmagne@TUDublin.ie}


\maketitle

\SItext

\section{Data of r/place and compositions}
\subsection{r/place data and generalities}
\label{secSI:rplacedata}

\begin{figure}
\includegraphics{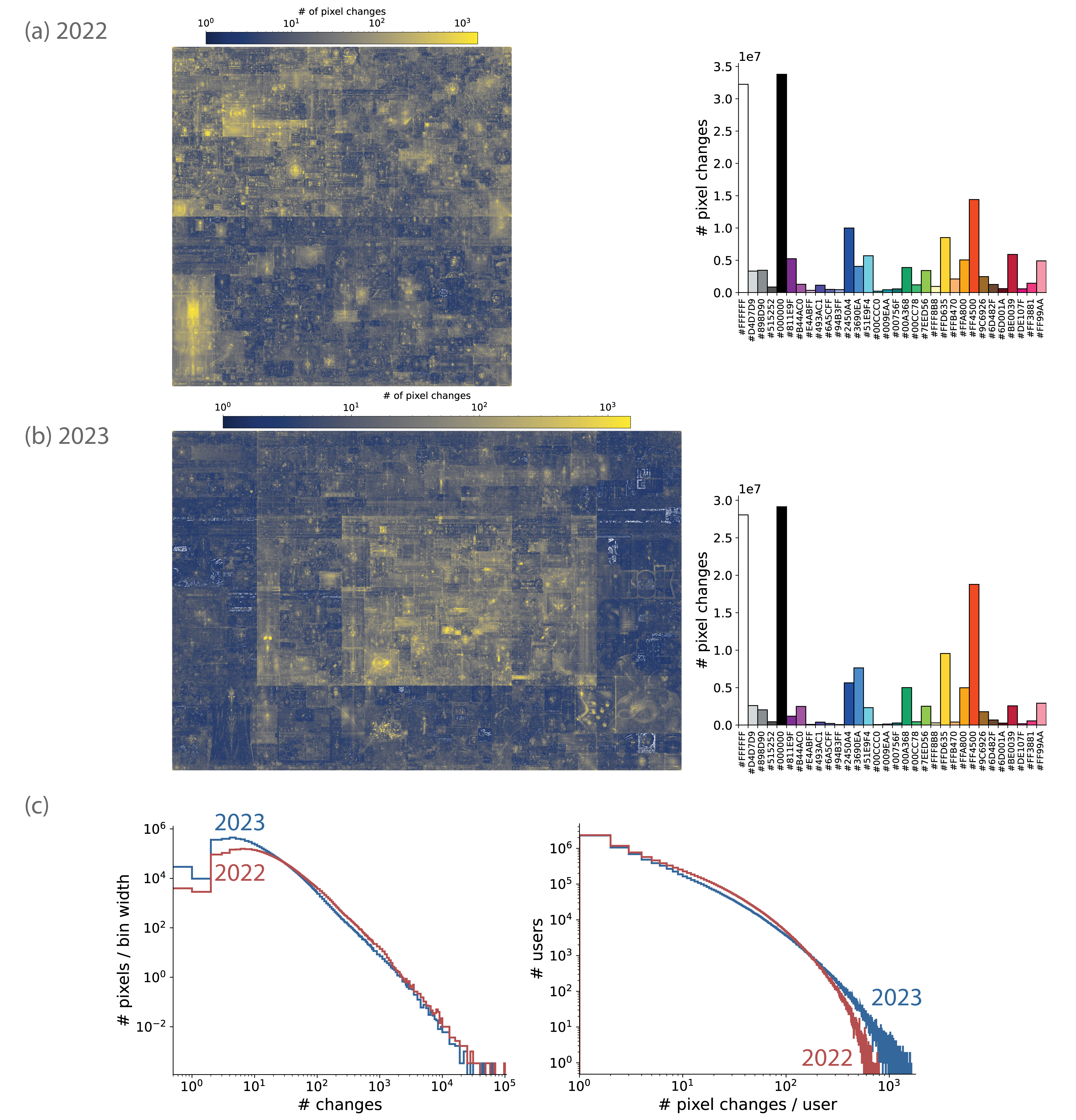}
\centering
\caption{General statistics of the r/place pixel change data from 2022 and 2023. \textbf{(a), (b)} For 2022 (a) and 2023 (b), the heat map of the number of pixel changes over the course of the entire event (left), and the distribution of pixel changes across available colors (right); note that some colors and canvas regions were only available later in the game. \textbf{(c)} For 2022 (red) and 2023 (blue), distributions of the number of changes per pixel (left), and distributions of the number of pixel changes per user (right).}
\label{figSI:rplace}
\end{figure}

\paragraph{Canvas basic rules and statistics.}
We use the data from the 2022 and 2023 editions of the r/place game-experiment. These events offered an initially white canvas to all registered Reddit users without stating an explicit goal. Users could choose a new color for any pixel, but had to wait for at least a 5-minute \textit{cooldown} before modifying another pixel. Due to this time constraint, the users had to collaborate and organize to create drawings carrying cultural significance---which we call \textit{compositions}. The 2022 event featured 160.4 million pixel changes from 10.4 million users across 3.5 days; the 2023 event featured 133.8 million pixel changes from 8.6 million users across 5.4 days. It is important to note that different pixels attracted widely different attention; a few thousand pixels stayed white for the whole event, while the top left pixel in 2022 was modified about 98 thousand times. Similarly, about 2 million users performed a single pixel change, while some users performed more than 500 changes. General statistical distributions of the 2022 and 2023 canvas are shown in Fig.~\ref{figSI:rplace}.

\paragraph{Raw data.}
The data from all r/place editions (2017, 2022 and 2023) is publicly available from Reddit. We extract the data for 2022 and 2023 from \url{https://placedata.reddit.com/data/canvas-history/index.html}. The dataset contains, for each pixel change, the pixel location, the used color, the time it was performed (at millisecond precision), and an anonymized user tag---allowing us to link changes performed by the same Reddit user. We first convert this data into a structured numpy array that reduces information redundancy and compacts the data by replacing user tag string with integers and removing duplicate changes.

\paragraph{Canvas extension history.}
New colors and pixels were added at times unknown to the users. The canvas initially showed a million pixels, but reached 4 million pixels in 2022 (6 million in 2023). In addition, there were from 8 to 32 available colors to choose from. For reference, we show in Table~\ref{tabSI:extensions} the discrete changes in available pixels and colors and the times at which they occurred. These extensions are not considered to significantly change the dynamics we studied, as the compositions are relatively localized, so compositions on new canvas regions behave as if it were the beginning of a new game, and the number of colors did not seem to limit user activity.

Toward the end of both events, only changes to the color white were allowed for about an hour, so compositions appeared to rapidly vanish before the event closed---the so-called \textit{white-out}. An hour before the white-out in 2023, only five grayscale colors were allowed. All times where only white or greyscale colors were available are excluded from the analysis, as this reduction of available colors did not allow for maintenance of existing compositions. 

\begin{table}[h]
    \centering
    \textbf{2022}
    \vspace*{3mm}
    
    \begin{tabular}{r|c|c|c|c|c|c|c}
    time of extension (in s)& 0 & 99635& 99646& 195574& 195583 & 295410 & 300590\\
    \hline
    number of pixels & 1M & & 2M&& 4M&&end\\
    number of colors & 16 & 24 & & 32&& 1 (white) & end
    \end{tabular}

    \vspace*{3mm}
    \textbf{2023}
    \vspace*{3mm}

    \setlength{\tabcolsep}{4pt}
    \begin{tabular}{r|c|c|c|c|c|c|c|c|c|c|c|c|c}
    time of extension (in s)& 0 & 97208& 116999& 155754& 208964& 277192  & 277200 & 305994 &361992 & 362003 & 456228 &459029 & 463090\\
    \hline
    number of pixels & 1M &1.5M && 2M&3M && 4M&5M&&6M&&&end\\
    number of colors & 8 &      &16& &&24& &&32&& 5 (grayscale) &1 (white) & end
    \end{tabular}

    \vspace*{2mm}
    \caption{Time points at which the canvas was extended or offered more color choices.}
    \label{tabSI:extensions}
\end{table}

\paragraph{Cooldown.}
In general, the enforced cooldown between two consecutive pixel changes by the same user was 5 minutes. However, some users were capped at a 10 or 20-minute cooldown, notably if their Reddit account was not verified. Other changes in rules took place at the end of the 2023 event, where the cooldown was reduced incrementally: part of the users, chosen based on credibility criteria by moderators, were offered a 4-minute cooldown from second 273600. We neglect the effects of this change on composition dynamics, as it is small and concerns only a fraction of the users. Right after the white-out in 2023, the cooldown was gradually reduced for all users to 3 minutes, then 2, then 1, and finally 30 seconds. Times with these cooldown values were excluded already by excluding the white-out period.

\paragraph{Redundant changes.} 
Sometimes, users ``changed'' pixels using the same color the pixel was already in, potentially by mistake, out of boredom, or to have their user name attached to that pixel. There were 26.4 million such \textit{redundant} changes in 2022 and 17.7 million in 2023; they form an essential part of the canvas dynamics and should not be removed. Interestingly, 46 thousand (1.1 million) redundant changes in 2022 (in 2023) are also redundant in the user name, but we do not assign a special status to them since the motivations for making these changes may be similar to other redundant changes.

\paragraph{Moderators.}
Moderators from the Reddit team were given rights to modify many pixels at a time to erase inappropriate content. The changes made by moderators in a filled rectangle or circle of pixels are labeled in the raw data. However, the single-pixel changes by moderators are not labeled. We identify a portion of these by tagging as moderator changes those whose user tags are the same as for other explicit moderator changes; this is efficient only in 2022 as moderators in 2023 usually changed user tags at every action.
There were about 100 thousand explicit moderator changes in 2022 and 1.6 million in 2023. Though these changes break the 5-minute cooldown rules and could show as sudden peaks of activity in our variables, they are still part of the dynamics seen by users and constitute a negligible part of the user activity, and were therefore not discarded from the computation of our variables.

\paragraph{Multiple accounts and bots.}
Highly engaged users created multiple accounts to bypass the 5-minute cooldown. These accounts are listed as separate users in the data and considered separate in our analysis, as there is no way to distinguish them using the anonymized data. Though this is considered ``cheating'' and was fought against by the moderation team, it is still part of the canvas dynamics observed by all users, and thus should be included in our analysis. Some changes were also made by ``bots'', meaning without a human presence. Users programmed these bots to continue contributing toward their goals even when they were not actively playing. Bots also often relied on numerous ad-hoc Reddit accounts. This practice was frowned upon by the Reddit community, and was actively fought against by the r/place organizing team. However, this team could not detect all bots and their changes, so there was still some significant bot activity, in particular in 2023. An estimated 5 million pixel changes did not use the graphical interface of r/place in 2023. This number was likely smaller in 2022, when there were fewer widely-available tools to do so. Again, there is no infallible way to exclude these changes, but they are still part of the dynamics as they affected the canvas seen by users.

Even when excluding flagged moderator changes and the times with intentional reduced cooldown values in 2023, some changes were made by the same user within less than 5 minutes: 108 thousand changes in 2022 and 1.6 million in 2023. One source of these changes are moderator changes that are not identified, which also explains the much higher number in 2023 as there were more moderator changes that year. Other possible explanations include communication lags between user actions and Reddit servers where two changes from a user were recorded at the same time despite occuring with more than 5 minutes between them; or there could have been lags of Reddit servers themselves, allowing for two very close changes from the same user to be both recorded. Bots might have exploited server lags to perform faster changes, although we cannot verify this with the available dataset. Again, we neglect the effects of these cooldown-cheating changes as they are few and are part of the canvas dynamics visible to users. 

\subsection{Data of compositions}
Users in r/place spontaneously formed communities, each of which built pixel drawings they defended. These compositions form the subsystems we study and compare.

\label{secSI:compodata}
\paragraph{Crowdsourced data.}
We use the Atlas of all compositions referenced by users after each r/place edition ended~\cite{haagmansPlaceAtlasInitiative2024}, extracted from \url{https://raw.githubusercontent.com/placeAtlas/Atlas/master/web/Atlas.json} for 2022 and \url{https://raw.githubusercontent.com/placeAtlas/Atlas-2023/master/web/Atlas.json} for 2023. It contains canvas coordinates for the borders of the image over various time ranges with up to 30-minute resolution for each composition, as well as a name, a description, and often links to social media channels where the corresponding community organized---normally a subreddit forum, a Discord server, Youtube channel, or Twitch channel. Despite imperfections, this Atlas contains essential cultural information that an automatic detection algorithm would miss, so we emphasize that this crowdsourced data is key to this work. 

A majority of compositions have fixed borders, but some have different borders over different time ranges; in this case, we record pixel changes in the union of all borders, but we designate as \textit{active} only the pixels that are in the correct borders at a given time. Only changes of active pixels are included in the calculation of time series describing each composition.

\paragraph{Data cleaning.}
In practice, we build data objects that contain information about all pixel changes within the borders and time ranges of each composition. This pixel change information includes the time, user identifier, color identifier, position on the canvas; we also include more high-level information, such as whether the change is on an active pixel, from a moderator, redundant, or cheated the cooldown time. These objects are the input of the calculation of the time series describing each composition.

We performed some cleaning operations on the definition of compositions and their borders with minimal assumptions:
\begin{itemize}
  \setlength\itemsep{0.5mm}
    \item We remove duplicate compositions, defined as having
    more than 99\% spatial overlap and identical time ranges. We define \textit{spatial overlap} as the intersecting area of the two images divided by the area of the largest image. In practice, we kept the largest composition, merging metadata from both.
    \item We change the name of compositions that have different borders but the same name in the Atlas. 
    \item Rectangles are a common composition shape. To correct obvious mistakes in the drawing of the borders by contributors to the Atlas, we transform the compositions with a single time range and whose shape is close to rectangular into the smallest rectangle containing all pixels with the original border. We define a shape as close to a rectangle when both its dimensions exceed 10 pixels and its corners need to be moved by two pixels or less for the shape to be a rectangle.
    \item Compositions are mostly referenced thematically in the Atlas, such that the work of some communities that have been erased in one place but reappeared elsewhere is sometimes listed as a unique composition. 
    We consider `moved' compositions (that do not have any spatial overlap in different time ranges) as separate compositions. This significantly increases the number of studied subsystems.
    \item In compositions listing areas that do have a spatial overlap but are active on discontinuous time ranges, we include the gap between these time ranges so that the composition is defined on a continuous time range. In combination with the previous point, this means that compositions are made continuous in canvas space and in time.
\end{itemize}

Within a composition cleaned as above, there can be significant changes in the borders of the image, as encoded in the Atlas. Due to a border change, a composition's time series can show a discontinuity that is unrelated to canvas activity dynamics. We therefore define \textit{stable-area time ranges} over which the spatial overlap between images before and after any border changes always exceeds 90\%. We generally ignore times close to the start of a stable-area time range. When the computation of time series depends on multiple time steps, the necessary information is recorded over the union of all image borders. 

In 2022, there were 10,890 compositions before separating non-overlapping areas and 14,160 compositions after separation. In 2023, there were 6,667 compositions before separation and 6,930 compositions after. Additional requirements are set for the training dataset: Only the 8,871 in 2022 (5,417 in 2023) clean compositions with more than 100 pixels are kept, to avoid small-statistics noise in the time series. From these, only 6,291 (3,546 in 2023) compositions have at least one time instance that passes all requirements for the training data.

\section{Transitions}
\label{secSI:transitions}
This work predicts transitions in a coherent way over thousands of subsystems, which first requires defining transitions in these subsystems. Most variables describing the state of each composition, which are our subsystems of choice, vary along a continuum. These variables often change substantially during a transition, but not in a binary way. Therefore, defining transitions necessarily involves setting arbitrary thresholds on a state variable. To our knowledge, no universal method exists to define transitions on a continuous state variable, so we rely on a system-specific appreciation of how users perceive a transitioned composition. This translates into an absolute and a relative threshold on \texttt{diff pixels reference}, the fraction of pixels differing from the reference image. The image must have 35\% of pixels differing from a proxy of what users perceive as the typical image of the composition, and that fraction must be 6 times higher than its average over the past sliding window. 

Another requirement for transitions is that they must begin at least one sliding window width after the start of a stable-area time range. This ensures that \texttt{diff pixels reference} is computed excluding jumps unrelated to canvas dynamics when the image borders change. 

\begin{figure}[h!]
\centering
\includegraphics[width=0.75\textwidth]{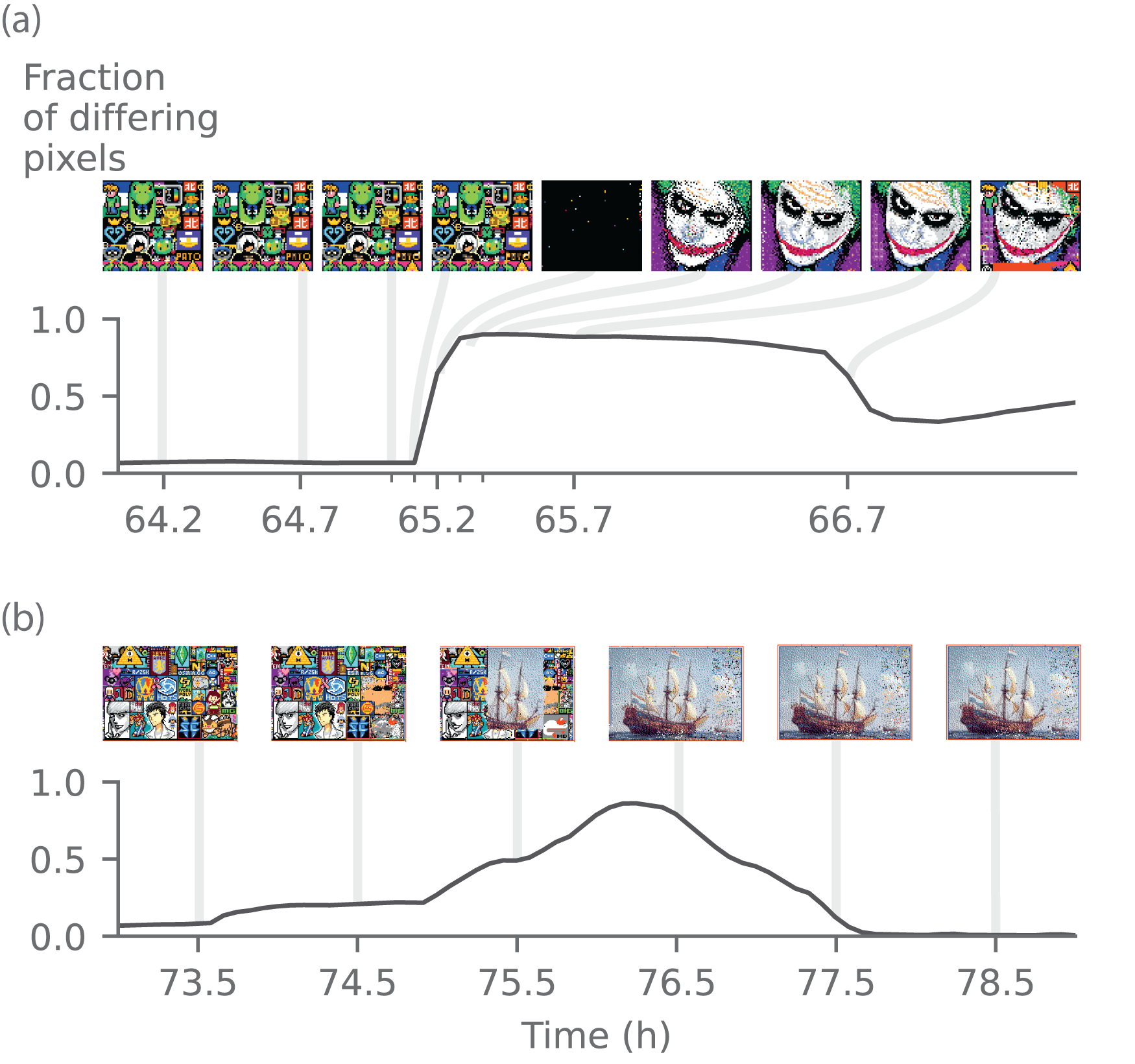}
\caption{Examples of compositions where no transitions are tagged despite a large image change. \textbf{(a)} The fraction of differing pixels (\texttt{diff pixels reference}) passes the transition threshold requirements, but the pre-transition image is a patchwork of compositions. \textbf{(b)} Fraction of differing pixels shows
too slow of a change to a new composition. In addition, the pre-transition image is not an Atlas composition, hence not a monitored subsystem.}
\label{figSI:notrans}
\end{figure}

\paragraph{Stability of pre-transition systems.}
We defined transitions to represent what human players would perceive as transitioning systems. However, to resemble a practical, real-world warning system, we had to reject some periods that might look like transitions to users; see two examples in Fig.~\ref{figSI:notrans}. First, requiring a relative threshold of 6 on the \texttt{diff pixels reference} ensures that the pre-transition system shows significant stability. 
A real-world warning system would not monitor unstable systems, as it would be unclear if they are already in transition or whether there is a clear state or equilibrium to transition from. This relative threshold combined with the absolute threshold of 0.35 ensures the abruptness of the regime shift; it would be difficult in our system to include smooth transitions, as nearly any gradual evolution of a composition image could then qualify as a transition. We checked that the quality of our machine learning predictions does not change significantly when varying these thresholds; this sensitivity study is shown in Fig.~\ref{figSI:sensitivity}a. 

\paragraph{Transitions from a patchwork.}
For motivations of applicability to a live warning system, predicting transitions makes sense only in a system that is already monitored and considered of interest. 
Therefore, we exclude transitions from a patchwork of pre-existing compositions into one new composition. We could technically include these transitions in our study because in the full dataset, we already know this canvas area will become a composition; but this would not be known in a live warning system. For similar reasons, we also exclude the times before the start of an Atlas composition from the training data. 
Note that not labeling these periods as transitions reduces our apparent prediction performance: in addition to reducing the available training data, the dynamical properties of this type of transition would be easier to discriminate than those of the transitions we retain, as there are specific properties of composition patchworks that the algorithm could associate with the incoming creation of a composition. 

\paragraph{Time of the transition.}
The variable we aim to predict with our algorithm is the time remaining before the next transition, if one exists. To this end, we define the exact time of the transition using the values of \texttt{diff pixels reference} at the time step the transition was identified and at the previous time step. The time of transition \ttrans is the moment at which a linear interpolation of \texttt{diff pixels reference} crosses both transition thresholds. 

\section{Composition-level time series and variables}
\label{secSI:variables}

For each composition, we define many time series variables that describe various aspects of its evolving state. These variables are computed at 5-minute time steps; we call the set of feature values of these variables at a given time step a \textit{time instance}. The 5-minute width of the time steps coincides with the cooldown, which is the most relevant time scale in the data. Let us describe in more detail all the time series and variables that we have defined with the expectation that they could contribute to predicting the next transition. They are organized by types of variables, including efforts to reproduce variables commonly used to assess critical slowing down. As explained in Section~\ref{secSI:varselect}, only 19 times series and 5 time-independent variables are selected to be input in the training, for their higher contributions to the predictions and lower correlations. 

\subsection{Normalizations}

Here, we describe standard transformations that we apply to some variables input in the training. 

First, some variables are normalized by time and area units. The time unit is 5 minutes, meaning that variables are effectively already time-normalized considering our 5-minute time step. The area unit is the number of active pixels in the composition image at a given time step, 
which may change as the composition’s borders evolve in the Atlas.

Second, some variables are computed over or averaged over the past sliding window. This is a moving average whose range always ends at the considered time step.

Third, some variables have absolute values with different typical values in different compositions. To make values of these variables comparable among compositions, we instead consider relative changes in these variables. This is computed as the ratio of the variable at this time step to its average over the past sliding window---except for the autocorrelation where the sliding window average is instead subtracted, as the autocorrelation can have value zero.

Last, the averages over pixels or over the top decile of pixel values use only the pixels active at the considered time, \textit{i.e.} those summing to the current area. 

\subsection{State variable}

A major challenge in interpreting our results within the framework of early warning signals literature is the absence of direct equivalents in our system for the variables typically used as indicators. These indicators include the variance, lag-1 autocorrelation and rate of return after perturbation, computed on a state variable that represents the system’s overall health. Because our systems are described by a number of possible image states that increases exponentially with the number of pixels and number of color, they are difficult to summarize with a single state variable. Some variables can adequately describe the stability of the system, such as the \texttt{diff pixels reference} variable we chose to define transitions. However, they are not simple scalar quantitative indicators, like population in ecology or price in economics, and they often do not conform to the paradigm of small perturbations around a well-defined equilibrium (see Discussion in main text and Section~\ref{secSI:toymodel}).

We therefore design multiple equivalents for variance, autocorrelation, and return rate, each offering a different approximation of standard early warning indicators which could capture distinct properties of the pre-transition periods. For the purpose of reaching higher performance of the algorithm, only the most predictive variables are kept. However, when interpreting the results, it should be noted that these variables do not precisely correspond to their conventional definitions in the literature.

\subsection{Variance}

The variance can be defined in many ways in our system. We explored two main approaches: estimating the variations of a scalar variable describing the image change or directly of the image. Let us consider first the latter. We start with variables that estimate linear variations as they have simpler definitions in the context of r/place, before considering variables with quadratic variations. Most of the proposed variance variables are very correlated, though they can still capture slightly different dynamical properties. 

\paragraph{Differing pixels and other linear variations.}

Multiple variables are based on the fraction of pixels that differ between images at two different times. The considered pixels are always the ones active at the considered time step, even when they differ from those in the borders of the past image.

The variable \texttt{diff pixels reference} compares the instantaneous image at time $t$, as seen by users on the canvas at the end of the considered time step, to the reference image in the sliding window as defined in the main text; this is normalized by area. The variable \texttt{diff pixels instant} compares the instantaneous image at time step $t$ to the image composed of the mode colors of the previous time step $t-1$; this is normalized by time and area. Another possibility is to compare to the instantaneous image at $t-1$, but this showed lower contribution to the predictions.

Another route relies on a measure of how stable the pixel colors are within a single time step. We define the instability of a pixel by the fraction of this time step that it spends in a color different than the mode color:
\begin{equation*}
    \textrm{instability} = \frac{1}{N_p}\sum_{p}\sum_{c\neq \textrm{mode}(p)} f_{p,c}
\end{equation*}
where $f_{p,c}$ is the fraction of this time step that pixel $p$ spent in color $c$, and the sum runs over the $N_p$ considered pixels and all colors that are not the mode color for pixel $p$. Averaging over pixels with the top decile of values results in the \texttt{instability top} selected variable (which is time normalized), while averaging over all pixels resulted in slightly lower performance.

A similar variable, which we excluded for its high correlations with instability and the fraction of differing pixels, is the cumulative attack time. It represents the average, over all pixels, of the fraction of this time step that a pixel spent in a color different from the reference color.

\paragraph{Quadratic pixel variations.} 

Though the image is not a scalar continuous variable, we can still get closer to the standard variance definition by computing quadratic variations of the color of each pixel. The following definitions are meant to estimate the spread of the distribution of colors that a given pixel has been in, relying on the quantity $S_p = \sum_c f_{p,c}^2$, instead of a linear sum of the $f_{p,c}$ fractions we discussed above. This quadratic sum equals $1$ when a single color is used for pixel $p$ this time step, while its theoretical minimum, occurring when the pixel spends an equal time in all colors, is $1/N_c$, where $N_c$ is the number of colors. 

As $S_p$ is anti-correlated with the spread of the color distribution, a first option for the variance is 
$$\frac{1}{N_p}\sum_p \frac{1}{S_p}.$$ 
Another option is to consider first the sum $S_p$ for all pixels before inverting it, which amounts to computing the spread of colors directly on the entire image: 
$$\frac{N_p}{\sum_p S_p}$$ 
Both these variance variables equal $1$ in fully inactive time steps, and $N_c$ when all pixels stay an equal time in all colors over the time step. 
We used the maximum $N_c=32$ along the whole time of the event because the highest reached values are far from the theoretical maximum, showing that the time-dependent number of available colors does not act as an upper bound. 

How widely time is distributed across a limited set of colors for a given pixel is similar to the dispersion of independent categorical variables, which is well represented by the variance of a multinomial distribution. In this context, it translates into $\sum_c f_{p,c} (1-f_{p,c}) = 1 - S_p$. Therefore, we define the multinomial-like variance as
$$1 - \frac{1}{N_p}\sum_p S_p.$$
This is also called the Gini impurity index in some contexts. A quantity of very close interpretation is the Shannon entropy of the color distribution of a pixel
$$-\sum_c f_{p,c} \log f_{p,c},$$
that we set aside as it is very highly correlated with the multinomial version.

A last similar possibility is to sum the squared color frequencies only for non-mode colors. For this, we define a \textit{non-mode dot product}: $$D(\vec{a},\vec{b}) = \left(\sum_i a_i b_i\right) - \max_i(a_i b_i),$$
where $\vec{a}$ and $\vec{b}$ are arbitrary vectors. In our usage, the vector components correspond to the time a pixel spent in each color, meaning there are 32 components. This results in the variance
$$ \frac{1}{N_p}\sum_p D(\vec{c}_p,\vec{c}_p) = \frac{1}{N_p}\sum_p [S_p - \max_c(f_{p,c}^2)].$$

When computed over all time steps and compositions, all the above quadratic variables showed very high correlations ($\rho>0.95$) with each other and with another more intuitive variable: the number of colors that a pixel is assigned within this time step. We end up using the pixel average \texttt{n colors} and top decile average \texttt{n colors top}, as they outperform all these variables. 

Another approach to close the gap with the standard variance formula for a 32-color pixel results in the seemingly linear variables explained earlier. Considering $t_i$ as infinitesimal time steps within the time step $\Delta t$ under study, we take every color at $t_i$ as a single point in a sample for which we calculate the variance:
$$\frac{1}{\Delta t}\sum_{t_i} (c_i - \langle c\rangle)^2 = \sum_{c} f_{p,c} (c - \langle c\rangle)^2 = \sum_{c\neq \langle c\rangle} f_{p,c}$$
The ``mean'' color $\langle c\rangle$ is not well defined as the color $c$ is a categorical and not a 
continuous variable, so we define it following two different assumptions. For the same reason, the difference of colors is not well defined, but we assume it is $1$ where the two discrete colors are different, and $0$ otherwise (neglecting that some color pairs could be considered more different than others). 
if we define the ``mean'' color $\langle c\rangle$  as the mode color or the reference color, we recover the instability or the cumulative attack fraction, respectively.

\paragraph{Variations of an existing time series.} 

A different approach is to summarize the state of the image in a scalar time series $v(t_i)$, before calculating at time $t_n$ the standard variance of its values over a past sliding window $(t_1,...,t_n)$. 
Here, we try to avoid the assumptions due to defining a mean image over the past sliding window, which we take in some cases to be the reference image. We can avoid the need for a mean in the variance via this approximated variance:
$$\sum_{t_i} \frac{n}{2(n-1)} v(t_i)^2,$$
which converges to the standard variance $\sum_{t_i} \frac{n}{n-1} (v(t_i)-\langle v\rangle)^2$ at large $n$. 
A variable that effectively describes the differences at the image level \textit{between each time step} (and not the difference from the mean) is \texttt{diff pixels instant}, which we use for $v$.
This is therefore the \texttt{variance} we use in the training, with a sliding window of 10 time steps (50 minutes). Out of the variance variables we calculate, this one is the most similar to those calculated to evaluate critical slowing down in the ecology literature. 

\subsection{Autocorrelation}

We define multiple variables that are analogs, for an image, to the lag-1 time autocorrelation of a scalar quantity. A first route relies on the same quadratic differences used to define the variance variables, applied to the color distributions of a pixel $f_{p,c}(t)$ in this time step and $f_{p,c}(t-1)$ in the previous time step. Using the non-mode dot product once again, a first autocorrelation variable, \texttt{autocorr non-mode}, is 
$$ \frac{1}{N_p}\sum_p D(\vec{c}_p(t),\vec{c}_p(t-1)) = \frac{1}{N_p}\sum_p \left[\left(\sum_c f_{p,c}(t)f_{p,c}(t-1)\right) - \max_c(f_{p,c}(t)f_{p,c}(t-1))\right].$$
Now using an analogy with the variance of a multinomial distribution, another autocorrelation variable is:
$$\frac{1}{N_p}\sum_p \left[\sum_c f_{p,c}(t)(1-f_{p,c}(t-1))\right].$$
Lastly, in analogy with the $\chi^2$ dissimilarity between two distributions, we also define the following autocorrelation:
$$\frac{1}{N_p}\sum_p  \left[\sum_c \frac{1}{2} (f_{p,c}(t)-f_{p,c}(t-1))^2\right].$$
These three variables are highly correlated over all time instances, so we kept only the best-performing one.

We also define a per-pixel autocorrelation variable 
that compares the color distributions in this time step $c_t$ and in the previous one $c_{t-1}$, with respect to a ``mean'' color distribution $c_{\textrm{ref}}$---that of the reference image. For each pixel, the case-by-case autocorrelation is: 
\[    
\begin{cases}
      0 & \text{ if } c_{t}=c_{t-1}=c_{\textrm{ref}} \\
      1 & \text{ if }  c_{t}=c_{t-1} \text{ and }  c_{t}\neq c_{\textrm{ref}} \\
      0 & \text{ if }  c_{t}\neq c_{t-1} \text{ and } (c_{t}=c_{\textrm{ref}} \text{ or } c_{t-1}=c_{\textrm{ref}}) \\
      -1 & \text{ if }  c_{t}\neq c_{t-1} \text{ and } c_{t}\neq c_{\textrm{ref}} \text{ and } c_{t-1}\neq c_{\textrm{ref}}
    \end{cases}
\]
\texttt{autocorr by case} is the average of the above over all pixels. We need this case-by-case definition because standard correlations require a notion of directionality lacking in the color space. This issue did not appear in the above discussion on variance, as the sign of the deviations do not matter there.

\subsection{Attack duration and return rate}

In addition to the variance and autocorrelation, the time or rate of return to equilibrium after a perturbation is another quantity used as evidence for critical slowing down. We take the reference image as our proxy for the equilibrium that the instantaneous image should return to. 
The return rate therefore measures the fraction of attacked pixels that are changed back to the reference color within one time step. We also introduce a related variable, the attack duration, which we initially designed as a return time on a longer time scale but then attributed a more nuanced interpretation.
 
The attack duration of a pixel is defined based on two cases. 
If the pixel is in an attack color, the attack duration is the time since the moment the pixel was attacked, meaning when the pixel left the reference color. For a pixel in the reference color, it is the duration of the last attack on this pixel, if this attack started within the sliding window. Pixels not attacked within the sliding window are ignored. We use as training variables both \texttt{attack duration}, the average over recently-attacked pixels, and \texttt{attack duration top}, the average of the top decile of pixel values. 

It is important to note that certain attacks against the reference might be innovations in the image, agreed upon by the community defending this composition. This is a problem when the time scale of innovations is shorter than the sliding window, meaning the reference image is not truly the equilibrium to return to. The consequence is that the \texttt{attack duration} can increase without indicating a lower performing community or a more likely transition---which is why we do not name this variable ``return time''. 

We define the \texttt{return rate} as a variable on a shorter time scale to partially avoid the issue of innovations introducing lingering differences from the reference. It considers all pixels that are in an attack color at the end of the previous time step and computes the fraction of these pixels that returned at least once to the reference color. 
There can still be attacked pixels that are actually innovations, but they contribute only a single pixel in the fraction, so they do not dominate over the defense reactions. On the contrary, innovations dominate in the \texttt{attack duration} variables because they can reach values of the order of the sliding window width, which is much larger than the typical reaction time of the defense. 

\subsection{User activity}

Here, we present the variables associated with the activity of users changing pixels in a composition. 

First, we consider two variables related to the type of pixel change. We use the \texttt{redundant changes} as defined in Section~\ref{secSI:rplacedata}. Then, we measure the \texttt{attack fraction} as the fraction of pixel changes in the time step that differ from the pixel's reference color. It is set to $0.5$ when there are no pixel changes in the time step. 

We then consider variables that count users. The number of users contributing to a composition in the sliding window, \texttt{n users sw}, is normalized by the number of time intervals (60) contained in the sliding window range and by its area. Then, \texttt{new users} is the fraction of users active in a composition and time step but not in the past sliding window. Another possibility is to consider the fraction of new users compared to those of the previous time step, which we excluded as it is often very close to $1$: only a tiny fraction of users consistently changes pixels in consecutive time steps. Lastly, \texttt{changes/users sw} measures the engagement of individual users; it is the ratio of the number of changes to the number of users over the sliding window.

\subsection{Image complexity}
\label{secSI:imagecomplex}
We use multiple variables to estimate the complexity of the image composition. Complexity is key in the dynamics of the community, as simpler images may be easier to coordinate but might attract fewer supporters.

\paragraph{Entropy.}
We compute a proxy for an entropy measure known as the computable information density, introduced in Ref.~\citenum{martinianiQuantifyingHiddenOrder2019}, which is the ratio of the total  length of losslessly compressed binary data to the original length of the data. We tested a variety of different lossless compression algorithms: Lempel-Ziv 1977 (LZ77), Lempel-Ziv 1978 (LZ78), and Deflate. Due to a combination of computation speed and compression factor, we chose the LZ77 algorithm. We use the implementation of LZ77 from the open-source Sweetsourcod library~\cite{martinianiSweetsourcod2019}. Note that we are not the first to use image compression to characterize image complexity on r/place; M\"uller and Winters used compression to characterize the evolution of certain canvas compositions in the 2017 experiment, as well as the entire canvas~\cite{mullerCompressionCulturalEvolution2018}.

These compression algorithms require a one-dimensional representation of the data. We must therefore flatten our two-dimensional images. The naive approach is a simple row-by-row 
raveling, where rows are sequentially concatenated one after the other. An approach that aims to conserve the locality in the image is to use the Peano-Hilbert space-filling curve, also known as a Hilbert Scan~\cite{martinianiQuantifyingHiddenOrder2019}. Though the Hilbert scan can in theory lead to better compression rates than a simple ravel, we did not observe this consistently. Because the Hilbert scan showed limited compression benefits and significantly increased computation time, we opted instead to use a simple ravel.  

To obtain comparable values among compositions and favor dynamical signals within a composition, we use the \texttt{entropy change}, which is the ratio of the entropy to its sliding window average. Because of this normalization of the entropy measure, we only need to calculate the numerator of the computable information density, which is the length of the losslessly compressed linearized data describing the image. The denominator, the uncompressed data length, is proportional to the number of pixels. Since scaling factors will cancel out during the normalization, we can simply use the pixel number as the denominator. 

When calculating entropy values for a composition over the sliding window without normalizing by a per-composition reference value, as in our \texttt{entropy} variable, we instead need to introduce a new normalization to make sure that entropy values are comparable across different compositions. Since this entropy measure is known to scale with the size of the image (see SI appendix of Ref. \citenum{martinianiQuantifyingHiddenOrder2019}), we must normalize as a function of the image size. We therefore introduce the normalization:
$$
\textrm{Entropy}_\textrm{norm} = \frac{\textrm{Entropy}_{\textrm{comp}} - \textrm{Entropy}_{\textrm{min}}(A)}{\textrm{Entropy}_{\textrm{max}}(A) -\textrm{Entropy}_{\textrm{min}}(A)}
$$
where $\textrm{Entropy}_{\textrm{comp}}$ is the entropy of a composition; $\textrm{Entropy}_{\textrm{min}}(A)$ is the minimum entropy for an image of the same area $A$ (or number of pixels) as the composition; $\textrm{Entropy}_{\textrm{max}}(A)$ is the maximum entropy at a given area. 
The minimum entropy is simply that of a blank image of a given size, which is proportional to $1/A$. 
To find the maximum entropy, we generate 10 images of the given image size by randomly choosing from 8, 16, 24, or 32 available colors depending on the number of available colors at that time. We then take the mean compressed length of the 10 images, and use this value as our maximum. 
Some images could be less compressible than a random image, but they are uncommon in our dataset, and the amount by which they are less compressible is not significant. 

\paragraph{Fractal dimension.}
The computation of the fractal dimension starts with separating the image into the 8, 16, 24, or 32 colors that are currently available to users. We use the reticular box counting method~\cite{bisoiCalculationFractalDimension2001}, separately for images composed of each single color of the original image: a grid of boxes is superimposed on top of the single-color image, and the number of boxes touched by the image is counted as a function of box size. At each box size, the grid is superimposed in four ways---by aligning the first box with each of the four corners. We take the average of these four box counts to minimize finite-size effects of the grid. We calculate the box counts for box size from 1 pixel up to 10\% of the smaller dimension of the bounding rectangle of the composition, requiring a maximum box size of at least 3 pixels, with a cap at 10 pixels. For each color, a power law is fitted to the mean box count as a function of the box size. The absolute value of the negative exponent is taken to be the fractal dimension for each color, and we calculate a scalar color-weighted fractal dimension by summing the dimensions of each color weighted by the fraction of pixels of that color in the composition. The weighted fractal dimension is $2$ when the color is uniform and decreases as the colored patch perimeters become more fractal. For the same reasons as for the entropy, we consider the \textit{change} in fractal dimension, \texttt{fractal dim change}, by calculating the ratio of the fractal dimension to its sliding window average.

\paragraph{Multiscale complexity.}
To explicitly measure complexity at different scales, we use a multiscale entropy-based measure introduced by Zhang~\cite{zhangComplexity1Noise1991}. In this measure, complexity is defined based on the cumulative disorder present across multiple spatial scales. To implement this, we perform a coarse-graining procedure appropriate for color images. We first downsample the image at a range of dyadic scales. Instead of taking the mean of pixels in a downsampled box, which would be meaningless for color values, we take the mode, as explained in \textit{Downsampled images} in Section~\ref{secSI:additionalVars}. We then compute the Shannon entropy of the distribution of pixel values in the downsampled image. Following Zhang, we calculate a cumulative complexity by integrating over the scale-dependent entropies weighted by scale length (for 2D images). Finally, we normalized by the total number of pixels in the original image to obtain the dimensionless \texttt{multiscale complexity}, taken as the ratio to its sliding window average. For this measure, highly regular or purely random patterns give lower complexity, while images that contain meaningful structure at both fine and coarse scales yield higher values. 

\paragraph{Levenshtein complexity.}
As another measure of spatial complexity, we implement the Levenshtein-based complexity index introduced by Papadimitriou~\cite{papadimitriouModellingSpatialLandscape2009}. The Levenshtein distance, which is widely used in computer science and bioinformatics, measures the minimum number of insertions, deletions, or substitutions required to transform one string into another, and provides a simple structural complexity metric. Following Papadimitriou’s approach, we treat each row and each column of the image as a string of symbols, where each pixel is represented by an integer value of 0–31 encoding its color. We compute Levenshtein distances between all adjacent rows and all adjacent columns, yielding vectors of row-wise and column-wise distances. Taking the row-wise vector as a column vector and the column-wise vector as a row vector, we perform an outer product, giving a 2D matrix; the mean of this matrix is taken as the overall complexity score. The variable \texttt{Levenshtein complexity} is that score divided by its sliding window average. Intuitively, this measure captures how dissimilar neighboring rows and columns are in their sequences of colors, with higher values indicating more dissimilarity, a proxy for more complex spatial organization. 

\subsection{Image characteristics}

\label{secSI:spatialvars}
Spatial metrics of the composition image

\paragraph{Wavelet variables.}
Wavelets provide a way to characterize features at multiple scales in an image because they decompose a pattern into contributions from different spatial scales, quantifying both fine detail and larger patterns. Our images are not grayscale intensity images but maps of 32 possible colors, represented by pixel values 0-31. Changing the numbering could change the absolute size of wavelet energies, but because the transform is linear this only rescales all values uniformly. In our analysis, we use ratios of wavelet energies across scales, so these rescalings cancel out. As a result, these measures of the structure of boundaries between colors are not affected by the arbitrary numbering of colors. 

We calculate wavelet energy measures from multi-level 2D discrete wavelet transforms (DWT)~\cite{grapsIntroductionWavelets1995}. We use the Haar wavelet to iteratively decompose each image into approximation (LL for low frequency on both dimensions) and detail (HH for high frequency, LH, HL) subbands. At each level, the DWT acts as a band-pass filter whose effective wavelength is proportional to \(2^L\), where L is the number of times the image has been downsampled. Each time the DWT is applied, the image is further downsampled, and the characteristic feature size gets larger. 

We define three frequency bands: (i) high-frequency energy, from the first-level details (finest scale, no downsampling), (ii) mid-frequency energy, from the level whose band-center wavelength was closest to 10\% of the smaller image dimension, and (iii) low-frequency energy, from the level nearest 40\% of the smaller dimension. For each band, we sum the squares of the detail subband coefficients (HH, LH, HL) to obtain scale-specific energy values. The variables used in our analysis are ratios of these three energy measures, capturing the relative balance of fine-, intermediate-, and coarse-scale structure that together reflect image complexity: the high-frequency to low-frequency (\texttt{wavelet high/low}), the mid-frequency to low-frequency (\texttt{wavelet mid/low}) and the high-frequency to mid-frequency (\texttt{wavelet high/mid}). The ratio to the sliding window average is used for these three variables, as well as the sliding window average itself (\texttt{wavelet high/low sw} and \texttt{wavelet high/mid sw}). 

Because wavelet analysis can also be performed on a time series as opposed to over space in an image, we perform another wavelet analysis on \texttt{diff pixels instant}, a variable close to our nominal state variable. At each time step, we apply a 1D DWT with a Daubechies-4 wavelet to the time series within the past sliding window, decomposing it into approximation coefficients and detail coefficients. We square the sums of these coefficients to give the low- and high-frequency wavelet energies, and take the ratio of high to low-frequency energies as a new variable, \texttt{wavelet high/low time}. This quantifies how much of the temporal variability is characterized by short-timescale fluctuations versus long-timescale trends. 

\paragraph{Clustering of changes.}

Another way spatial dynamics could provide predictive information is through the level of clustering of the positions of pixel changes.
Pixel changes can be more or less spatially clustered on the composition image, which might reflect differences in attack or defense dynamics. Coordinated attacks could target specific parts of the composition, whereas noisy attacks and defense changes could be more spread out.

The simplest way to 
quantify clustering is to measure how far changed pixels are from each other on the canvas. Within a time interval, we consider all pixel changes and compute the distances $d_{i,j}$ between each possible pair of changes $i$ and $j$. We then define a clustering measure as the average \texttt{distance between changes} $\frac{\sum_{i>j} d_{i,j}}{n(n-1)/2}$, where $n$ is the number of pixel changes.

The Ripley's K function~\cite{dixonRipleysFunction2014} gives a more fine-grained estimate of clustering, since it is a function of the clustering scale $d$. The causes for clustering can indeed vary with scale; for example, short-distance clustering may reflect an attack on an image feature, while long-distance anti-clustering could come from defense across the whole image. 
To calculate the K function, we start from the fraction of pairs of pixel changes whose distance is less than $d$: $\frac{\sum_{i>j} [d_{i,j} < d]}{n(n-1)/2}$. We then normalize this by a reference value for that composition, which is the average of the K function for 10 arrangements of randomly placed pixel changes. Contrary to a simple normalization with the image area, this accounts for the impact of various shapes on K values. We use this normalized Ripley's K function at three different scales: $d=2$, $d=\max(3, 0.2\sqrt{A})$ and $d=\max(5, 0.5\sqrt{A})$, measured in pixels, where $A$ is the number of pixels of the image. We name these variables \texttt{ripley d=2}, \texttt{ripley d=20\%} and \texttt{ripley d=50\%}.

\paragraph{Spatial autocorrelation.}
Rapid movement or transformation of features of the image could indicate pre-transition dynamics. The spatial correlation between images at the current time instance ($I_t(x,y)$) and the preceding one ($I_{t-1}(x,y)$) can capture these image changes. For each color $c$, we use the cross correlation of boolean images $I^c$ composed of $1$ for pixels containing $c$ and $0$ elsewhere, then sum for all colors:
\begin{equation}
    C(u,v,I_1,I_2) = \sum_c \sum_{x,y} I_1^c(x,y) I_2^c(x+u,y+v)
\end{equation}
where $(u,v)$ for which $C$ is high can be interpreted as relative coordinates of some similar features of $I_1$ and $I_2$. Because we care about analyzing the changes of image features with time rather than static image properties, we subtract the cross correlation between the images at $t$ and $t-1$ from that of the image with itself:
\begin{equation}
    C^t(u,v) = C(u,v,I(t),I(t)) - C(u,v,I(t),I(t-1))
\end{equation}

This cross correlation is still a two-dimensional array that we need to reduce to a few predictive scalars at each time $t$. As there should not be preferred directions across compositions, we first reduce to one dimension via the radial average 
\begin{equation}
    C^{t,1D}(r_i) = \frac{\sum_{u,v; \sqrt{u^2+v^2} \in [r_i,r_{i+1}[} C(u,v) }{
    \sum_{u,v; \sqrt{u^2+v^2} \in [r_i,r_{i+1}[} 1}
\end{equation} where ${r_i}$ is a binning of distances. The denominator simply counts the number of pixels of $C^t(u,v)$ at radius $r_i$ from its center. Then, we normalize $C^{t,1D}(r_i)$ by $C^{t,1D}(r_0)$ to get relative values that are comparable among compositions. The resulting function relates to possible shifts of features of the image between a time step and the next: negative values at radius $r_i$ indicate that image regions separated by a distance $r_i$ are more similar when comparing images at $t$ and $t-1$ than when comparing identical images. 

Finally, we extract three characteristic quantities from this one-dimensional function 
of the radial shift: \texttt{image shift min}, the minimum value across $r_i$ bins; \texttt{image shift min val}, the value of $r_i$ at the position of the minimum; and \texttt{image shift slope}, the slope of a linear fit of that function.

\paragraph{Downscaled images}
Large compositions might be perceived by users at a coarser resolution than the pixel scale. This means that some user decisions, and therefore some transitions, might be best understood at these coarser scales. Therefore, we consider downscaled versions of a variable close to our state variable: the fraction of differing pixels between instantaneous images at times $t$ and $t-1$. 

To downscale by a factor $s$, we draw boxes of size $s$x$s$ on the smallest rectangle containing the whole image, starting from the top left. Each pixel of the downsampled image contains the mode color of one of these boxes, meaning the color most present in the pixels within the box. For boxes that are not fully filled by pixels active in this composition, the mode color is computed only from the active pixels.

We then look at three scales: 2-pixel boxes, 4-pixel boxes, and the largest power of $2$ that results in a rectangle of at least 16 downscaled pixels. We name these variables \texttt{diff pixels inst downscaled 2}, \texttt{diff pixels inst downscaled 4} and \texttt{diff pixels inst downscaled 16px}.

\subsection{Time-independent per-composition}

We include in the training data five variables that are only trivially time dependent and, for four of them, have a typical value per composition:
\vspace{-2mm}
\begin{itemize}[noitemsep]
    \item \texttt{area} is the base 10 logarithm of the active area of the composition; 
    \item \texttt{entropy} is the entropy proxy computed as in Section~\ref{secSI:imagecomplex} averaged over the sliding window;
    \item  The \texttt{age} and \texttt{border corner center} as defined in the main text.
    \item The \texttt{canvas quadrant}, which is the number of the canvas extension the composition is placed in, ordered by time of opening of this part of the canvas.
\end{itemize}

\subsection{Variables selected in the training}
\label{secSI:varselect}
As stated in the main text, only 19 time series from those listed in this section are kept in the training, as well as 5 variables with trivial time-dependence; their full names, short names, and short descriptions are listed in Table~\ref{tabSI:variables}. 
An extended set of variables was also trained on, for the sensitivity analysis described in Section~\ref{secSI:additionalVars}; the additional variables compared to the nominal set of Table~\ref{tabSI:variables} are listed in Table~\ref{tabSI:additionalvariables}. The time series are chosen for training so as to maximize their combined predictive power, using criteria based on SHAP values and correlations. 
Our process for selecting variables and features is described in Section~\ref{secSI:featureselec}. The distribution of SHAP values over all time instances for each training variable is shown in Fig.~\ref{figSI:shapdistr}. 

\setlength\cellspacetoplimit{4pt}
\setlength\cellspacebottomlimit{4pt}
\begin{table}[hp!]
    \centering
\hspace*{-2mm}
    \begin{tabular}{c|c|Sc|c}
        \textbf{Full name} & \textbf{Short name} & \textbf{Short description} & \makecell{\textbf{Coarse} \\\textbf{time} \\\textbf{ranges?}}
        \\
\hline
\multicolumn{4}{l}{\textit{With memory}}\\
\hline
    fraction of differing pixels vs reference & \texttt{diff pixels reference}  & fraction of pixels that differ between $t$ and the reference image  & \\
    fraction of differing pixels vs previous time & \texttt{diff pixels instant}
    & fraction of pixels that differ between $t$ and $t-1$ &\\
    fraction of attack changes & \texttt{attack fraction} & \makecell{fraction of pixel changes whose color \\does not fit that of the reference image} &\\
    number of pixel changes & \texttt{n changes} &\makecell{number of pixel changes, \\normalized by time step size and image active area}&\checkmark\\
    instability for top decile &\texttt{instability top} & \makecell{fraction of time step spent in non-mode colors, averaged over \\the decile of pixels with highest values, normalized by time step size} 
    \\
    variance of diff pixels instantaneous & \texttt{variance} & \makecell{variance of the \textit{diff pixels instantaneous} variable \\over last ten time steps: burstiness of image activity} & \checkmark \\
number of colors per pixel & \texttt{n colors} & average number of colors shown by a pixel within this time step \\
number of colors per top decile pixel &\texttt{n colors top} &\makecell{number of colors shown by a pixel within this time step, \\averaged over the decile of pixels with highest values}&\checkmark\\
change of autocorrelation case by case & \texttt{autocorr by case} & \makecell{autocorrelation comparing the colors that are both different from \\the reference between $t$ and $t-1$, minus its sliding window average} \\
autocorrelation from non-mode product & \texttt{autocorr non-mode} & \makecell{autocorrelation comparing within-time-step deviations \\to the mode color between $t$ and $t-1$} \\
duration of last attack & \texttt{attack duration} & \makecell{average duration for which \\pixels showed their most recent non-reference color} &\checkmark\\
duration of last attack for top decile & \texttt{attack duration top} &\makecell{duration for which pixels showed their latest non-reference \\color, averaged over the decile of pixels with highest values}& \checkmark\\
return rate&\texttt{return rate} & \makecell{average fraction of pixels in non-reference colors at $t-1$ \\that returned to the reference at $t$}
&\\
number of users in sliding window & \texttt{n users sw} & \makecell{number of active users in the past sliding window \\normalized by time step size and image active area} &\checkmark\\
changes per user in sliding window&\texttt{changes/user sw} & \makecell{number of changes per user, averaged over the past sliding window}&\checkmark\\
fraction of new users vs sliding window& \texttt{new users} & fraction of users that were not active in the past sliding window &\\
fraction of redundant changes & \texttt{redundant changes} & fraction of pixel changes that repeat the existing color &\\
change of entropy vs sliding window & \texttt{entropy change}& image entropy divided by its sliding window average &\\
change of fractal dimension vs sw &\texttt{fractal dim change}&image fractal dimension divided by its sliding window average&\\
\hline
\multicolumn{4}{l}{\textit{Without memory}}\\
\hline
active area & \texttt{area} &logarithm of the number of pixels in the active image&--\\
age from Atlas birth & \texttt{age}& time from the composition birth from the Atlas &--\\
entropy in sliding window &\texttt{entropy}& \makecell{entropy of the image, normalized via the max and min entropy \\of an image of this area and averaged over past sliding window} &--\\
canvas quadrant &\texttt{canvas quadrant}& canvas expansion number for this part of the canvas &--\\
special canvas position &\texttt{border corner center}& \makecell{if the composition is placed close to \\a border, a corner, or a center of the canvas} &--
\end{tabular}

\vspace{3mm}
    \caption{Variables used in the nominal training. The short name is that used in figures and in the text when referring to this exact variable. The last column is checked for variables whose memory is recorded in 9 rather than 12 time range features.}
    \label{tabSI:variables}
\end{table}

\setlength\cellspacetoplimit{4pt}
\setlength\cellspacebottomlimit{4pt}
\begin{table}[h!]
    \centering
\hspace*{-2mm}
    \begin{tabular}{c|c|Sc|c}
        \textbf{Full name} & \textbf{Short name} & \textbf{Short description} & \makecell{\textbf{Coarse} \\\textbf{time} \\\textbf{ranges?}}
        \\
\hline
\multicolumn{4}{l}{\textit{With memory}}\\
\hline
    Levenshtein complexity index & \texttt{Levenshtein complexity} & \makecell{mean of the matrix of Levenshtein distances \\between adjacent rows and columns} &\\
    wavelets high to low frequency  & \texttt{wavelet high/low}  & \makecell{ratio of high and low frequencies from a 2D\\ discrete wavelet transform, divided by its sliding average}  & \\
    wavelets high to mid frequency & \texttt{wavelet high/mid}& \makecell{ratio of high and mid frequencies from a 2D\\ discrete wavelet transform, divided by its sliding average}  & \\
    time wavelet high to low frequency & \texttt{wavelet high/low time} & \makecell{ratio of high and mid frequencies from a 1D\\ discrete wavelet transform of \texttt{diff pixels instant} } &\checkmark\\
    min of normalized spatial autocorrelation & \texttt{image shift min} &\makecell{Minimum, across distances, of the difference\\ between the spatial autocorrelation of the image \\with itself or between $t$ and $t-1$ images}& \checkmark\\
    distance between pixel changes & \texttt{distance between changes} &\makecell{average distance between pairs of pixel changes}& \\
    Ripley's K at 2-pixel scale &\texttt{ripley d=2} & \makecell{Ripley's K function for pixel changes positions, at a scale\\ of 2 pixels, normalized by the value for random positions}\\
    Differing pixels downscaled by 2 &\texttt{diff pixels inst downscaled 2} &\makecell{fraction of pixels differing between $t$ and $t-1$ \\images downscaled by a factor $2$   }&\\
    Differing pixels downscaled to 16 px &\texttt{diff pixels inst downscaled 16px} &\makecell{fraction of pixels differing between $t$ and $t-1$ \\images downscaled to $\simeq16$ pixels}&\\
\hline
\multicolumn{4}{l}{\textit{Without memory}}\\
\hline
high/low wavelet frequencies in sw  & \texttt{wavelet high/low sw} & \makecell{sliding window average of the\\ high/low ratio of spatial wavelets} &--\\
high/mid wavelet frequencies in sw  & \texttt{wavelet high/mid sw} & \makecell{sliding window average of the\\ high/mid ratio of spatial wavelets} &--\\
\end{tabular}

\vspace{3mm}
    \caption{Additional spatial variables added to those of Table~\ref{tabSI:variables} in a version of the training using an extended set of variables, detailed in Section~\ref{secSI:additionalVars}. }
    \label{tabSI:additionalvariables}
\end{table}

\begin{figure}[p]
\includegraphics{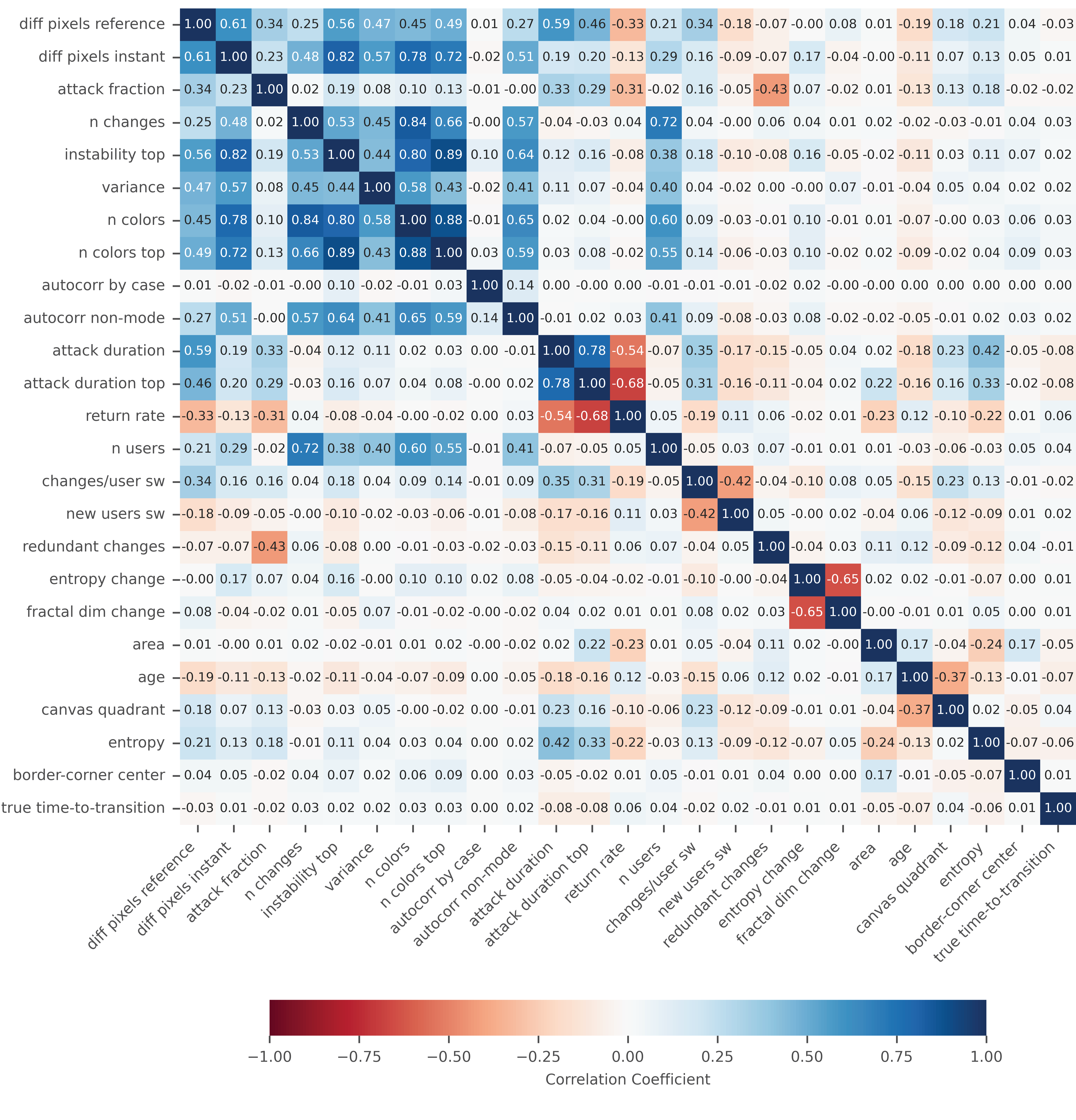}
\centering
\caption{Correlation coefficients between all training variables and the  time-to-transition, over all time instances used in the training.}
\label{figSI:correlations}
\end{figure}

\section{Data preparation and training of the machine learning algorithm}

We use the variables listed in the previous section computed for all compositions and time steps as input for training a gradient-boosted decision trees algorithm to predict the time-to-transition \ttt.

\subsection{Memory of time series embedded in the features}
\label{secSI:memory}

Many time series prediction algorithms train on values of the variables in the latest or the few latest time steps. However, we reach higher performance by considering a significant memory preceding each time instance. We choose to record a 7-hour memory for all time instances, using 9 or 12 scalar features per variable. This applies to the first 19 time series variables listed in Table~\ref{tabSI:variables}; the last 5 variables depend only trivially on time, so only the value at the time instance is used. 

Keeping the information about all 5-minute time steps in the 7-hour memory would result in 85 features per variable and high correlation among features; this would make the training much longer for a performance that might be similar, or worse, due to overfitting on non-discriminant features. Therefore, we use features that each correspond to the average over a time range within the 7-hour memory, with exact coverage of the memory by all features. Table~\ref{tabSI:timeranges} lists the time range covered by each feature, both in absolute time and in the number of time steps over which the variable is averaged. 
A finer resolution of time ranges is used for the more recent past, as more information about an incoming transition is expected there; features further in the past encode typical variable values in this composition, while the recent features describe more dynamical aspects---for example, a recent increase of this variable compared to its typical value.

To reduce the number of highly correlated variables, we use 9 coarser time features, rather than 12, for 7 of the 19 dynamic variables that evolve more slowly. Variables that include an average over the sliding window tend to show slower changes, such as \texttt{n users sw} or \texttt{attack duration}. We also use coarse time ranges for variables significantly correlated with other variables, such as \texttt{n changes} and \texttt{variance}. The last column of Table~\ref{tabSI:variables} specifies the time range features  used for coarse and non-coarse variables.

\begin{table}[h!]
    \centering
    \begin{tabular}{Sc|c|c|c|c}
    \makecell{\textbf{Time steps}\\ \textbf{to integrate}} &
    \makecell{\textbf{Integrated}\\ \textbf{time range}} & 
    \makecell{\textbf{Name for figures}\\\textbf{(average)}} & \makecell{\textbf{coarse} \\\textbf{time ranges?}} & \makecell{\textbf{non-coarse} \\\textbf{time ranges?}}  \\
    \hline
         0& 0--5~min& -- &\checkmark & \checkmark\\
         1& 5--10~min& 5~min & & \checkmark\\
         3-1& 5--20~min& 10~min &\checkmark & \\
         3-2& 10--20~min& 15~min & & \checkmark\\
         5-4& 20--30~min& 25~min & & \checkmark\\
         7-4& 20--40~min& 30~min &\checkmark & \\
         8-6& 30--45~min& 35~min & & \checkmark\\
         12-9& 45--65~min& 55~min & & \checkmark\\
         13-8& 40--70~min& 55~min &\checkmark & \\
         17-13& 1h5--1h30& 1.25~h & & \checkmark\\
         24-14& 1h10--2h05& 1.5~h &\checkmark & \\
         24-18& 1h30--2h05& 1.75~h & & \checkmark\\
         39-25& 2h05--3h20& 2.7~h 
         &\checkmark & \checkmark\\
         54-40& 3h20--4h35& 3.9~h 
         &\checkmark & \checkmark\\
         69-55& 4h35--5h50& 5.2~h 
         &\checkmark & \checkmark\\
         84-70& 5h50--7h05& 6.4~h 
         &\checkmark & \checkmark
    \end{tabular}
    
    \vspace{3mm}
    \caption{Time ranges of the memory stored for each feature and time instance. Larger times correspond to earlier times in the memory. The central column is the midpoint of the time range, which is used as shortcuts in figures. The last two columns indicate whether a given time range feature is used for coarse or non-coarse variables, which are listed in Table~\ref{tabSI:variables}.}
    \label{tabSI:timeranges}
\end{table}

\subsection{Filtering of training dataset}
\label{secSI:filter}

To circumvent some data limitations, we filter the time instances that are kept in the training and testing dataset:
\begin{itemize}[noitemsep]
    \item As explained in the main text, instances must be preceded by a 3-hour stable period that fulfills our stability requirement for transitions. The effect of this filter, as well as the one described in the following item, is evaluated in Section~\ref{secSI:sensitivity} and Fig.~\ref{figSI:sensitivity}c.
    \item Some variables compute information over the 3-hour past sliding window. We require the time instance to be at least a sliding window width after the birth of the Atlas composition, ensuring variables are computed using information only from after the composition birth. As explained in the main text and in Section~\ref{secSI:transitions}, this excludes transitions from a patchwork into an Atlas composition; we choose to reject these areas that would not be monitored beforehand. Also note that this requirement allows for an overlap of the 7-hour memory with times preceding the birth of the composition; knowledge of a recent transition into the current composition could help predictions, without being unrealistic in real-world applications. 
    \item The memory should not overlap with significant changes in the borders of the image as it would result in artificial jumps in some variables. Therefore, the time instance should be at least 7 hours after the start of a stable area time range. We use a margin of 10 hours rather than 7 hours so that the 3-hour sliding window of variables evaluated 7 hours ago does not overlap with a change of image borders. A stable area time range cannot start before the opening of this canvas region.
    \item The time instance should precede the restrictions to grayscale or white colors (see Table~\ref{tabSI:extensions}) and the end of the composition as indicated in the Atlas. A 1-hour margin after the Atlas death is given when a transition is identified after the Atlas death.
    \item Transition periods are excluded: the time instance must not fall within a sliding window width after the start of a transition. In addition, \texttt{diff pixels reference} must not exceed transition thresholds, meaning its absolute and relative values should not both exceed 0.35 and 6, respectively. 
    \item Finally, to prevent near-duplicate instances in the training data, we do not include multiple time instances describing similar compositions at the same time. Using only the Atlas information on all compositions, we compile a list of compositions and times to reject. When two compositions have a spatial overlap exceeding 90\% and share some time steps, we exclude these time steps for the composition with the later Atlas birth. 
\end{itemize}
This results in 1.53 million instances in 2022 and 1.34 million in 2023.

\subsection{Selection of variables and features}
\label{secSI:featureselec}

\paragraph{Why and how to prune}
The gradient-boosted decision trees algorithm can be overtrained and perform worse when many highly correlated features are included for each instance. As a simple example, let us consider two identical (100\% correlated) features in the training: half of its predictive information will be extracted by the algorithm from the first feature, and half from the second. This would reduce the statistical significance of an extracted predictive property, which could drive its importance measure below the noise threshold and therefore make it indistinguishable from fluctuations of the training dataset. 

Pruning the input features is therefore key to reliable predictions. We perform this in two main steps: we first select the time series variables to keep in the training, then reject the lowest-performance time features. Our pruning criteria rely on:
\begin{itemize}
    \item the mean absolute SHAP values for each variable or feature. SHAP values estimate, for each instance, how much a given input feature contributes to the predicted target value (defined in Section~\ref{secSI:target} as a transformation of the time-to-transition). A high absolute SHAP value therefore indicates that this feature significantly affects the prediction for this instance. The average of these absolute SHAP values over instances measures the importance of this feature in the prediction. 
    \item the correlations between variable values, shown in Fig.~\ref{figSI:correlations}. This indicates redundancy of variables in the training;
    \item the correlations between the SHAP values of instances of pairs of variables. The value correlation (previous item) estimates how similar variables are, but the similarity of the predictive information of variables could be even more important. The SHAP correlation indeed quantifies the redundancy between the predictive power of two features; features that bring the same predictive information to the algorithm can be pruned without performance reduction. 
\end{itemize}

\paragraph{Variable selection}
When using the SHAP values for a time series variable, we first arithmetically sum the SHAP values for each of its time features, representing the total effect of this variable on the predictions. The distribution of SHAP values over instances for each variable in Fig.~\ref{figSI:shapdistr} gives insight into how different variables contribute to the algorithm. Variables are ordered by the mean absolute SHAP so that the variables higher in the figure are the higher-performance ones; this ordering is used to isolate the variables with lowest mean absolute SHAP. 

Variable pairs with correlation coefficients above 0.9 are rejected by removing the variable with lowest mean absolute SHAP. Some variables with lower correlations but very low mean absolute SHAP are also removed. We also remove variables with much lower SHAP than other variables derived from the same quantity---for example two complexity variables, or two ways of computing the fraction of differing pixels. In practice, we run the training, then assess correlations and SHAP values to remove the worst-performing variables, then repeat until no obvious removal can be made. The first of these preliminary trainings contains all the variables defined in Section~\ref{secSI:variables}, except the spatial variables of Section~\ref{secSI:spatialvars} and \texttt{multiscale complexity} and \texttt{Levenshtein complexity}. As explained in Section~\ref{secSI:additionalVars}, these excluded variables are only used to train a more extensive version of the algorithm, which was rejected because it increases complexity (hence decreases interpretability) without improving the global performance. 

A similar selection of variables was performed on the extended set of variables, though keeping all nominal variables to aid comparison. The SHAP distributions for the nominal pruned variables, the extended set of variables, and the pruned extended variables are shown in Fig.~\ref{figSI:shapdistr}a, b and c. 

\paragraph{Subsequent feature selection}

The 19 time series variables with 9 or 12 time features plus the 5 memory-less features (listed in Table~\ref{tabSI:variables}) result in 212 features. At a late stage of algorithm tuning, we remove 37 features among those with the lowest mean absolute SHAP, from which we obtain the 175 final features. 

After training the algorithm using the pruned input variables, we consider each feature in increasing order of mean absolute SHAP. For each feature, we compute the following SHAP selection index:
\begin{equation}
    \frac{S-S_{min}}{ \max(\rho_{shap}/(1.01-\rho_{value}))^{\alpha}}
\end{equation}
where: 
\begin{itemize}
    \item $S$ is the mean absolute SHAP of this feature
    \item $S_{min}=4\times 10^{-5}$ is a minimal SHAP value that matches the typical level of noise in the SHAP signal. To obtain this value, we tried training the algorithm with 20 purely random features, of which a few could reach a mean absolute SHAP of $5$ to $9\times 10^{-5}$.
    \item $\rho_{shap}$ and $\rho_{value}$ are the correlation coefficients between SHAP values or between the feature values themselves. The quantity $\rho_{shap}/(1.01-\rho_{value})$ is computed for correlations between the feature at hand and all features except the twenty least performing ones; the maximum value of this quantity across features is kept in the index.
    \item $\alpha=0.7$ gives more importance to SHAP values compared to correlations when correlations are high.
\end{itemize}
We add this feature to the pruning list if this index is below $10^{-4}$ and if the correlation with other features already in that list is below $0.85$. This correlation threshold ensures that two features with low performance but the same predictive information are removed together, as training with only one of them may bring that feature above the threshold. We stop the procedure either when 20 features have been added to the pruning list or when all features have been tested. We then train again without the pruned features, and apply this selection procedure again. We stop iterating when, after pruning, the training does not show any feature passing the above rejection criteria. 

The extended set of variables---not used in the nominal results---contains, after variable selection, 9 additional time series variables and 2 additional memory-less features (listed in Table~\ref{tabSI:additionalvariables}), amounting to 316 features; these are reduced to 231 features after the feature selection with SHAP. 
The extended set before selection of the additional variables contains 35 variables (including the 19 variables of the nominal set) or 395 features.

We also explored other ways to reduce dimensionality of the training data, notably through a Principal Component Analysis (PCA) applied to the data before feature pruning, and either before or after variable pruning. The transformed features are then input into the training. When applying PCA after variable pruning, keeping the features necessary to preserve 95\% of the variance of the dataset leads to a similar number of features as after the nominal pruning. However, the performance degrades substantially compared to our nominal results. In addition, it would be necessary to unfold the SHAP values to the original features to extract the type of interpretations we provide; but this unfolding is only approximate after pruning. This lesser performance means that our human readable input variables provide more readily available predictive information to the algorithm than complex linear combinations of these variables. 

\subsection{Target value, weights and loss term}
\label{secSI:target}

The goal of the algorithm is to predict \ttt, the time to the next transition. However, we expect predictive power only up to a few hours before a transition, whereas \ttt could be as long as the duration of the experiment or be undefined where there are no identified transitions in that composition. Therefore, we make some structural decisions aimed at reducing the importance of higher \ttttr values. These decisions each bring important performance improvements.

First, \ttt is set to 12 hours when there is no future transition or when it is more than 12 hours away. Next, the target value we set to predict for each instance is not exactly \ttt, but 
$$\tfour = \log_{10}(100\xspace\textrm{ s} + \ttt) - 1.5$$
where \ttt is counted in seconds. Adding the 100 seconds in the logarithm prevents its argument from approaching 0, while maintaining a sufficient resolution of a third of a time step. Incorporating the logarithm favors the precision of predictions in lower \ttt values, as we do not expect generalizable properties that can discriminate \ttt values larger than a few hours. 

We then give lower weights to instances with a high \tfour. The weight is set to $1$ for $\ttttr<3$h, 0.4 for $\ttttr\geq12$h, and interpolated with a power law between these two values for the \ttttr values in between. 

Finally, we implement a preference for relative rather than absolute deviations in the algorithm minimization itself. The squared \textit{relative} deviation between the true and expected values rather than the more standard squared \textit{absolute} deviation is used as the loss term of the objective function for each instance in the objective function that is minimized in the training. 
The introduction of $-1.5$ in the above definition of \tfour is a last element favoring lower \ttt values, as the proximity of lower \ttt values to zero amplifies their arithmetic deviations when comparing them geometrically.

\subsection{Algorithm setup, training, and hyperparameters}
\label{secSI:algo}

The input dataset must now be split into a training and a testing sample. When training and testing on 2022 data, 
we randomly select compositions for the training dataset that amount to 80\% of time instances and attribute the rest to the testing dataset. This ensures, as is standard, that the trained trees have no prior knowledge of the instances they are evaluated on, but also that no correlations across time within a given composition can be exploited by the training. In real-world scenarios, this precaution means that warning systems are trained on certain subsystems but must make predictions for entirely new ones. When testing on the 2023 data, the full 2022 data is used in the training. 

We then train gradient-boosted decision trees using XGBoost 1.7.5.~\cite{chenXGBoostScalableTree2016} in Python. The hyperparameters of the trained models directly impact the algorithm performance. They were loosely optimized on the test sample. The optimization criteria, based on the areas under the ROC and PR curves explained in Section~\ref{secSI:ROCPR}, are the PR area for the $\ttt<1$~h and $\ttt<3$~h binary targets, the ROC area for the 1-hour target, and the ROC area at false positive rate less than 0.2, for the 1-hour and 3-hour targets. 

No independent validation sample was kept apart for the hyperparameter optimization, to avoid reducing the data available for training or testing. Instead, the random seed for selecting the training and testing samples was modified at a very late stage of the analysis, thereby breaking any excessive optimization on the evaluation of a specific testing sample. We also do not perform an extensive automated search in hyperparameter space, to avoid fluctuations in performance evaluations from being interpreted as genuine improvements in hyperparameters. Lastly, other design choices for the training data and algorithm significantly affect performance more than fine-tuning the hyperparameters.

Some of the hyperparameters add complexity in the trained model to minimize the loss term of the objective function, while others regularize the model to reduce overtraining. In the first category, we use a maximum tree depth (\texttt{max\_depth}) of 8, and 160 trees (\textit{a.k.a.} rounds, \texttt{num\_boost\_round}) in 2022 (140 trees when testing on 2023). In the second category, the learning rate (\texttt{learning\_rate}) is 0.035; the minimum summed weight in an end node of a tree (\texttt{min\_child\_weight}) is 8; the sub-sampling ratio of the instances (\texttt{subsample}) is 0.8 at each round; and the sub-sampling ratio of the features is 0.75 both at each round (\texttt{colsample\_bytree}) and at each tree level (\texttt{colsample\_level}). Other XGBoost parameters are kept at their default values. We also consider the weight of high true \tfour instances, 0.4, as a hyperparameter to optimize on.

\section{Testing and performance evaluation}
\label{secSI:testing}

\subsection{Calibration of model output}
\label{secSI:calib}
The actual predictive power of the algorithm lies only in the ranking of instances according to their predicted \tfour value. However, the typical values of the predicted target can be offset or scaled compared to true target values, and depend on irrelevant algorithm details and parameter choices.

Therefore, we use the predicted and true \tfour values in the testing set to calibrate the output of the algorithm. No optimization is performed on the chosen calibration method, so there is no bias introduced by using the testing set. For the calibration, we first keep only the instances with $\tttpr< 3600~\textrm{s} + 3.5 \, \ttttr $. 
This rejects low $\ttttr$ instances that have high $\tttpr$, which are clear false negatives (undetected signal) that would excessively drive up the calibrated predictions. Then we perform a fit  in the $(\tfour_{\textrm{true}}, \tfour_{\textrm{predicted}})$ logarithmic space. We compute the median value of $\tfour_{\textrm{predicted}}$ versus $\tfour_{\textrm{true}}$, that we fit 
to continuous piecewise linear function with one breakpoint. This function has start and end points fixed at $(\ttttr, \tttpr) = (0,0)$ and $(12~\textrm{h},\, 20~\textrm{h})$. The fitted piecewise function is the calibration function, whose inverse is applied to the predicted \ttt shown in the evaluation results.

\subsection{ROC and PR curves}
\label{secSI:ROCPR}
The evaluation of the performance of an algorithm on a testing sample is usually reduced to quantities relating to a binary classification into signal and background instances, which is why we convert our results into binary warning systems of various \ttt warning ranges as explained in the main text. This allows us to count the true and false positives and negatives. The true (false) positive rate is the fraction of signal instances---those with $\ttttr$ inside the warning range---that have $\tttpr$ below (above) the set tolerance on predicted values. The precision, also known as purity, is the fraction of identified warnings that are true positives---in other words, the number of true positives divided by the total number of positives.

\begin{figure}[h!]
\includegraphics{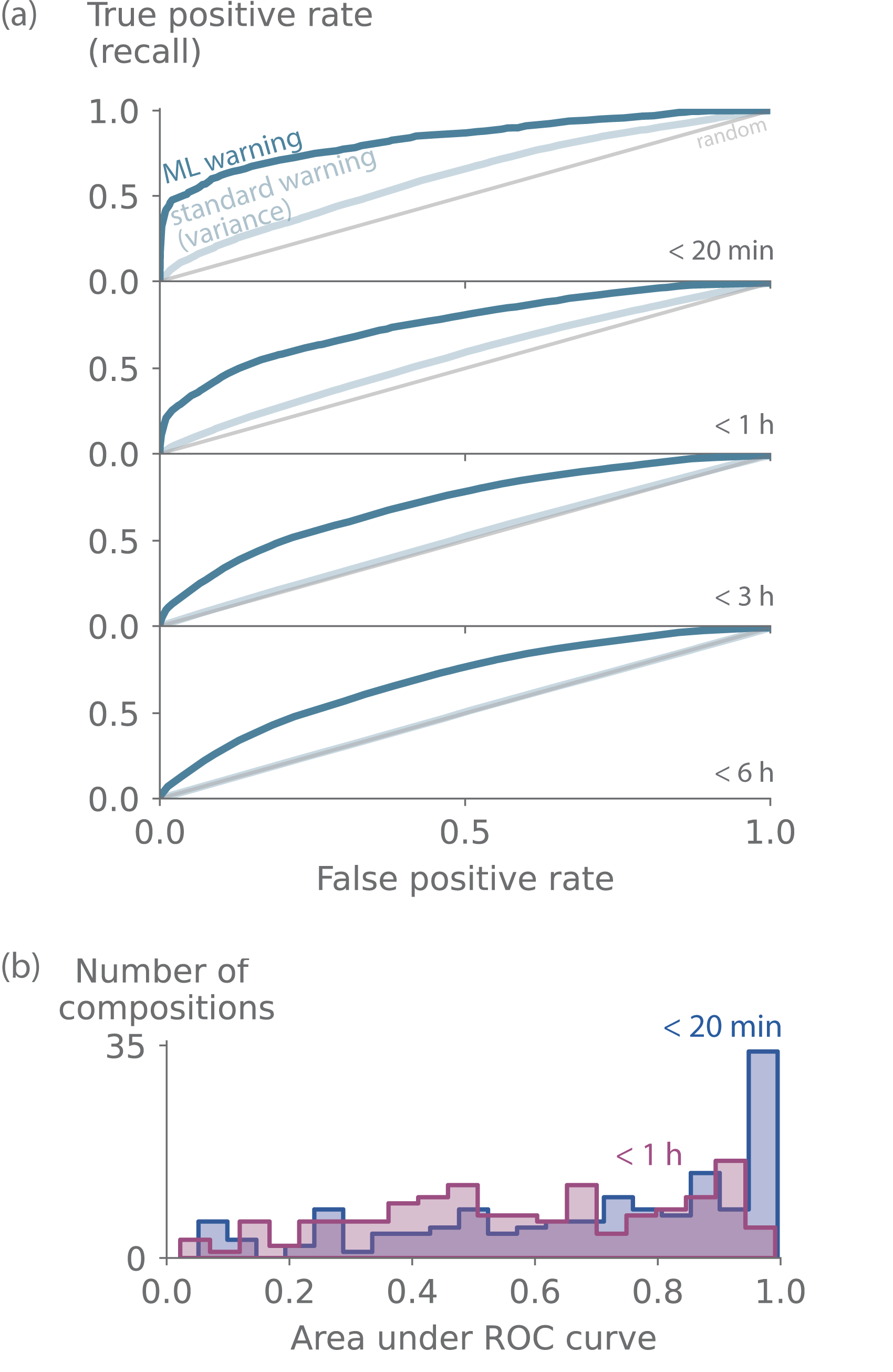}
\centering
\caption{\textbf{(a)} ROC curves for our machine learning algorithm trained and tested on 2022 r/place data (dark blue) and for a standard single-variable warning signal of variance (light blue), for four different warning ranges. \textbf{(b)} Histogram of areas under the ROC curves for time instances grouped by composition, for warning ranges of $20$~min (blue) and $1$~h (purple).}
\label{figSI:ROC}
\end{figure}

\begin{figure}[p]
\includegraphics{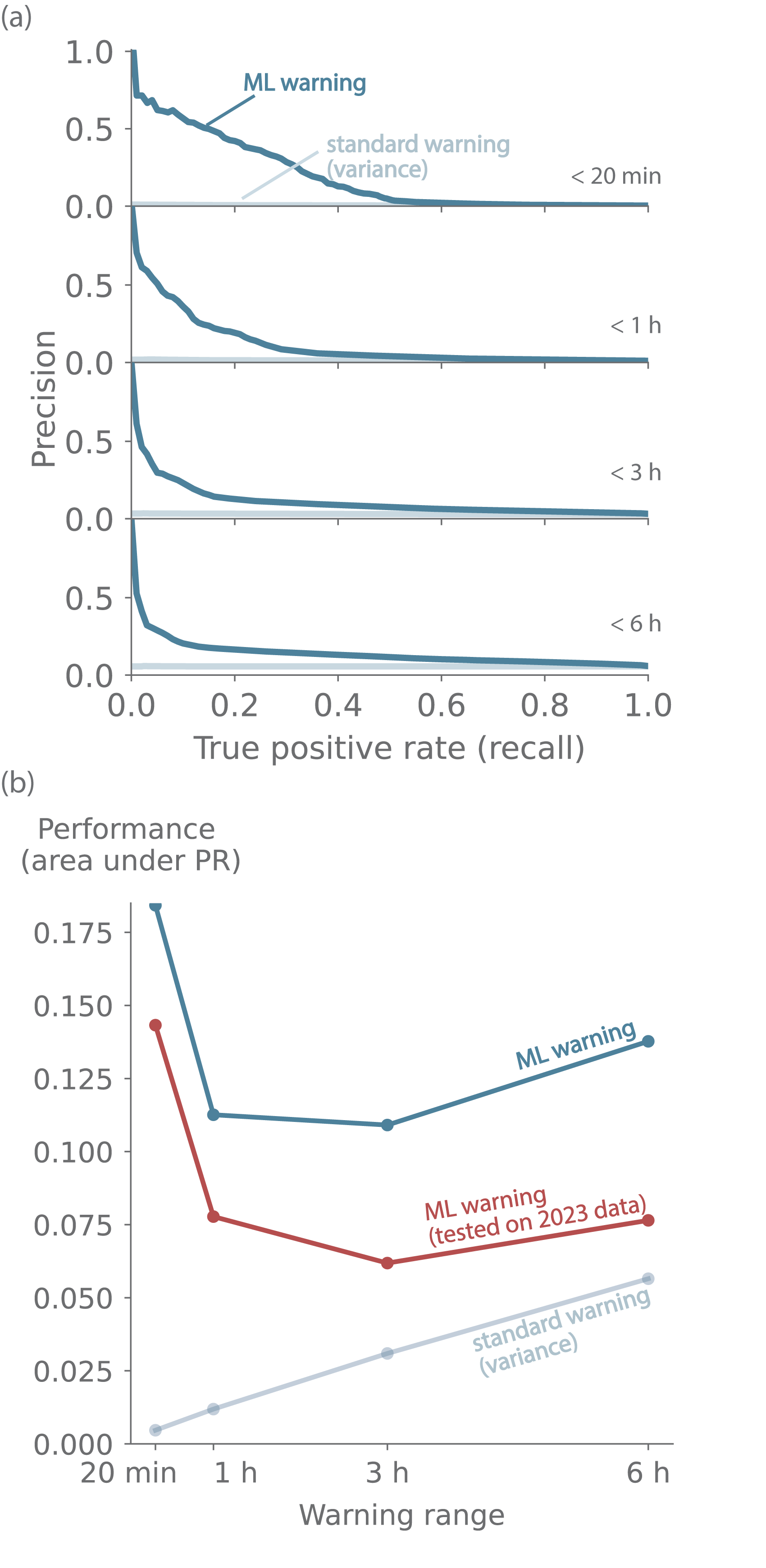}
\centering
\caption{\textbf{(a)} Precision-Recall (PR) curves for our machine learning algorithm trained and tested on 2022 r/place data (dark blue) and for a standard single-variable warning signal of variance (light blue), at four different time thresholds for warnings. \textbf{(b)} Area under the PR curves as a function of warning threshold time for the machine learning warning signal (dark blue), a standard single-variable warning signal of variance (light blue), and for the machine learning warning signal tested on 2023 r/place data (red).}
\label{figSI:PR}
\end{figure}

The receiver operator characteristic (ROC) curve shows the true positive rate versus the false positive rate. For a random classification, it follows the diagonal of the unit square of the plot and the area under the ROC curve equals 0.5, while it equals 1 for a perfect classification. The precision-recall (PR) curve shows the precision as a function of the true positive rate, also known as the the recall. It also equals 1 for a perfect classification, but specifies the fraction of signal in the sample, which, for example, is 0.3\% for a warning range of $20$~min. The ROC curves and areas under them are plotted for four warning ranges (20~min, 1h, 3h, and 6h) in Fig.~\ref{figSI:ROC}a, while the PR curves and areas under them for the same thresholds are plotted in Fig.~\ref{figSI:PR}.

When using such a warning system in practice, a system manager would set a threshold on the $\tttpr$ values under which a warning is issued. This corresponds to a certain point on the ROC and PR curves, meaning that modifying this threshold is a direct trade-off between the true and false positive rates desired by the system manager.

\begin{figure}
\includegraphics[height=.88\textheight]{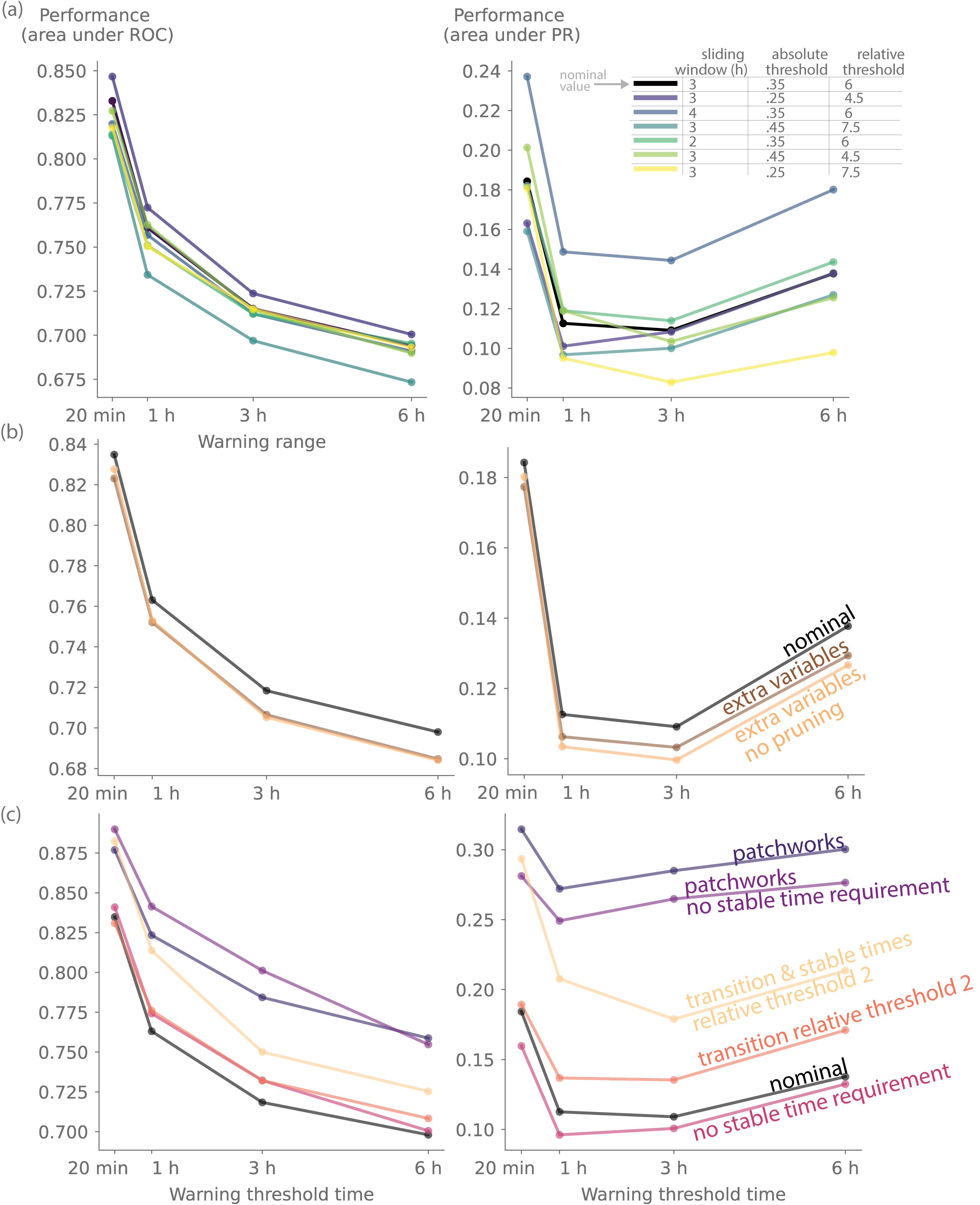}
\centering
\caption{Sensitivity of prediction performance to design choices. Performance is shown as area under the ROC curve (left) and precision–recall curve (right) for different warning threshold times (20 min, 1 h, 3 h, 6 h). 
The nominal version of the training is shown in black in all panels. 
(a) Sensitivity to parameter values: sliding window length, absolute and relative thresholds on the state variable required for transition. (b) Effect of including extra variables, leading to 231 features (brown) and extra variables without pruning, leading to 395 features (peach). 
(c) Effect of removing various training data filters: including patchworks preceding Atlas compositions (blue), removing the requirement for a stable period preceding each time instance (magenta), including patchworks and removing the stable times requirement (purple), relative threshold on transitions of 2 (orange), relative threshold of 2 and looser stable times requirement fitting that new stability threshold (peach). 
}

\label{figSI:sensitivity}
\end{figure}

\begin{figure}[h!]
\includegraphics[height=.93\textheight]{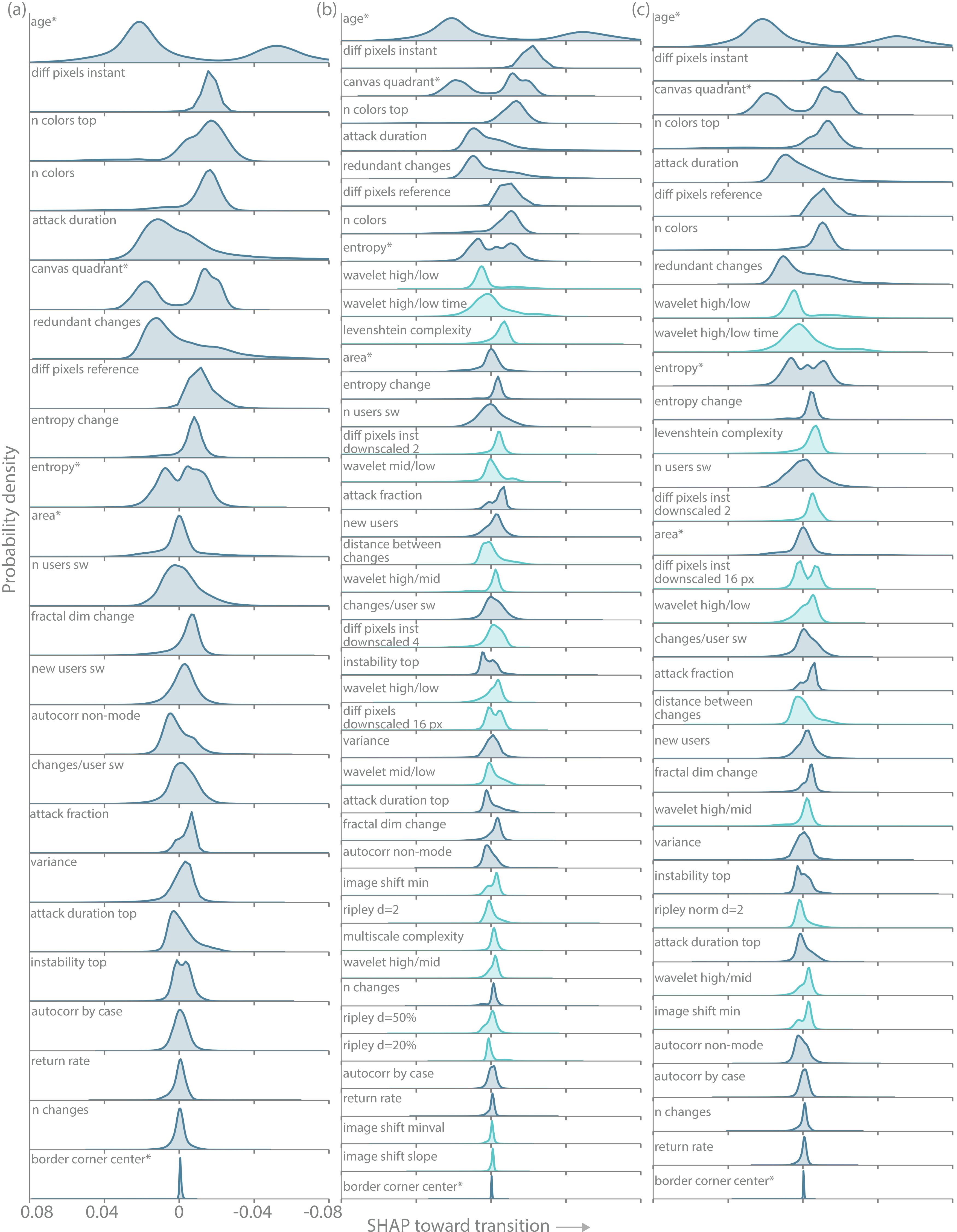}
\centering
\caption{Probability density distributions of SHAP values over time instances for each variable, calculated using kernel density estimation. The SHAP for a variable is the sum of the SHAP values of all time features associated with this variable. Variables are ordered according to the mean of the absolute values of the SHAP values. Distributions are shown for algorithms trained with three different sets of variables: (a) the set of variables for the nominal training, also listed in Table~\ref{tabSI:variables}; (b) the nominal set in (a), plus all the additional spatial variables we tried in training; (c) a pruned subset of the set in (b), composed of all the variables in Tables~\ref{tabSI:variables}~and~\ref{tabSI:additionalvariables}. The * indicates variables without memory, meaning they are used as a single time feature. The lighter blue color indicates variables not included in panel (a).}
\label{figSI:shapdistr}
\end{figure}

\subsection{Per-composition results} 
\label{secSI:percompo}

We also compute binary warning signals for individual transitions of individual compositions in the 2022 test data. A time instance is signal if it falls within the warning range of a transition. Note that only time sequences preceding a transition are considered so that there are possible true warnings, thereby excluding 
compositions with no transitions. To have sufficient precision, we keep only the 139 transitions with at least 30 time instances outside the warning range and 4 instances within it. There can be multiple transitions considered in a single composition. The distribution of ROC AUC values for each of these compositions is shown in Fig.~\ref{figSI:ROC}b.

This per-composition perspective is valuable when a warning system can be tested beforehand, allowing the threshold on the predicted quantity to be tuned to achieve the desired trade-off between true and false rates. However, it might be necessary to set a threshold before encountering a new system within the studied class of systems. Therefore, we also computed, at a given threshold on the predicted \ttt, the time of the first warning for each transition versus the false positive rate for all compositions, in a similar manner as for the ROC curve. This allows a system manager to estimate how early a transition can typically be warned for, versus the acceptable false positive rate.

\subsection{cooldown warning system}
\label{secSI:cooldownWarn}
We also considered a warning system based on a cooldown, rather than considering each time instance separately, closer to the method of Ref.~\citenum{hylandEarlyPredictionCirculatory2020} and arguably closer to a real-world system management strategy. We then evaluate a ROC curve equivalent for this system for various warning ranges. There is no one-fits-all choice in the method, which must ultimately correspond to the need of a manager with respect to a specific system. For a certain warning range, we choose to count each true positive as the ratio of how early the transition was warned for (considering only warnings that are within the warning range, as premature warnings are false positives) to the width of the warning range; to calculate the rate, the sum of these scores is divided by the number of transitions. A false positive is a warning outside of the warning range preceding a transition, when this warning is not preceded by another warning closer than a warning range width; in the rate, the count is divided by the maximum number of false warnings (close to the cumulative time of all compositions divided by the warning range width). 

We show in Fig.~\ref{figSI:continuous-warn} the resulting curves; note that they are not strictly ROC curves due to our definition of the true warning score. They perform worse than our original warning system, which is expected as it is a harder task than predicting the time-to-transition for single time steps; however, we still obtain significant predictive power for the 20-minute warning range.
The same 139 pre-transition periods as in Section~\ref{secSI:percompo} are used for the true positive rate, while the 227 compositions from the 2022 testing sample with more than 30 background instances are used to compute the false positive rate.

\subsection{Sensitivity of predictions to warning system design choices} 
\label{secSI:sensitivity}

\subsubsection{Removing training data filters} 
To keep the warning system realistic, we filtered out cases that no real-world system would monitor (Section~\ref{secSI:filter}). First, we excluded patchworks of partial images, as in the sequence of Fig.~\ref{figSI:notrans}a, since no system would track arbitrary boundaries within which several compositions meet. Second, we removed time ranges where the dynamics are too noisy and unstable to form a system that a manager would want to preserve against an incoming transition. 
These filters create a more realistic dataset, but we hypothesized that they sacrifice performance. For example, the only reason a patchwork would be included in the data is because it precedes a composition in the Atlas, so the algorithm can exploit the presence of patchworks as signals for transition. To test the effect of these filters on performance, we rerun the full training and analysis without these filters (Fig.~\ref{figSI:sensitivity}c). Allowing patchwork transitions substantially increases performance, both for ROC and PR AUC. This gain is not just from having more data; when we downsample instances to match the nominal dataset, the performance does not change significantly. In contrast, allowing unstable times made 
a more subtle difference: a slight increase in ROC shows that noisy instances are easy to classify as far from transition as they cannot pass the relative threshold for transitions, and a decrease in PR reflects the higher number of background instances without more signal in the sample (demonstrated by a relative increase of the signal fraction of 9\% to 24\%). 

\subsubsection{Changing the transition definition}
We next test how performance changes when transitions are defined with a lower relative threshold on \texttt{diff pixels reference}. Reducing the relative threshold from 6 to 2 increases the number of instances classified as being close to transition (meaning the signal fraction is higher), and performance improves (Fig.~\ref{figSI:sensitivity}c). The gain is even larger when we also relax the requirement for stable times, allowing a threshold of 2 instead of 6. Together, these looser definitions capture more gradual transitions from unstable starting points: for example, the sequence of Fig.~\ref{figSI:notrans}b would now be considered a transition. 

As with the filters in the previous section, the performance boost is not explained by the larger sample size, since downsampling to the nominal dataset size gave similar improvements. Instead, the performance rise likely comes from two effects: first, more instances are counted as in-transition, increasing the proportion of signal in the dataset, not just raw dataset size; second, noisy dynamics often precede transitions (simply because instances with high \texttt{diff pixels reference} are already close to passing the absolute threshold for transition), so including more unstable instances gives the model an easy signal. However, these are not the transitions we aim to predict. Our focus is on sharp, sudden shifts from stable states, where early warning signals are needed the most. For this reason, we adopt a stricter definition that reduces apparent performance, but provides predictions that are more applicable in practice.

\subsubsection{Adding additional spatial variables}
\label{secSI:additionalVars}
Another possible concern is to what extent the predictive power depends on the specific variables included. To test this, we expanded our set of variables to include additional variables that capture spatial characteristics of compositions: those listed in Section~\ref{secSI:spatialvars} as well as \texttt{Levenshtein complexity} and \texttt{multiscale complexity}. The variables and features of this extended set are pruned as explained in Section~\ref{secSI:featureselec}; see Table~\ref{tabSI:additionalvariables} for the variables included after pruning. 

Performance evaluations show that including additional spatial variables, even when pruned, does not significantly improve predictions (Fig.~\ref{figSI:sensitivity}b). If anything, the added complexity tends to reduce performance slightly while hindering interpretation (see Section~\ref{secSI:interpretation}). Meanwhile, individual variables do perform similarly to the nominal variables, as shown in the SHAP distributions of Fig.~\ref{figSI:shapdistr}b and c, and correlations with the nominal variables are limited. This could suggest an inherent unpredictability of part of this dataset, likely related to the dynamics being driven by human decisions, which cannot be overcome with more complex models.

\subsubsection{Varying parameter values}
Some parameters are of structural importance to this work: the width of the sliding window, used notably in defining the reference image and for obtaining typical values of time series in the past; and the relative and absolute thresholds on \texttt{diff pixels reference} for defining transitions, which determine how much sudden change in the images can qualify as transitions. Here, we check that our main results, in particular the quality of the predictions of our machine-learning algorithm, hold when varying the values of these parameters. The training and analysis is fully rerun with 6 variations of these parameters. The actual transitions and their predictions will inevitably differ, but our ability to successfully predict these incoming transitions should remain mostly unchanged.

This sensitivity study is shown in Fig.~\ref{figSI:sensitivity}a, with sliding windows of 2 and 4 hours, absolute transition thresholds 0.25 and 0.45, and relative transition thresholds 4.5 and 7.5; the definitions of the six parameter variations are in the legend of the figure. 
The differences in performance across these variations are generally small, though two variations show differences worth noting:
\begin{itemize}
    \item The lower ROC AUC and PR AUC performance with the lower absolute ($0.25$) and higher relative (7.5) thresholds on transitions is     explained by the much lower number of transitions in the sample: a smaller signal fraction directly reduces performance.  
    \item The 4-hour sliding window with nominal transition thresholds shows a marginally higher ROC AUC but a significantly higher PR AUC. This combination would typically indicate a change in the ratio of close-to-transition to far-from-transition instances, but the signal (meaning close-to-transition) fraction shows only a 2 to 7\% relative increase compared to the nominal configuration---too small to explain the difference. The 4-hour window also excludes about 10\% of the data because more time margin is required before each selected instance. The removed instances likely come from shorter stable periods that follow noisier periods, potentially in newly formed compositions that are not yet fully stabilized. These excluded instances may be harder to predict, which could explain the apparent performance gain. In addition, several features depend directly on the sliding window length, and a longer sliding window can carry more information about past dynamics. Apart from shifts in signal fraction, a higher PR AUC with stable ROC AUC indicates that the model ranks instances similarly but predicts values closer to the true time-to-transition. This is likely how the removed instances and longer-term features increased performance. Despite this, we keep the 3-hour sliding window to preserve a larger, more representative sample of instances and compositions.     
\end{itemize}

\section{Interpretation of predictions}
\label{secSI:interpretation}
\subsection{SHAP trends}
\label{secSI:shap}
To uncover the set of 12 pre-transition behaviors from the set of 175 SHAP percentile curves, we systematically go through a series of steps. First, we classify variables based on the aspect of composition dynamics that they describe. These categories are: strength of the attack or defense changes, image activity---which is related to the variance in the framework of critical slowing down---, user activity and engagement, image complexity, innovation, and coordination versus noise. Many variables fit into multiple categories. 
We then look for trends shared by multiple time features of variables in a given category. We often consider trends at only at high or low variable values to be distinct. We also separate the trends of recent time features from those of time features in the past memory. Finally, we assign an interpretation to each of these shared trends based on our knowledge of the game rules and system dynamics. This leads to 12 interpretative messages, each associated to a group of time features that describe similar dynamics. These behaviors are shown in Fig.~6 
of the main text and Fig.~\ref{figSI:SHAPinterp}. We do not include all the time features of a given variable that follow a given behavior, for plot readability purposes. 

Despite our systematic approach, these interpretations are still quite qualitative. There are contradictory interpretations of the variables and trends we observe. We discuss some of these potential contradictions below and provide further reasoning for why we present our chosen interpretations rather than alternatives. 
We also explain how we relied only on trends that were robust against our requirement of stability before transitions, which could affect what the algorithm identifies as warning signs.

\paragraph{Contradictory variable interpretations.}
One prominent example of an ambiguous variable interpretation is in distinguishing innovations versus attacks. In our system, there is no sure way to identify whether a pixel change that does not match the reference image is an attack on the image, or an innovation that will eventually be adopted into the image by the defending community. We therefore must use context to interpret whether a trend in variables that can track either innovations or attacks represents one or the other. For instance, in Fig.~6
d of the main text, we present a series of features that decrease as a predicted transition nears. These features are both recent and past time features, and the variables they originate from could be interpreted as either attacks or innovations. We conclude that if a recent time feature of such a variable decreases as a transition comes closer, it is most likely picking up on a decrease in innovations rather than a decrease in attacks. An increase in attacks would indeed be necessary for a transition to take place. Another argument for this interpretation is that the \texttt{attack duration} variable computes how long attack pixels stay in the image over a timescale longer than a typical attack, so it does capture what changes are approved by the defense.

For past time features, however, a lower attacking force could signal a coming transition (Fig.~\ref{figSI:SHAPinterp}c). If a composition is sufficiently neglected by defenders as well as attackers, it may signify that the composition is of low interest to its own community, which means it would be an easy target for an attack. Interestingly, a low past \texttt{attack duration} could signal either attacker neglect or a lack of image innovation; both are reasonable interpretations, so we classify past time-feature SHAP curves for this variable in both messages, rather than choosing a single interpretation. 

\paragraph{Contradictory trend interpretations.}
In some cases, the contradictory interpretations come not from interpreting what the 
feature and its associated variable represent, but from opposite trends in the same feature being both associated to a coming transition. For instance, we find that both low past activity and increased past activity can be transition signals (Fig.~\ref{figSI:SHAPinterp}c and f). However, these seemingly contradictory signals can take place in different regimes of the feature values. For instance, the curve labeled ``\texttt{n users sw} 6.4 h ago,'' 
present in both Fig.~\ref{figSI:SHAPinterp}c and d, shows that an increase from a low percentile value brings a predicted transition closer, but increasing from a high to very high value pushes the transition further away. In this way, both positive and negative changes in the number of active users can signal a coming transition, depending on the activity level at which they start. Also note that these two contradictory signals can exist in different compositions, meaning that they do not necessarily appear together in the same subsystem.

\paragraph{Robustness versus stability requirement and prediction quality.}

We manually extract the trends that are significant in the SHAP curve of each feature. To verify the robustness of these trends, we run our pipeline with integer values from 1 to 6 for the relative threshold on \texttt{diff pixels reference} with which transitions are defined; we consider and discuss only the trends that hold when changing our fundamental definition of transitions. These tests ensure that the mentioned trends do not depend on the requirement of a stable period before the transition. We do see a substantial critical speeding up signal in the most recent feature of activity-related variables only at large relative thresholds, which we interpret as an artifact of the required pre-transition stable period: even though we consider only time instances whose preceding period is sufficiently stable on average (see Section~\ref{secSI:filter}), the necessary onset of the increase of activity right before transition can be slightly compensated by lower activity before this onset. This could increase the frequency of decreasing trends---meaning speeding up---in these variables before a transition. However, there is still a mild critical speeding up signal, in particular in features 30 to 60 minutes before the time instance, as shown in Fig.~6c; these trends are robust against changes of the relative transition threshold. 

Another possible concern could be that the SHAP values originate from predictions that are not perfect. However, the algorithm still picks up on these trends to make better-than-random predictions, meaning that these trends hold at least part of the true dynamics of the pre-transition periods. Testing that the trends hold when varying the transition threshold is also a strong evidence that these trends are not overfitting artifacts.

\paragraph{Drawbacks of SHAP.}
First, SHAP assumes feature independence; our variables have significant correlations, but we controlled their magnitude using feature and variable pruning (see Section~\ref{secSI:featureselec}). The power of a variable, quantified by its SHAP values, can then be split between correlated variables, affecting the magnitude of the effects shown in Fig.~6 
of the main text and Fig.~\ref{figSI:SHAPinterp}. Second, SHAP values explain the algorithm's behavior as learned from the data, meaning that if the algorithm is overfitting, the explanations could reflect fluctuations of the training data rather than general patterns. 

\subsection{Toy model of canvas}
\label{secSI:toymodel}
Variables related to critical slowing down may be ambiguous in our system, as explained in Section~\ref{secSI:variables}, and their dynamics close to transition can depend on modeling choices and on the type of transition. In addition, these variables often do not slow down close to a transition outside the context of small perturbations around a well-defined equilibrium, and they sometimes behave in ways that seem to contradict each other. We design a toy model of hypothetical dynamics of a composition on the canvas to illustrate the plausibility of such difficulties in our system.

The model consists of a two-dimensional differential equation system describing the fraction $f(t)$ of pixels differing from a \textit{fixed} (and not sliding as in our system) reference image, as well as the change in the number $n(t)$ of users defending the reference image:
\begin{eqnarray}
\frac{\textrm{d}f}{\textrm{d}t} &=& - r \, n(t) \frac{f(t)}{f(t)+f_m} + A (1-f(t)^{\beta}) \, \, ,\\
\frac{\textrm{d}n}{\textrm{d}t} &=& R ( f(t) - f(t)^{\alpha} ) - L \, \frac{n(t)}{n(t)+n_0} \, \, .    
\end{eqnarray}
The first term of $\frac{\textrm{d}f}{\textrm{d}t}$ is the decrease of the image differences from the defense users restoring the image their community aims for; $r$ is the rate of pixel changes per defense user, and $f_m$ is a small threshold below which users are not alarmed and do not make efforts to reduce the image differences. The second term represents the attacks on the image at rate $A$. These attacks decrease with the number of pixels available to attack, with exponent $\beta=1$ for indiscriminate attacks uniformly distributed on the canvas, and $\beta\gg1$ for more targeted attacks that aim at fully replacing a composition. The first term of $\frac{\textrm{d}n}{\textrm{d}t}$ represents the recruitment of users when the community is alarmed by large changes in the image, including a desertion term $- f(t)^{\alpha}$ when $f(t)$ is too close to $1$; $R$ is the recruitment rate; the exponent $\alpha$, constrained by $\alpha>1$, denotes how late the community switches from recruitment to abandonment, with higher values signifying a community remaining engaged until the image is nearly completely replaced. The motivation behind this recruitment term is to obtain an \textit{increase} of defense activity and return rate when the system approaches transition, rather than a critical slowing down. The last term of $\frac{\textrm{d}n}{\textrm{d}t}$ is a slow loss of users  as they gradually lose interest in the composition; $L$ is the rate of user loss and $n_0$ is the number of users in a core community that do not leave with time.

Using time units of minutes, we fix the following parameters: $\alpha=4$; $\beta=10$; $r=10^{-4}$ for the fraction of the image changed per user per minute, which corresponds to one pixel change per 5-minute interval in a 2000-pixel composition; $n_0=10$ users;  $f_m=0.01$; and $L=20$ users lost per minute when $f=0$. We consider the attack rate $A$ and the defense strength $R/L$ as two possible order parameters governing transitions of the composition from $f=0$ to $f=1$.

We then scan the equilibrium values, the return rate from an instantaneous perturbation 
(probed as a single small modification of $f(t)$), and the variance of $f(t)$ when a small Gaussian noise is applied to it, over $A$ and $R/L$ values. It should be noted that the variance is computed around the equilibrium value of $f(t)$, whereas in our system, we can only directly measure deviations from $f(t)=0$, which is how some of our variance variables are computed. Effectively taking the equilibrium to be zero would mean the equilibrium cannot evolve to differ from the long-term reference image. This is a possible source of confusion when comparing our r/place system to this model, and underscores the difficulty of defining appropriate variance, return rate, and state variables in a system without a clear single-variable descriptor of the system state.

When slowly decreasing $R/L$, so that $f(t)$ always follows its non-zero equilibrium, a second-order critical transition is reached: the equilibrium smoothly yet rapidly reaches $f=1$. In this case, we find a decreased return rate and increased variance, as the standard critical slowing down predicts. 

When we instead slowly increase $A$ from a low $f(t)$ equilibrium, the user recruitment adapts accordingly and always compensates for the increased attack; there is no bifurcation and the equilibrium of $f$ never reaches $1$. However, if $A$ undergoes a sudden increase, the number of users $n(t)$ fails to catch up fast enough and the defense fails, causing a rapid transition to $f=1$. This is a textbook example of rate-induced tipping~\cite{ritchieEarlywarningIndicatorsRateinduced2016}. The parameter space where tipping takes place, of course, depends on the initial value of $f$, and on the rate of increase of $A$ and its initial and final value, but the qualitative conclusions remain: the return rate increases because the number of users does. The variance decreases accordingly, when calculated on $f(t)$ itself over a short sliding window, that is, using the sample mean; however, the variance can increase when calculated based on deviations from the equilibrium or from zero, as $f(t)$ is drastically separating from that equilibrium. This can result in confusion as to how to interpret the contradictory behavior of such variables in terms of critical slowing down. This issue is even more pronounced when the variance is defined as deviations from $0$, as is the case in a few definitions of the variance in our system: the equivalent of the state variable $f(t)$ is then positively correlated with the variance, which therefore increases while the return rate also increases---resulting in contradictory critical slowing down interpretations.

\begin{figure}[!htb]
\includegraphics{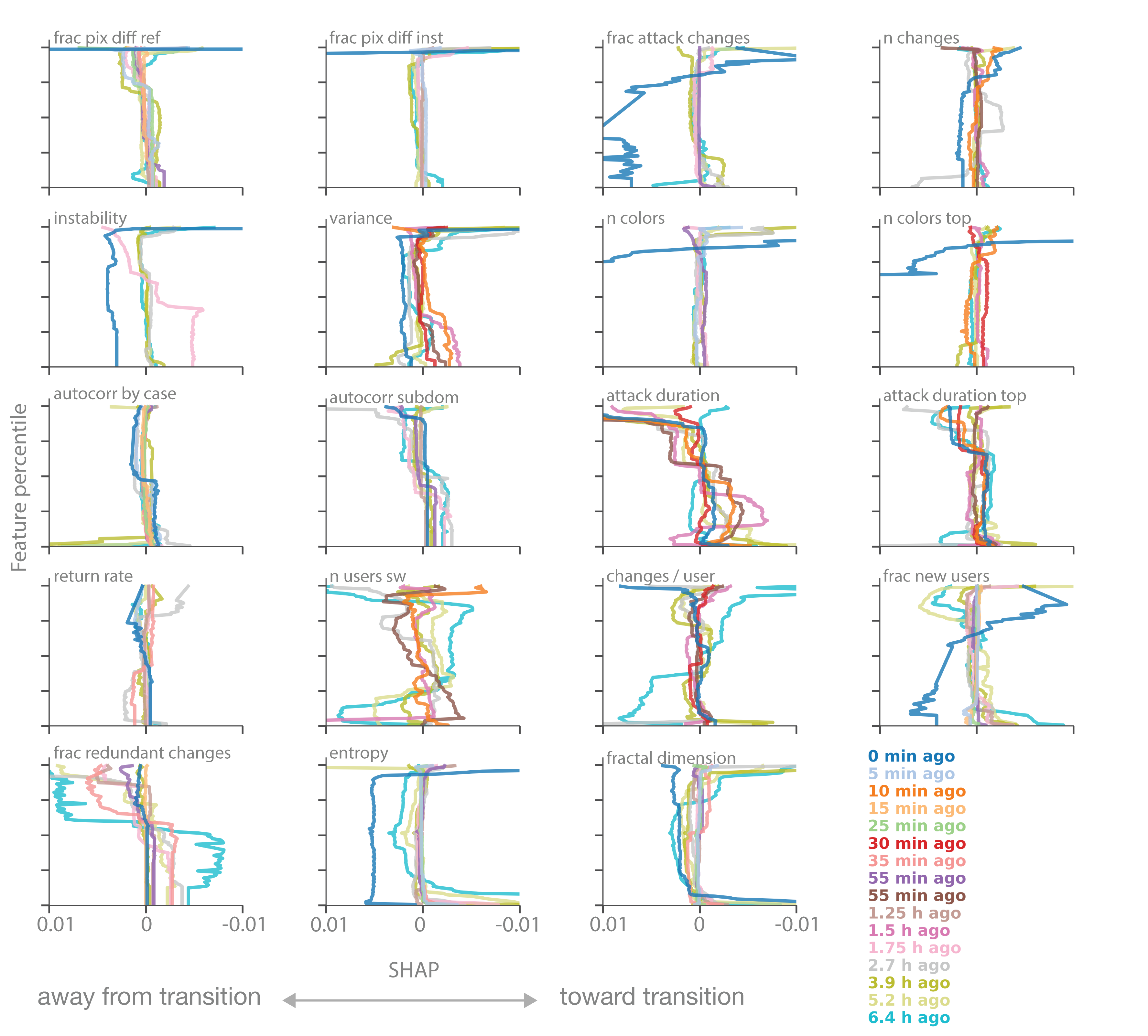}
\centering
\caption{All SHAP curves for each time-dependent variable and corresponding time features. Plots show the variable percentile against the SHAP value averaged across time instances. Plot limits were chosen to show detail at smaller values. The legend identifies time features using color.}
\label{figSI:SHAPall}
\end{figure}

\begin{figure}[!htb]
\includegraphics{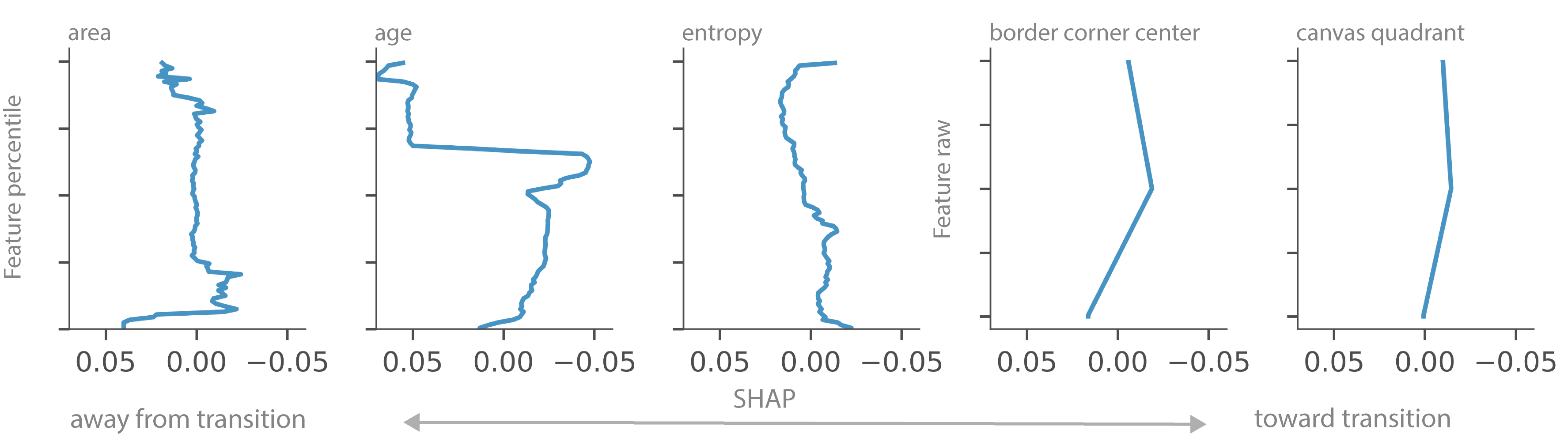}
\centering
\caption{SHAP curves for the variables that are recorded without memory. The variable percentile is plotted on the y-axis for \texttt{area}, \texttt{age} and \texttt{entropy}. The raw variable is plotted for \texttt{border corner center} and \texttt{canvas quadrant} since these two variables span only three distinct values.}
\label{figSI:SHAPnotime}

\bigskip
\bigskip
\bigskip
\bigskip

\includegraphics{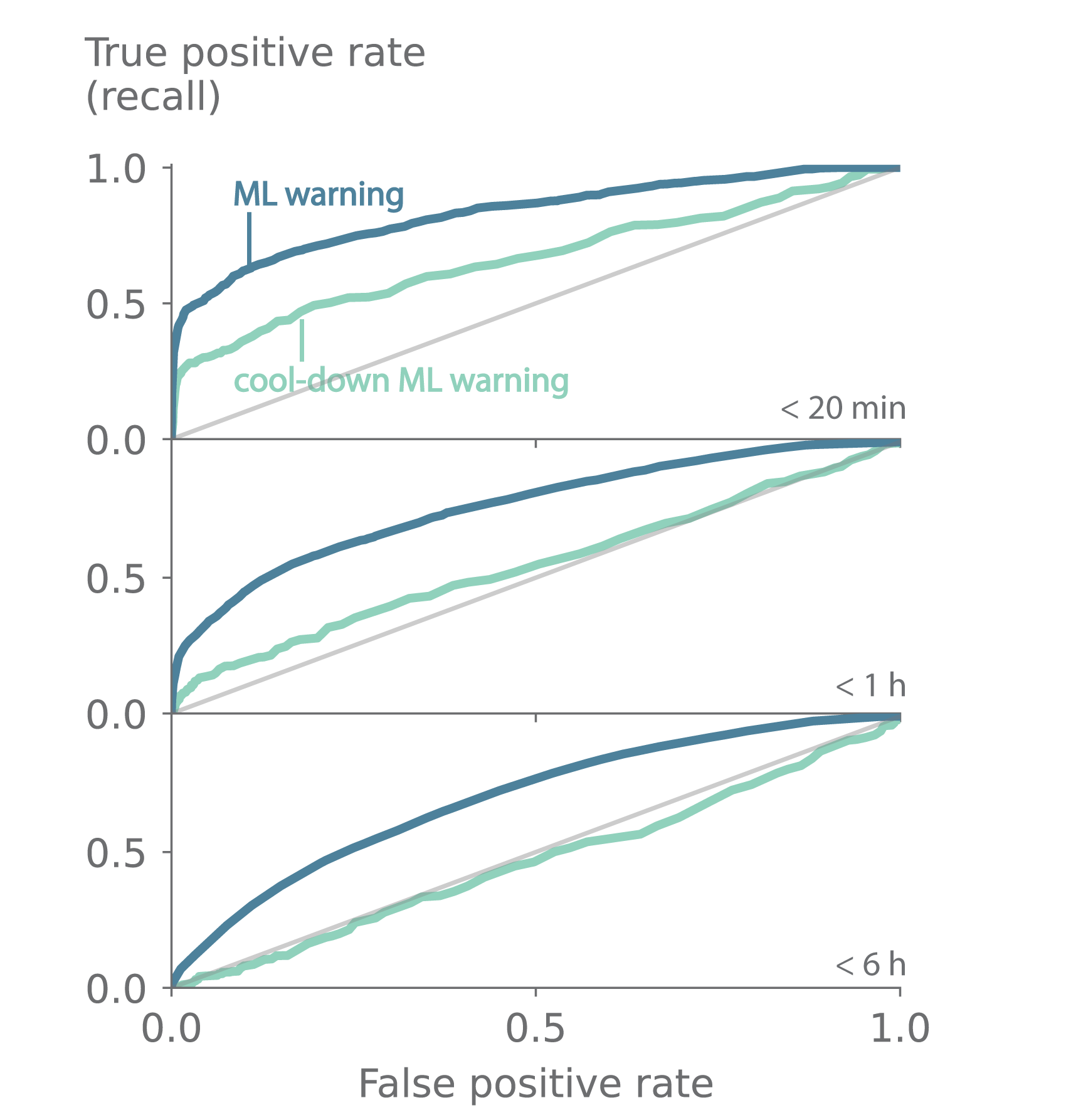}
\centering
\caption{Performance significantly decreases when using a cooldown warning system. Plots show receiver operating characteristic (ROC) curves for our cooldown machine learning warning system as compared to the machine learning warning system presented in the main text. Both warning systems are trained and tested on 2022 r/place data, at three different time thresholds for warnings.}
\label{figSI:continuous-warn}
\end{figure}

\begin{figure}[!htb]
\includegraphics{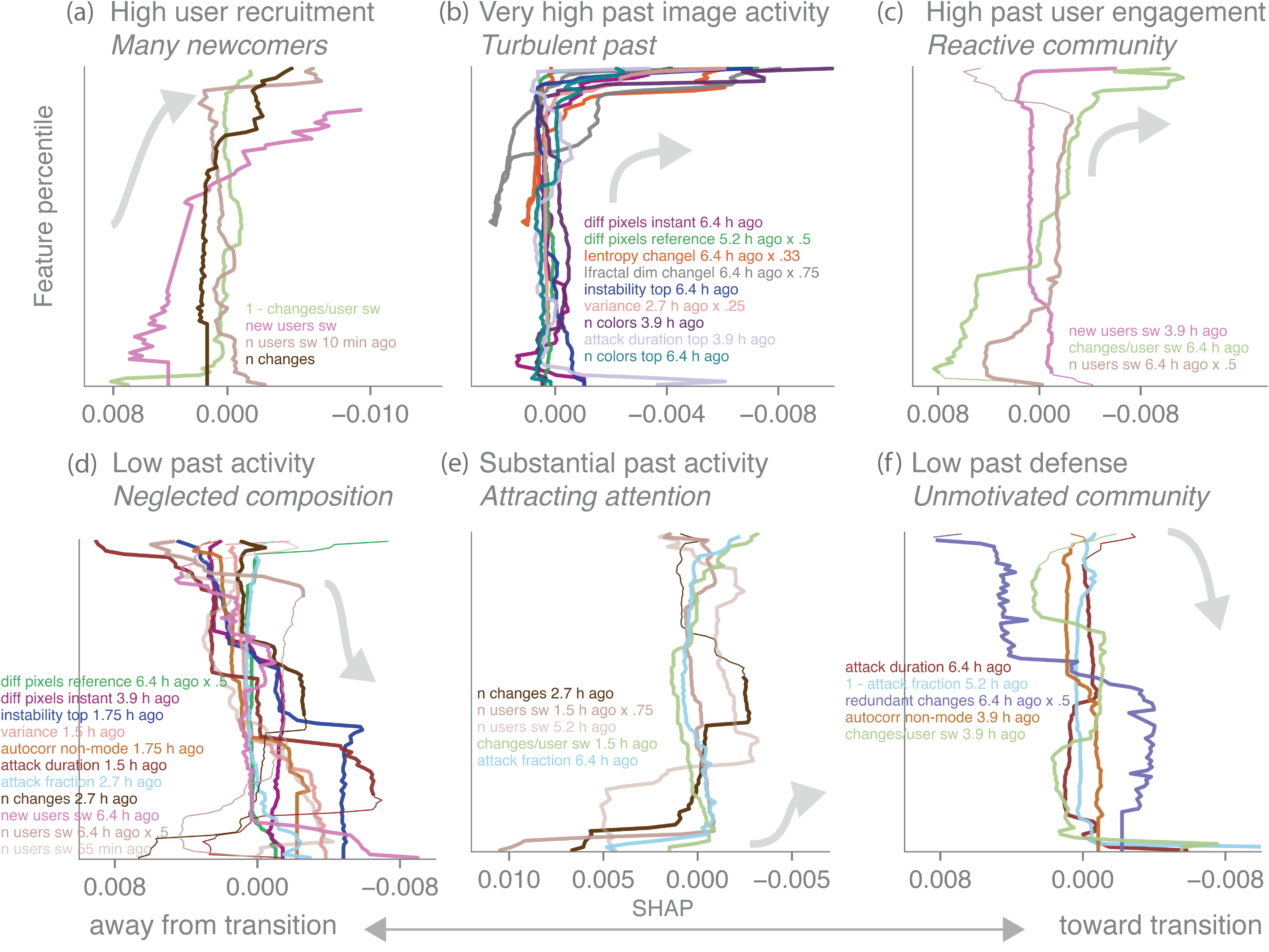}
\centering
\caption{Additional interpretation of model predictions with SHAP values. \textbf{(a-f)} Feature percentiles versus mean SHAP values, classified based on curve trends into six pre-transition behaviors, which are described in the panel title. Top titles describe the signal of a coming transition, and italicized subtitles provide 
an interpretation of the associated dynamics of users and images of the canvas. 
Grey arrows indicate the qualitative trend of the curves. The text in each legend label describes the curve of the same color as the label's text. 
Curves are drawn thinner when they show a trend that is not the focus of their respective panel.}
\label{figSI:SHAPinterp}
\end{figure}

\FloatBarrier


\bibliography{references}